\documentclass[twocolumn,preprintnumbers,amsmath,amssymb,superscriptaddress,longbibliography]{revtex4-2}
\usepackage{amsmath}
\usepackage{float}
\usepackage{amssymb}
\usepackage{subfiles}
\usepackage{bm}
\usepackage[colorlinks, linkcolor=blue, citecolor=blue, urlcolor=blue, breaklinks=true]{hyperref}
\usepackage[normalem]{ulem}
\usepackage[T1]{fontenc}
\usepackage{graphicx}
\usepackage{dcolumn}
\usepackage{amssymb,amsmath,color}
\usepackage{booktabs}
\usepackage{amsthm}
\usepackage{dsfont}
\usepackage{tcolorbox}
\usepackage{cases}
\usepackage{physics}
\usepackage{subfiles}

\usepackage{subcaption}
\usepackage{caption}
\usepackage{ragged2e}
\usepackage{braket}
\usepackage{xcolor}
\usepackage{ulem} 
\usepackage[normalem]{ulem}
\usepackage[utf8]{inputenc}

\newcommand{\commentold}[1]{}
\DeclareMathSymbol{:}{\mathpunct}{operators}{"3A}

\def\be{\begin{equation}}
\def\ee{\end{equation}}
\def\bea{\begin{eqnarray}}
\def\eea{\end{eqnarray}}

\newcommand\numberthis{\addtocounter{equation}{1}\tag{\theequation}}

\def\be{\begin{equation}}
\def\ee{\end{equation}}
\def\bal{\begin{align*}}
\def\eal{\end{align*}}
\def\bea{\begin{eqnarray}}
\def\eea{\end{eqnarray}}

\begin{document}
\date{\today}

\title{Unveiling Detuning Effects for Heat-Current Control in Quantum Thermal Devices}

\author{André H. A. Malavazi}
\email{andrehamalavazi@gmail.com}
\address{International Centre for Theory of Quantum Technologies, University of Gdańsk, Jana Bażyńskiego 1A, 80-309 Gdańsk, Poland}
\author{Borhan Ahmadi}
\email{borhan.ahmadi@ug.edu.pl}
\address{International Centre for Theory of Quantum Technologies, University of Gdańsk, Jana Bażyńskiego 1A, 80-309 Gdańsk, Poland}
\author{Paweł Mazurek}
\email{pawel.mazurek@ug.edu.pl}
\address{International Centre for Theory of Quantum Technologies, University of Gdańsk, Jana Bażyńskiego 1A, 80-309 Gdańsk, Poland}
\address{Institute of Informatics, Faculty of Mathematics, Physics and Informatics, University of Gdańsk, Wita Stwosza 63, 80-308 Gdańsk, Poland}
\author{Antonio Mandarino}
\email{antonio.mandarino.work@gmail.com}
\address{International Centre for Theory of Quantum Technologies, University of Gdańsk, Jana Bażyńskiego 1A, 80-309 Gdańsk, Poland}
\address{Department of Physics  Aldo Pontremoli, University of Milan, Via Celoria 16, 20133 Milan, Italy}

\begin{abstract}
Navigating the intricacies of thermal management at the quantum scale is a challenge in the pursuit of advanced nanoscale technologies. 
To this extent, theoretical frameworks introducing minimal models mirroring the functionality of electronic current amplifiers and transistors, for instance, have been proposed. 
Different architectures of the subsystems composing a quantum thermal device can be considered, tacitly bringing drawbacks or advantages if properly engineered.
This paper extends the prior research on thermotronics, studying a strongly coupled three-subsystem thermal device with a specific emphasis on a third excited level in the control subsystem. 
Our setup can be employed as a multipurpose device conditioned on the specific choice of internal parameters: heat switch, rectifier, stabilizer, and amplifier. 
The exploration of the detuned levels unveils a key role in the performance and working regime of the device. 
We observe a stable and strong amplification effect persisting over broad ranges of temperature. 
We conclude that considering a three-level system, as the one directly in contact with the control temperature, boosts output currents 
and the ability to operate our devices as a switch at various temperatures. 
\end{abstract}

\maketitle

\section{Introduction}
The extraction and harnessing of dissipated heat, an inevitable byproduct in energy conversion processes, is emerging as a matter of paramount concern for both governmental bodies and industries. This, driven by the constraints imposed by limited energy resources and the impact of global warming, is becoming a major issue also in the newly developing field of quantum technologies \cite{auffeves_qenergy}. In this sense, the power lost as uncontrolled heat flows poses a significant barrier to the practical implementation of quantum technologies. Despite recent signs of progress, their development is still in its infancy \cite{PRXQuantum.1.020101}, therefore major challenges taking into consideration the cost of energy sustainability still can be addressed \cite{berger2021quantum}. 
We expect more thorough thinking on how to optimize energy conversion. Therefore, 
the integration of quantum devices able to control heat fluxes 
in more complex technologies will soon become imperative \cite{PRXEnergy.1.033002, PRXEnergy.2.013002, PRXEnergy.2.033002}. Unwanted heat fluxes pose a challenge to the operation of nanoscale systems, such as superconducting quantum circuits \cite{senior2020heat}, hence the need for heat mitigation and control. Finally, the ability to design logical gates operating on heat fluxes in particular \cite{wang2007thermal, Kathmann_2020, lipkabartosik2023thermodynamic}, and the framework of Thermodynamical Computing in general \cite{conte2019thermodynamic} motivate research of heat transport in nanosystems.

An inspiration for it comes from the ability to manipulate electric signals with high precision. This capability has been of paramount importance for the development of modern classical computation and communication technologies, especially due to the possibility of designing and fabricating electronic components able to control the electric conduction at the single-electron level \cite{niu2023exceptionally}. More recently, the adept manipulation of quantum systems, coupled with the ascent of quantum thermodynamics in the creation of engines functioning at the quantum scale, has significantly fortified research endeavors aimed at exerting control over diverse forms of energy exchange \cite{binder2018thermodynamics}. Within this trajectory, the prospect of governing heat flows emerges as notably auspicious, prompting a myriad of concentrated efforts directed towards both the theoretical exploration and experimental realization of quantum heat transport \cite{pekola2015towards, pekola2019thermodynamics, pekola2021colloquium}. Thus, it is clear that understanding and modulating energy flows in realistic physical scenarios are of vital importance from a technological perspective \cite{garrido2001simple,dubi2011colloquium}. 

The above ideas lay the foundations for the new field of \textit{thermotronics}, focused on 
the design and development of quantum thermal devices \cite{bhattacharjee2021quantum,myers2022quantum}, e.g., heat valves \cite{ronzani2018tunable, PRL_heatValve}, heat rectifiers \cite{terraneo2002controlling,scheibner2008quantum, werlang2014optimal, Markos_2018,senior2020heat, PhysRevE.99.042121, PhysRevE.106.034116, PhysRevE.104.054137,poulsen2022entanglement}, quantum heat engines \cite{ahmadi2021irreversible,PhysRevLett.2.262,quan2007quantum,quan2009quantum,rossnagel2016single,peterson2019experimental,lobejko2020thermodynamics}, quantum batteries \cite{alicki2013entanglement,binder2015quantacell, campaioli2018quantum,lipka2021second, cruz2022quantum,rodriguez2023optimal,rodriguez2023catalysis,ahmadi2024nonreciprocal,quach2022superabsorption, PhysRevA.106.042601,kamin2020non,kamin2023steady}, etc. The overarching goal revolves around the meticulous control of energy fluxes, manifesting as both work and heat, with a specific focus on executing predefined tasks. This comprehensive pursuit signifies a paradigm shift in the manipulation and application of quantum phenomena for practical energy-related objectives. Along these lines, a transistor is a device designed for regulating, controlling, and amplifying electric currents through separated terminals. As an analogy, a device for regulating and controlling heat flows was proposed in \cite{10.1063/1.2191730}. More recently, the authors of \cite{joulain2016quantum} suggested a quantum thermal transistor. Since then, many theoretical proposals of architectures and contexts have been suggested and analyzed in the literature \cite{PhysRevE.99.032112, PhysRevA.97.052112,yang2020quantum, PhysRevB.101.245402, PhysRevB.104.045405,liu2021common,ghosh2021quantum,mandarino2021thermal,du2019quantum,yang2020quantum}. In particular, in \cite{guo2018quantum, PhysRevE.99.032112} it was shown how relatively simple architectures consisting of strongly coupled subsystems can operate with distinct functionalities, such as transistors, switches, valves, and rectifiers. As a first step to considering more realistic scenarios, one can assume more complex energetic structures. In this sense, the use of qutrits has already been considered in the literature for modeling quantum thermal transistors \cite{su2018quantum,guo2018quantum, PhysRevE.99.042102,wang2020polaron, majland2020quantum,wang2022cycle} and other devices, such as quantum heat engines \cite{PhysRevLett.2.262, PhysRevLett.98.240601,PhysRevA.96.063806,diaz2021qutrit}. 

Our work is motivated by the fact that all effective models for implementing two-level systems (TLS) are constrained to particular regions of the parameter space. This provides a limited description of the system under scrutiny, which can lead to inaccuracies, poor performance, and/or suppression of non-trivial effects, particularly in scenarios where the excitation of higher energy levels cannot be neglected. To address this, our investigation incorporates an additional level in the subsystem serving as a controlling element directly connected to a bath with variable temperature. In particular, our proposed architecture consists of two TLS and a qutrit strongly coupled working as terminals of the quantum transistor. The former constitute the left and right terminals of the device, while the latter plays the role of the middle one. The focus of our analysis will be the third excited level. Such increment will considerably increase the energetic intricacy of the device, allowing the emergence of some non-trivial effects. Our findings reveal the feasibility of attaining the desired transistor effect across a broad temperature spectrum, accompanied by an additional peak in amplification. Notably, we demonstrate the potential to engineer or modulate the device's operational range by leveraging the gap between the second and third excited levels. 
We also show that the addition of another degree of freedom can enhance the output heat currents and enable our device to function as a powerful heat switch with adjustable working temperatures. 

The structure of this paper unfolds as follows: Section \ref{MODEL} introduces the physical model for the quantum thermal device; Section \ref{Dynamics} looks into the system's dynamics, its steady state, and the heat currents coursing through the terminals; Section \ref{Performance} outlines the relevant figures of merit and general properties; Section \ref{Results} presents the numerical outcomes elucidating the system's behavior; while a comprehensive discussion of the results is presented in Section \ref{Discussion}.
\section{The Model}\label{MODEL}
\begin{figure}[H]
\includegraphics[width=10cm]{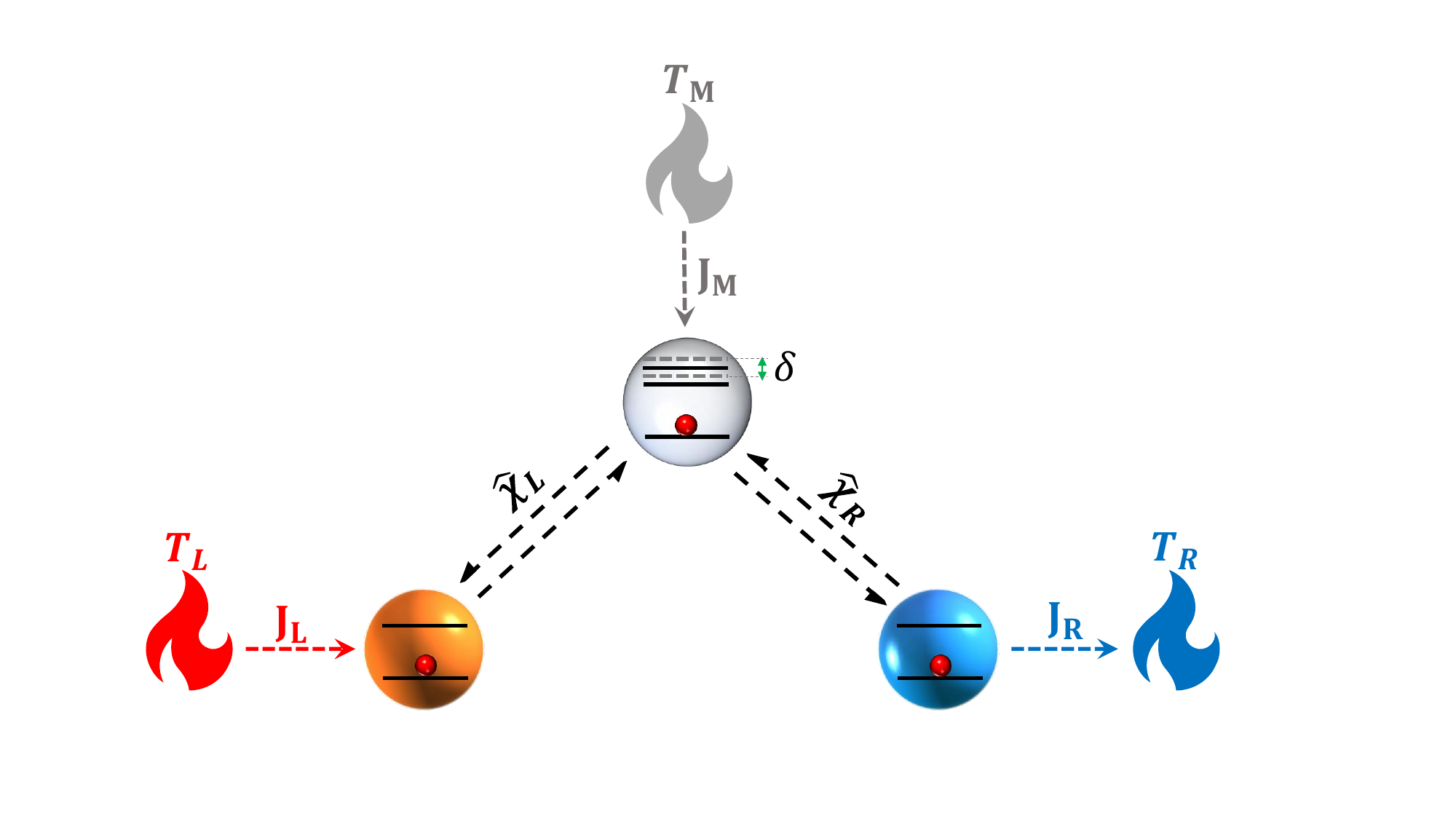}
\caption{Qubit-qutrit-qubit architecture for a quantum thermal device, consisting of two TLSs interacting independently with a qutrit. Each terminal is interacting with a different thermal reservoir with distinct temperatures $T_{\alpha}$.}
\label{FigModel}
\end{figure}

Our physical model is depicted in Fig. \ref{FigModel} and consists of three interacting subsystems, working as the device's terminals, such that each one is coupled to a different thermal reservoir with a distinct temperature $T_{\alpha}$ to avoid cross dissipation \cite{galve2017microscopic}. The left and right systems - referred to as $R$ and $L$ - are TLSs and the middle system is modeled as a qutrit. The total Hamiltonian describing the free evolution of our thermal device reads
\begin{equation}\label{Transistor Hamiltonian}
\hat{H}=\hat{H}_{L}+\hat{H}_{M}+\hat{H}_{R}+\hat{V},
\end{equation}
consisting of the sum of the local free Hamiltonians, $\hat{H}_{L, M, R}$, plus the coupling term $\hat{V}$ encompassing the interactions between the elements $L$ and $M$, and $R$. The free Hamiltonians of the two TLSs are written as
\begin{equation}\label{TLSHamiltonians}
  \hat{H}_{\alpha} = \hbar\omega_{\alpha}|e\rangle_{\alpha}\langle e|_{\alpha},\ \quad \alpha=L,R,
\end{equation}
where $|e(g)\rangle_{\alpha}$ is the excited (ground) state with energy $\hbar\omega_{\alpha} $($0$). The qutrit's Hamiltonian is given by
\begin{equation}\label{QutritHamiltonians}
\hat{H}_{M}	= \hbar\Omega|1\rangle\langle1|+\hbar(\Omega+\delta)|2\rangle\langle2|,
\end{equation}
where $\{|j\rangle, j=0,1,2\}$ is its basis with respective energies $\{0,\hbar\Omega,\hbar(\Omega+\delta)\}$. Finally, the TLS-qutrit coupling is modeled as
\begin{equation}\label{Coupling}
\hat{V}=\hbar\hat{\sigma}_{L}^{z}\otimes\hat{\chi}_{L}+\hbar\hat{\chi}_{R}\otimes\hat{\sigma}_{R}^{z},
\end{equation}
where $\hat{\sigma}_{L,R}$ are the standard $z$-Pauli matrices, and $\hat{\chi}_{L,R}=\sum_{j>i}\chi_{ij}^{L,R}\left(|i\rangle\langle i|-|j\rangle\langle j|\right)$ is the qutrit operator connecting the states $|i\rangle\leftrightarrow|j\rangle$, with internal coupling strengths $\chi_{ij}^{L,R}\in\mathbb{R}$. Thus, it is clear that $\forall \alpha:$ $[\hat{V},\hat{H}_{\alpha}]=0$.
Notice that, in general, the system contains 12 distinct energy levels and, by construction, $\hat{H}$ is diagonal in the composition of the local bases $|E_{ljr}\rangle\equiv|l\rangle_{L}|j\rangle|r\rangle_{R}$, with $l,r=g,e$ and $j=0,1,2$. Therefore, obtaining the 12 values composing the energy spectrum is straightforward (see Appendix \ref{Energy spectrum}) and one can easily compute all possible energy gaps $\omega$ relative to different transitions. Throughout this work, it will be assumed that there is no energy gap degeneracy (further details are given in Appendix \ref{Bohr frequencies}).
\subsection*{Thermal reservoirs}
It is assumed that each terminal $\alpha=L, M, R$ of the system of interest is weakly coupled to a thermal reservoir with temperature $T_{\alpha}$. All reservoirs are separated and modeled as infinite collections of non-interacting harmonic oscillators, such that $\hat{\mathcal{H}}_{\alpha}\equiv\hbar\sum_{k}\xi_{\alpha k}\hat{b}_{\alpha k}^{\dagger}\hat{b}_{\alpha k}$, where $\hat{b}_{\alpha k}$ and $\hat{b}_{\alpha k}^{\dagger}$ are the annihilation and creation bosonic operators satisfying $[\hat{b}_{\alpha},\hat{b}_{\beta }^{\dagger}]=\delta_{\alpha,\beta}$. From now on, it will be assumed that $T_{L}>T_{R}$. The interaction between the subsystem $\alpha$ and its relative bath is given by
\begin{equation} \label{BathInt}\hat{h}_{\alpha}=\hat{\sigma}_{\alpha}^{x}\otimes\mu_{\alpha}\hbar\sum_{k}\left(c_{\alpha k}^{*}\hat{b}_{\alpha k}^{\dagger}+c_{\alpha k}\hat{b}_{\alpha k}\right),
\end{equation}
where $\mu_{\alpha}c_{\alpha k}$ is the interaction strength between subsystem $\alpha$ and the $k$-th mode of its respective reservoir, $\hat{\sigma}_{L, R}^{x}$ are the $x$-Pauli matrices, and $\hat{\sigma}_{M}^{x}=\sum_{j>i}\left(|i\rangle\langle j|+|j\rangle\langle i|\right)$.
Moreover, when examining the time-evolution of the system of interest, these interactions are fully characterized by the spectral densities of the reservoirs. In this study, we consider Ohmic spectral densities with a soft cut-off for all $\alpha$, defined as $2\pi J^{\alpha}(\xi)=\xi\exp(-\xi/\kappa_{\alpha})$, where $\kappa_{\alpha}$ represents the cut-off frequency. It is important to note that different spectral densities, such as sub- or super-Ohmic, may influence the system's behavior (refer to Ref. \cite{mandarino2022quantum} for more details).
\section{Dissipative dynamics}\label{Dynamics}
The non-unitary time-evolution of the state of the total system $\hat{\rho}(t)$ can be fully described by the following GKLS master equation \cite{gorini1976completely}:
\begin{equation}\label{Master Equation}
\frac{d}{dt}\hat{\rho}(t)=\mathcal{L}[\hat{\rho}(t)]=-\frac{i}{\hbar}\left[\hat{H},\hat{\rho}(t)\right]+\sum_{\alpha=L,M,R}\mathcal{D}_{\alpha}[\hat{\rho}(t)],
\end{equation}
where the structure
\begin{align*}\label{Dissipator}
\mathcal{D}_{\alpha}[\boldsymbol{\cdot}]:=\sum_{\omega}\gamma_{\omega}^{\alpha}\left(\hat{S}_{\omega}^{\alpha}\boldsymbol{\cdot}\hat{S}_{\omega}^{\alpha\dagger}-\frac{1}{2}\left\{ \hat{S}_{\omega}^{\alpha\dagger}\hat{S}_{\omega}^{\alpha},\boldsymbol{\cdot}\right\}\right)
\end{align*}
is the dissipator relative to the interaction with the reservoir $\alpha$ and
\begin{equation}\label{decayR}
    \gamma^{\alpha}_{\omega}=\mu_{\alpha}^{2}\hbar^{-1}\omega\exp(-|\omega|\hbar^{-1}/\kappa_{\alpha})\left(1+\bar{N}_{\alpha}(\omega\hbar^{-1})\right)
\end{equation}
their respective induced decay rates, satisfying $\gamma^{\alpha}_{-\omega}=\exp(-\beta_{\alpha}\omega)\gamma^{\alpha}_{\omega}$, with $\bar{N}_{\alpha}(\xi)=\left(\exp(\beta_{\alpha}\hbar\xi)-1\right)^{-1}$. $\hat{S}_{\omega}^{\alpha}$ is the set of jump operators acting on the system of interest s.t. $\hat{S}^{\dagger}_{\omega}=\hat{S}_{-\omega}$.

Such an expression is derived under the assumptions commonly applied to open quantum systems, namely, weak coupling, Born-Markov, and full secular approximations \cite{Breuer,cattaneo2019local}. Taking this into account, it is important to emphasize that (i) as mentioned earlier, no energy gap degeneracy is assumed, i.e., all pertinent Bohr frequencies $\omega$'s (Appendix \ref{Bohr frequencies}) are distinct. This is guaranteed by proper tuning of the relevant parameters, and one should be aware of eventual crossing points; (ii) given the first assumption, the jump operators $\hat{S}^{\alpha}_{\omega}$ are computed assuming the so-called global approach (also referred to as strong-coupling formalism \cite{werlang2014optimal,joulain2016quantum}). 
In contrast to the local approach, the jump operators are computed from the full Hamiltonian and, therefore, act on the composite Hilbert spaces (see Appendix \ref{Jump operators} for further detail). Such an approach provides the proper formalism for considering non-negligible or strong internal couplings between the subsystems.
Also, it is worth mentioning that the Lamb-shift component is suppressed in Eq. \eqref{Master Equation}, since it does not play any determinant role in the energetic exchanges at the stationary state.
\subsection{Derivation of the nonequilibrium steady states}
All relevant quantities can be written in terms of the density matrix's populations in the energy basis, represented by
$P_{ljr}(t)=\langle|E_{ljr}\rangle\langle E_{ljr}|\rangle(t)$. 
Starting from the general form of the master equation in Eq. \eqref{Master Equation}, 
one can show that the dynamics of the diagonal (populations) and non-diagonal elements (coherences) of $\hat{\rho}$ are decoupled in such a way that:
\begin{align*}\label{Populations}
        \frac{d}{dt}P_{ljr}(t)&=\left(\delta_{l,e}-\delta_{l,g}\right)\Gamma_{gjr,ejr}^{L}(t)\\&+\sum_{\eta>\mu}\left(\delta_{j,\eta}-\delta_{j,\mu}\right)\Gamma_{l\mu r,l\eta r}^{M}(t)\\&+\left(\delta_{r,e}-\delta_{r,g}\right)\Gamma_{ljg,lje}^{R}(t),\numberthis
\end{align*}
where $\delta_{i,j}$ is the usual Kronecker delta and
\begin{equation}\label{transition rates}
    \begin{split}
        \Gamma_{gjr,ejr}^{L}(t)&=\gamma^{L}_{-\omega_{ge,j}^{L}}P_{gjr}(t)-\gamma^{L}_{\omega_{ge,j}^{L}}P_{ejr}(t),
        \\\Gamma_{l\mu r,l\eta r}^{M}(t)&=\gamma^{M}_{-\omega_{\mu\eta,lr}^{M}}P_{l\mu r}(t)-\gamma^{M}_{\omega_{\mu\eta,lr}^{M}}P_{l\eta r}(t),\\
        \Gamma_{ljg,lje}^{R}(t)&=\gamma^{R}_{-\omega_{ge,j}^{R}}P_{ljg}(t)-\gamma^{R}_{\omega_{ge,j}^{R}}P_{lje}(t)
    \end{split}
\end{equation}
quantifies the different transition rates between the states $|l\rangle_{L}|j\rangle|r\rangle_{R}\leftrightarrow|l^{\prime}\rangle_{L}|j^{\prime}\rangle|r^{\prime}\rangle_{R}$ induced by the thermal reservoirs. Notice that the expression (\ref{Populations}) can also be written in a more compact way as a continuous-time Markov chain, by defining the population vector $\mathbf{P}=(P_{g0g},..., P_{e2e})^{T}$ and the transition matrix $\mathbf{X}$, such that $\frac{d}{dt}\mathbf{P}=\mathbf{X}\mathbf{P}$, where $\sum_{j}\mathbf{X}_{ij}=1$ for all $i$.
Nevertheless, we are generally interested in the device's behavior at the long time limit, i.e., when the system of interest reaches a non-equilibrium stationary state (NESS) (one can also focus on the transient regime, see Ref. \cite{ghosh2021quantum}). This is characterized by the set of stationary populations $P_{ljr}^{SS}$, such that
\begin{equation}\label{SteadyState}
    \frac{d}{dt}P_{ljr}^{SS}=0
\end{equation}
for all $l$, $j$ and $r$, or simply $\mathbf{X}\mathbf{P}^{SS}=\mathbf{0}$.
\subsection{Heat Currents}
The system's internal energy is identified as the expectation value of its Hamiltonian, i.e., $\langle\hat{H}\rangle(t)=\Tr{\hat{H}\hat{\rho}(t)}$ \cite{Alicki, Gemmer, ahmadi2023work}. Since the changes in energies are necessarily induced by the interaction with reservoirs one can classify them as \textit{heat}. Thus, the energetics is fully characterized by
\begin{equation}\label{Total energy change}
\frac{d}{dt}\langle\hat{H}\rangle(t)=\sum_{\alpha=L,M,R}J_{\alpha}(t),
\end{equation}
where
\begin{equation}\label{Heat current}
J_{\alpha}(t):=\Tr\left\{ \hat{H}\mathcal{D}_{\alpha}[\hat{\rho}(t)]\right\} 
\end{equation}
is the heat current flowing through terminal $\alpha$. However, Eq. (\ref{Heat current}) can be cast as the sum of the individual contributions $\mathcal{J}_{ljr,l^{\prime}j^{\prime}r^{\prime}}^{\alpha}$ due to different state transitions $|l\rangle_{L}|j\rangle|r\rangle_{R}\leftrightarrow|l^{\prime}\rangle_{L}|j^{\prime}\rangle|r^{\prime}\rangle_{R}$, i.e.,
\begin{equation}
    \begin{split}
        J_{L}(t)&=\sum_{j=0}^{2}\sum_{r=g,e}\mathcal{J}_{gjr,ejr}^{L}(t),\\J_{M}(t)&=\sum_{j>i}\sum_{l,r=e,g}\mathcal{J}_{lir,ljr}^{M}(t),\\J_{R}(t)&=\sum_{j=1}^{2}\sum_{l=g,e}\mathcal{J}_{ljg,lje}^{R}(t),
    \end{split}
\end{equation}
where
\begin{equation}\label{local currents}
    \begin{split}
        \mathcal{J}_{gjr,ejr}^{L}(t)&:=\omega_{ge,j}^{L}\Gamma_{gjr,ejr}^{L}(t),\\\mathcal{J}_{lir,ljr}^{M}(t)&:=\omega_{ij,lr}^{M}\Gamma_{lir,ljr}^{M}(t),\\\mathcal{J}_{ljg,lje}^{R}(t)&:=\omega_{ge,j}^{R}\Gamma_{ljg,lje}^{R}(t).
    \end{split}
\end{equation}
Notice that once the steady state is reached, the total heat current flowing through the transistor must be null, i.e.,
\begin{equation}\label{tth}
    \sum_{\alpha=L,M,R}J^{ss}_{\alpha}=0.
\end{equation}
From now on, all heat currents will be assumed to be in the steady state thus, to simplify the notation, we will drop the $SS$ superscript shown above.
\section{Thermal properties: sensitivity, heat amplification, switch, and rectification}\label{Performance}
The evaluation of the performances of a thermal device acting as a multi-purpose tool requires the introduction 
of several figures of merit able to gauge the different regimes. In the following, we introduce and briefly discuss  
those that will employed in characterizing our device in its different modes of operation. 
\subsection{Differential thermal sensitivity}
The behavior of the physical system is characterized in terms of the control temperature $T_M$. To examine the response of the terminal's heat currents over small changes of $T_M$, we introduce the differential thermal sensitivity \cite{yang2019thermal, PhysRevB.104.045405, PhysRevB.108.235421}, defined as 
\begin{equation}\label{sensitivity}
    \mathcal{S}_{\alpha} := \partial_{T_{M}}J_{\alpha}, \qquad \alpha =L,M,R
\end{equation}
where $\partial_{T_{M}}J_{\alpha}:=\partial J_\alpha/\partial T_M$. 
%
%
%
%
%
%
\subsection{Amplification factor}
For an efficient thermal transistor operation, one ideally expects a working regime with (i) stable $J_{M}$ current and strong $J_{L, R}$ amplification, relative to the control parameter $T_{M}$, i.e., respective low and high thermal sensitivities; (ii) no ``wasted'' heat passing through terminal $M$, such that we have $J_{L}\approx-J_{R}$. Notice that condition (ii) can be achieved by controlling the qutrit-reservoir interaction, either by changing the coupling amplitudes $\mu_{m}$ or engineering the spectral density accordingly. Such control would directly affect the relevant decay rates, which would also influence the transition rates $\Gamma_{l\mu r,l\eta r}^{M}$ (Eq. (\ref{transition rates})) and, therefore, the individual heat currents $\mathcal{J}_{lir,ljr}^{M}$ (Eq. (\ref{local currents})). However, for fixed reservoir interaction and spectral density, one may still be able to find suitable parameters that render functioning devices, i.e., when the so-called \textit{transistor effect} is achieved. The transistor effect stands for a high response of the $J_{L}\approx -J_{R}$ currents to changes of the controlling temperature $T_{M}$, compared to the response of the $J_{M}\approx 0$ current, provided both responses are quasi-linear. Should these conditions be satisfied, the relevant figure of merit is the so-called amplification factor, defined as the ratio
\begin{equation}\label{AmpFactor}
    \alpha_{L,R}:=\frac{\partial_{T_{M}}J_{L,R}}{\partial_{T_{M}}J_{M}}=\frac{\mathcal{S}_{L, R}}{\mathcal{S}_{M}}.
\end{equation}
Thus, the thermal transistor effect is captured for $|\alpha_{L, R}|\gg1$. This happens whenever the qutrit's thermal sensitivity $\mathcal{S}_{M}$ reaches values close to zero for finite $\mathcal{S}_{L, R}$. Note that given Eq. \eqref{tth} the equality $\alpha_{L}+\alpha_{R}=-1$ is satisfied. Temperature regimes in which the transistor effect is described by high amplification factors are suitable for fine, continuous control of currents flowing through the system, with simultaneous minimization of heat leakage through the control.
\subsection{Heat switch}
Alternatively, the system may provide a mode of operation in which the currents $J_{L}\approx -J_{R}$ (with $J_{M}\approx 0$) take values that are either substantial or close to zero, for two control temperatures $T_{M}$, respectively. Switching between these two temperatures would turn on/off the heat flow, realizing the so-called \textit{heat switch} \cite{10.1063/1.2191730}. Provided this happens, the ratio of on/off state currents quantifies the switch performance.
\subsection{Rectification}
Another mode of operation of heat devices is the rectifier. It was proposed with the goal of efficient thermal management and potential applications in heat modulation technologies \cite{PhysRevE.99.032112, Sánchez_2015, Markos_2018, Tesser_2022, Palafox_2022}. It is characterized by strong suppression of the heat current when the temperature gradient between the two terminals of the device is reversed. In the limit $\hat{\chi}_{R}\rightarrow0$, the right TLS becomes effectively decoupled from the rest of the heat-amplifier system, and the device operates in the two-terminal mode. Anticipating the vanishing of the current $J_{R}$, we use rectification factors to quantify the asymmetry in heat currents between the remaining two terminals, when reversal of the temperature gradient is applied \cite{Markos_2018}: 
\begin{equation}\label{eta}
    \mathcal{R}(\Delta) := \frac{|J(\Delta)|-|J(-\Delta)|}{|J(\Delta)|+|J(-\Delta)|},
\end{equation}
where $J$ is the stationary heat current floating through the device, and we have $|J|=|J_{M}|=|J_{L}|$. $\Delta:=T_{M}-T_{L}$ is the temperature difference between the two terminals, and we will investigate two ways of imposing it. Namely, $\mathcal{R}_{T_{L}}$ 
refers to the situation when the temperature difference is imposed by fixing $T_{L}$ to a specific value, while for $\mathcal{R}_{\overline{T}}$ it is the average $\overline{T}:=\frac{T_{M}+T_{L}}{2}$ that is kept constant. While $\mathcal{R}_{T_{L}}$ may be more suitable for quantifying the potential of blocking heat flow in an unwanted direction resulting from temperature fluctuation in one of the baths, $\mathcal{R}_{\overline{T}}$ focuses on comparison of the device functioning in different temperature regimes. 
\section{Results}\label{Results}
As mentioned earlier, the introduction of the qutrit increases the complexity of the system. This is evident not only in the alteration of its spectrum and dimension but also in the significant expansion of the parameter space due to additional interaction possibilities. On one hand, this complexity leads to a more intricate dynamic and energetic behavior of the system. Consequently, characterizing and identifying the relevant transitions for achieving the desired outcome can be challenging. On the other hand, as demonstrated below, one can leverage this complexity to create various operational regimes and observe a broader range of effects.
In principle, this intricacy allows for tailoring the relevant parameters to meet specific needs and design requirements. In the upcoming sections, we will numerically solve the set of equations for the stationary state provided by Eq. \eqref{SteadyState}, calculating heat currents and relevant figures of merit. For these calculations, we assume $\hbar=1$, $k_{B}=1$, $\mu_{\alpha}=0.01$, and a cut-off of $\kappa_{\alpha}=30$ for all reservoirs.

It is essential to emphasize that the stationary effects presented in this study are anticipated to persist across the entire range of strong internal coupling. They are not confined to the parameter values chosen below for illustrative purposes. Specifically, we anticipate that heat currents will exhibit similar temperature characteristics, remaining invariant when both the couplings with the environment $\mu_{\alpha}$ and the system Hamiltonian $\hat{H}$ are simultaneously scaled, along with the rescaling of bath temperatures and cut-offs $\kappa_{\alpha}$. This rescaling, in turn, only influences the magnitudes of the heat currents.
\begin{figure}[H]
\includegraphics[width=8.5cm]{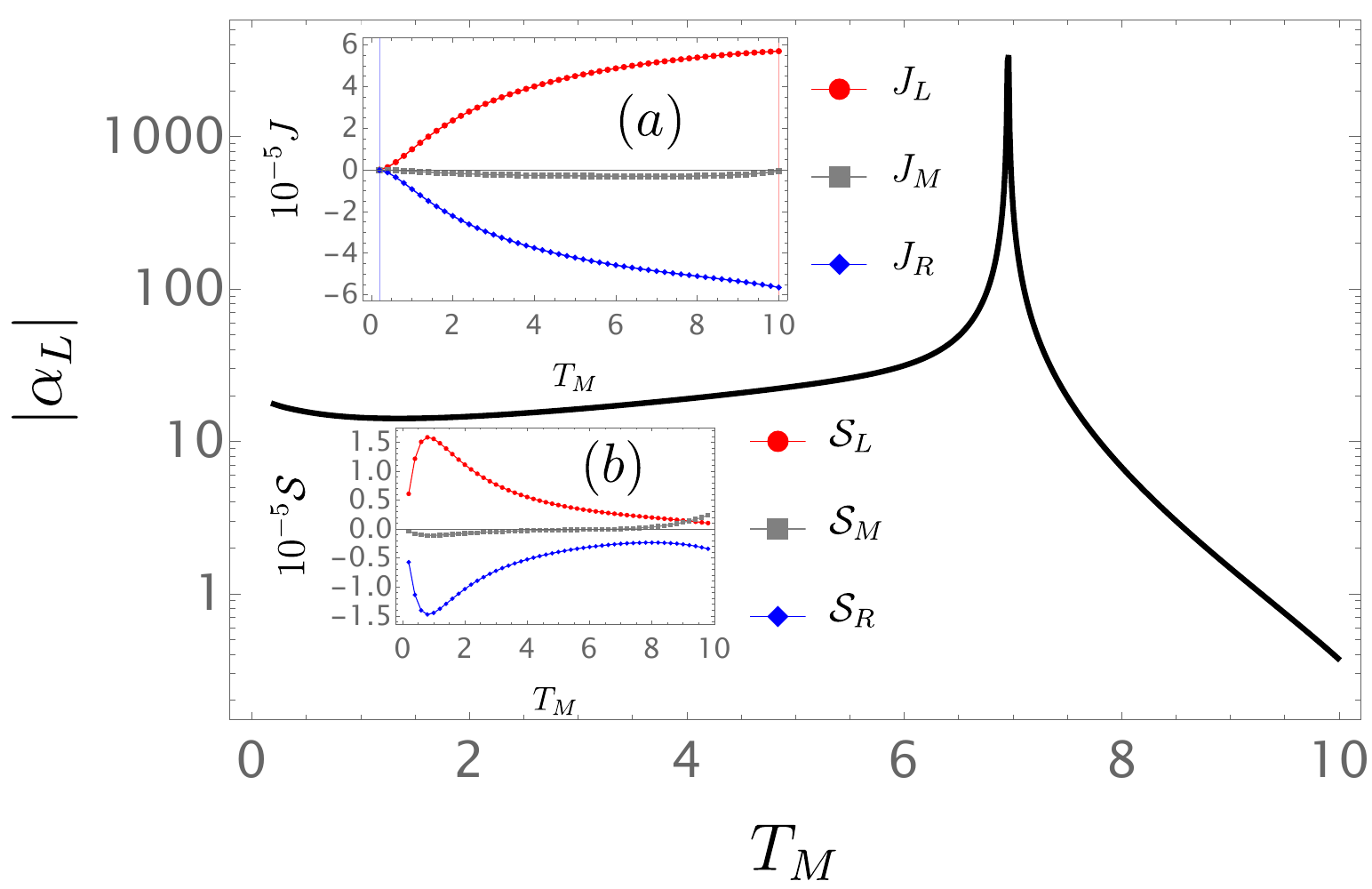}
\caption{Quantum thermal transistor behavior. The main plot depicts the amplification factor $|\alpha_{L}|$ in relation to the temperature $T_{M}$, showcasing the characteristic transistor effect. Inset: $(a)$ illustrates the stationary heat currents $J_{\alpha}$ through their respective terminals, with vertical blue and red lines denoting the temperatures of the right and left reservoirs; $(b)$ presents the differential thermal sensitivity for all terminals. Parameters: $\omega_{L}=1$, $\omega_{R}=2\omega_{L}$, $\Omega=3\omega_{L}$, $\delta=3\omega_{L}$, $\chi_{01}^{L}=15\omega_{L}$, $\chi_{01}^{R}=16\omega_{L}$, $\chi_{02}^{L}=18\omega_{L}$, $\chi_{02}^{R}=15\omega_{L}$, $\chi_{12}^{L, R}=0.1\omega_{L}$, $T_{R}=0.2$, $T_{L}=10$. We have $||\hat H||= 103.2 $}
\label{Fig1}
\end{figure}
\subsection{Thermal transistor behavior and heat switch}\label{Thermal transistor behavior}
%
%
%
%
As a proof of concept of our setting, Fig. \ref{Fig1} shows the standard expected behavior for a quantum thermal transistor assuming $T_{R}=0.2$ and $T_{L}=10$ for the reservoir's temperatures, and strong internal couplings, such that $||\hat{V}||>||\hat{H}_{\alpha}||\gg\mu_{\alpha}$. In the inset $(a)$ of Fig. \ref{Fig1} we present the stationary heat currents passing through the terminals for a wide interval of $T_{M}$, ranging from the right to the left reservoir's temperature (blue and red vertical lines, respectively), i.e., $T_{M}\in[T_{R}, T_{L}]$. As demanded for the effective functioning of the device as a transistor, one observes the stability for the thermal current $J_{M}$ close to zero, while $|J_{L, R}|$ increases. The inset $(b) $ illustrates such response of the currents over the control temperature through the differential thermal sensitivity $\mathcal{S}_{L, M, R}$, i.e.,  it shows a very low sensitivity of $J_{M}$ for most of the considered interval, while $\mathcal{S}_{L, R}$ presents an initial increase for low temperatures and a subsequent decrease, accompanied by a rise of $\mathcal{S}_{M}$ for temperatures closer to $T_{L}$. The main plot of Fig. \ref{Fig1} shows the characteristic signature of the transistor effect, i.e., $|\alpha_{L}|\gg1$ and a peak in the amplification factors (the analogous behavior of $|\alpha_{R}|$ is not reported). As suggested by Eq. \eqref{AmpFactor}, the peak appears exactly in the region where $\mathcal{S}_{M}\approx 0$ for $\mathcal{S}_{L }\neq 0$, around $T_{M}\approx 6.96$. The temperature region for which the amplification factor takes large values, and  $|J_{M}|\ll |J_{L}|, |J_{R}|$, is characterized by $T_{M}<T_{H}$. For $T_{M}\approx T_{H}$, the current $J_{M}$ starts to dominate (see Appendix \ref{Asymtotics} for details). For completeness, the behavior of the fully qubit transistor architecture is presented in Appendix \ref{QUBIT}. We conclude that, within the investigated range of parameters, the qutrit device outperforms its qubit analogous. In particular, the former steady heat currents are approximately twice the latter, providing a straightforward advantage.
%
%
%
%
%
%
%
%
%
%
%
%
%
%
%
%
%
%
%
%
\begin{figure}
\centering
\includegraphics[width=0.35\textwidth]{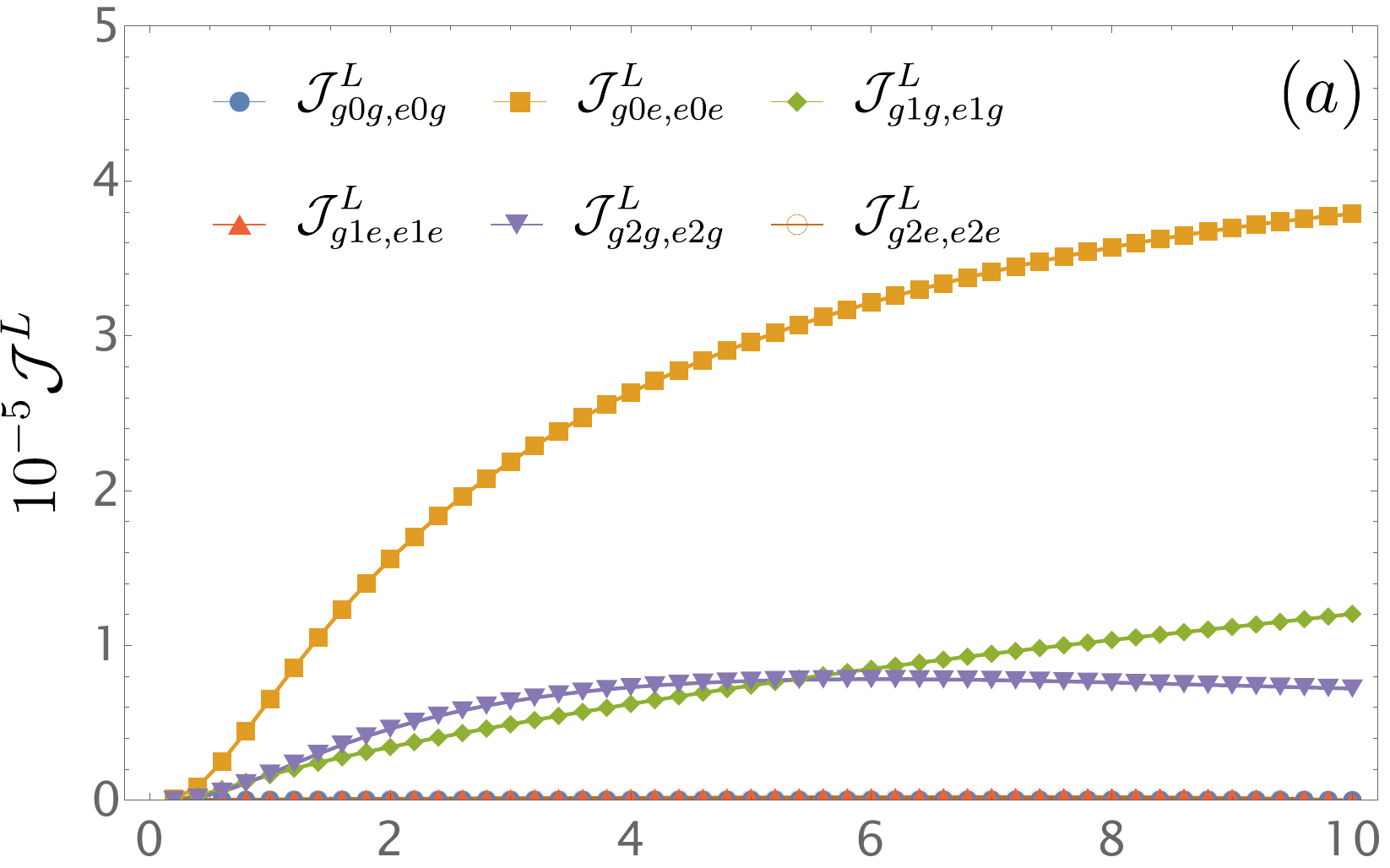} 
\includegraphics[width=0.35\textwidth]{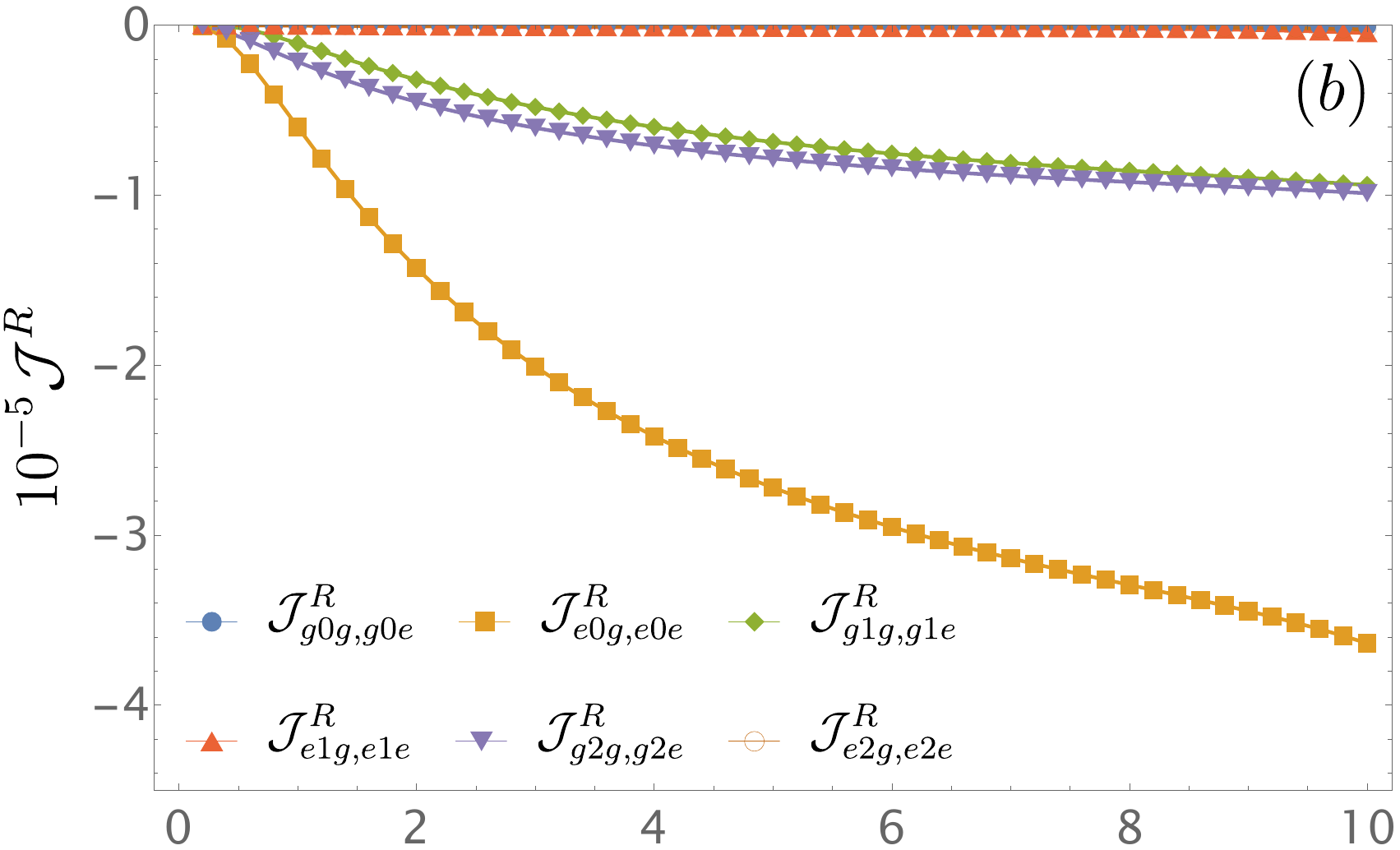} 
\includegraphics[width=0.35\textwidth]{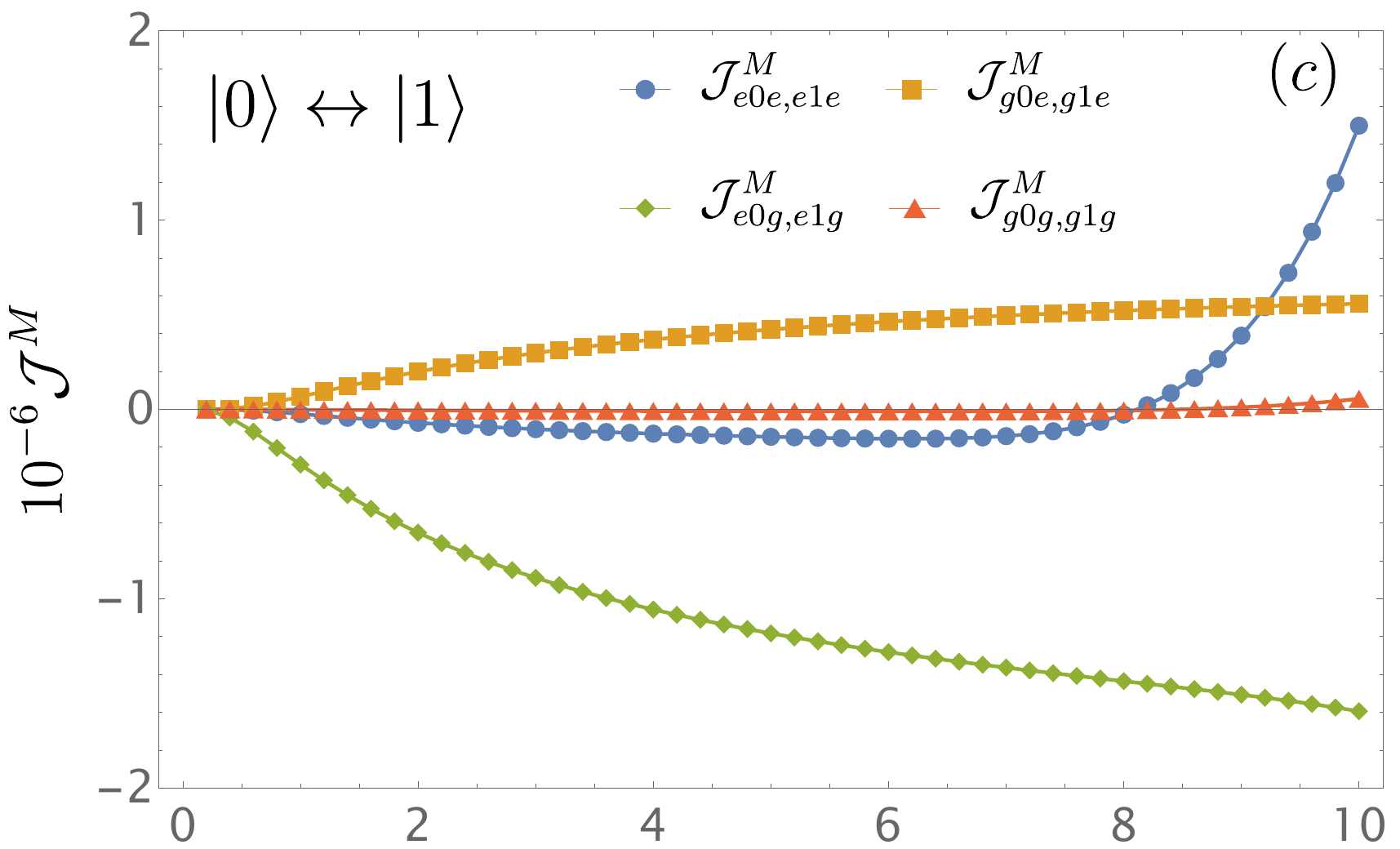}
\includegraphics[width=0.35\textwidth]{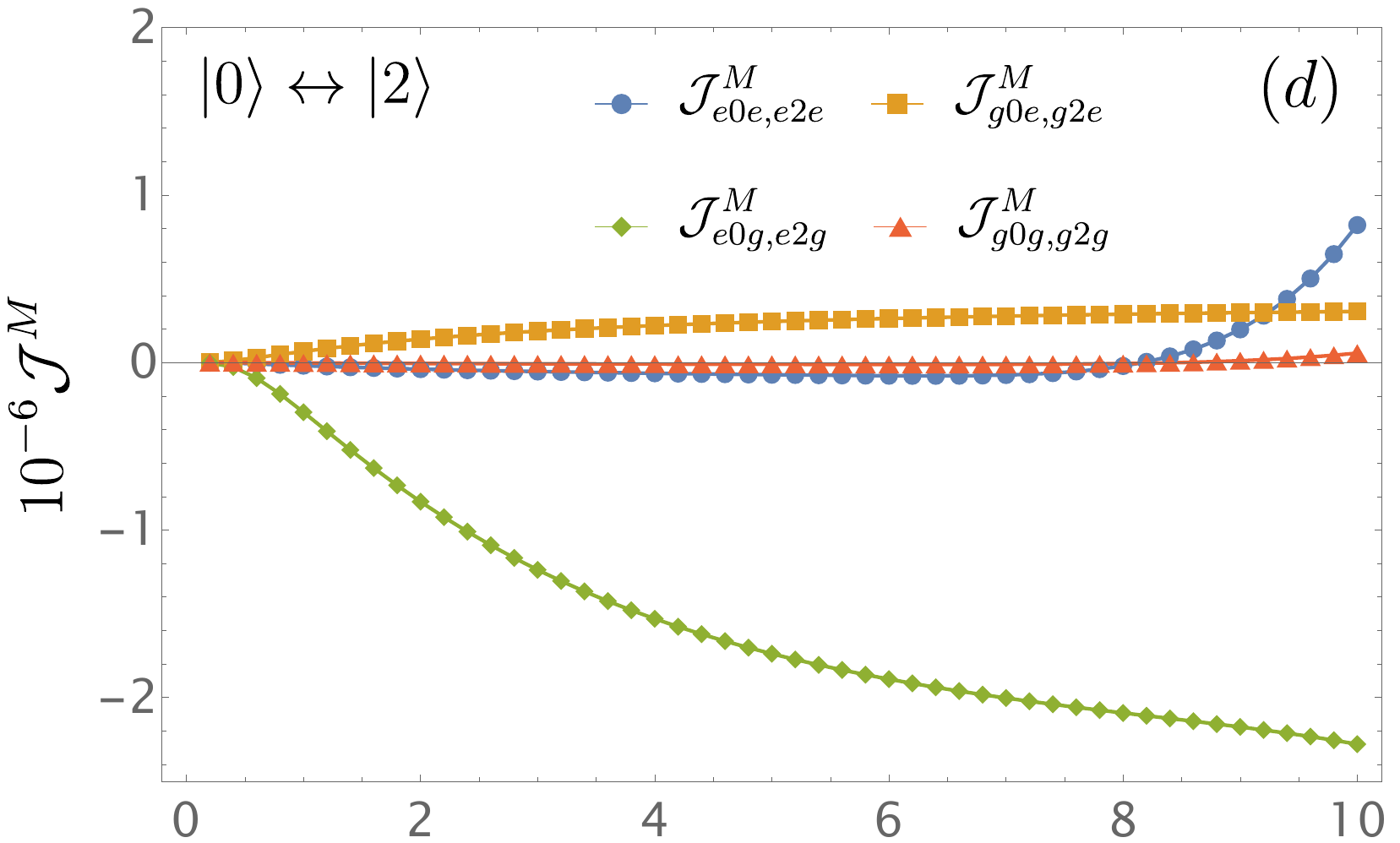}
\includegraphics[width=0.35\textwidth]{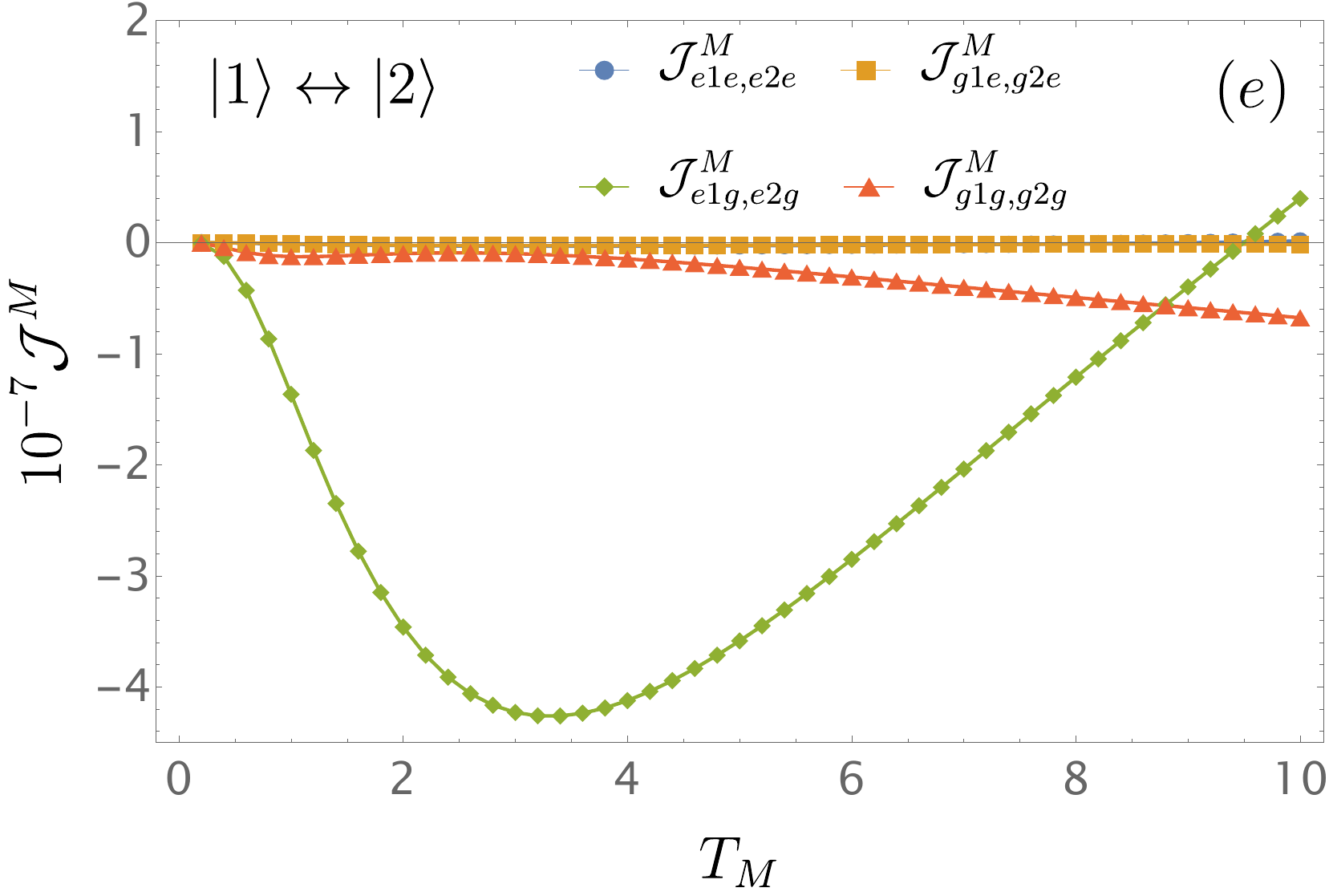}
\captionsetup{justification=justified}
\caption{Individual contributions to the heat currents, as given by Eq. \eqref{local currents}, relative to the qutrit's temperature $T_M$. (a) Terminal $L$; (b) Terminal $R$; (c) Terminal $M$: transitions $|0\rangle\leftrightarrow|1\rangle$; (d) Terminal $M$: transitions $|0\rangle\leftrightarrow|2\rangle$; (e) Terminal $M$: transitions $|1\rangle\leftrightarrow|2\rangle$. All parameters are consistent with those in Fig. \ref{Fig1}. \justifying}
\label{Fig2to6}
\end{figure}

Furthermore, Fig. \ref{Fig2to6} shows the individual steady heat currents associated with specific state transitions for the terminals $L$, $R$, and $M$. It is possible to see that, for the chosen parameters, some currents are negligible and not all transitions strongly affect the net heat currents. In particular, the heat fluxes due to the transitions $|1\rangle\leftrightarrow|2\rangle$ do not play a significant role, as a priori expected due to the relatively weaker internal coupling values. Nevertheless, given Eq. \eqref{local currents}, it is clear these currents are the result of the complex interplay between the energy gaps and the transition rates which, in turn, depends both on the decay rates and the stationary states. For instance, for fixed $j$ one also fixes $\omega_{ge,j}^{R, L}$, and the current amplitudes are determined by the transition rates, e.g., given the curves for $\mathcal{J}_{g0r,e0r}^{L}$ with $r=g,e$, it is possible to conclude that $|\Gamma_{g0e,e0e}^{L}|>|\Gamma_{g0g,e0g}^{L}|$ for the temperature range considered. However, for different $j$'s, the energy gaps might be completely different and, therefore, are expected to also play a major role, e.g., we have $\omega_{ge,0}^{L}=-65$, $\omega_{ge,1}^{L}=30.8$ and $\omega_{ge,2}^{L}=37.2$ for the chosen parameters. We observe that the local $L$ and $R$ heat fluxes are only indirectly affected by the middle terminal temperature $T_{M}$ via the steady states, while the current flowing through terminal $M$ is also directly influenced by it due to the dependence of $\gamma^{M}_{\omega}$ on $\bar{N}_{M}$. Thus, the temperature-dependent profile of the decay rates throughout the system's spectrum also plays a significant role in the current response. In this sense, transitions with energies much above the chosen cut-off are relatively independent of the $T_{M}$. This can be modulated by proper engineering of the spectral densities of the reservoirs by exploiting the so-called \textit{off-resonant} coupling \cite{PhysRevA.81.052103}, allowing for a potential enhancement of the performance of the device \cite{mandarino2022quantum}.

\begin{figure}[H]
\center
\includegraphics[width=8.5cm]{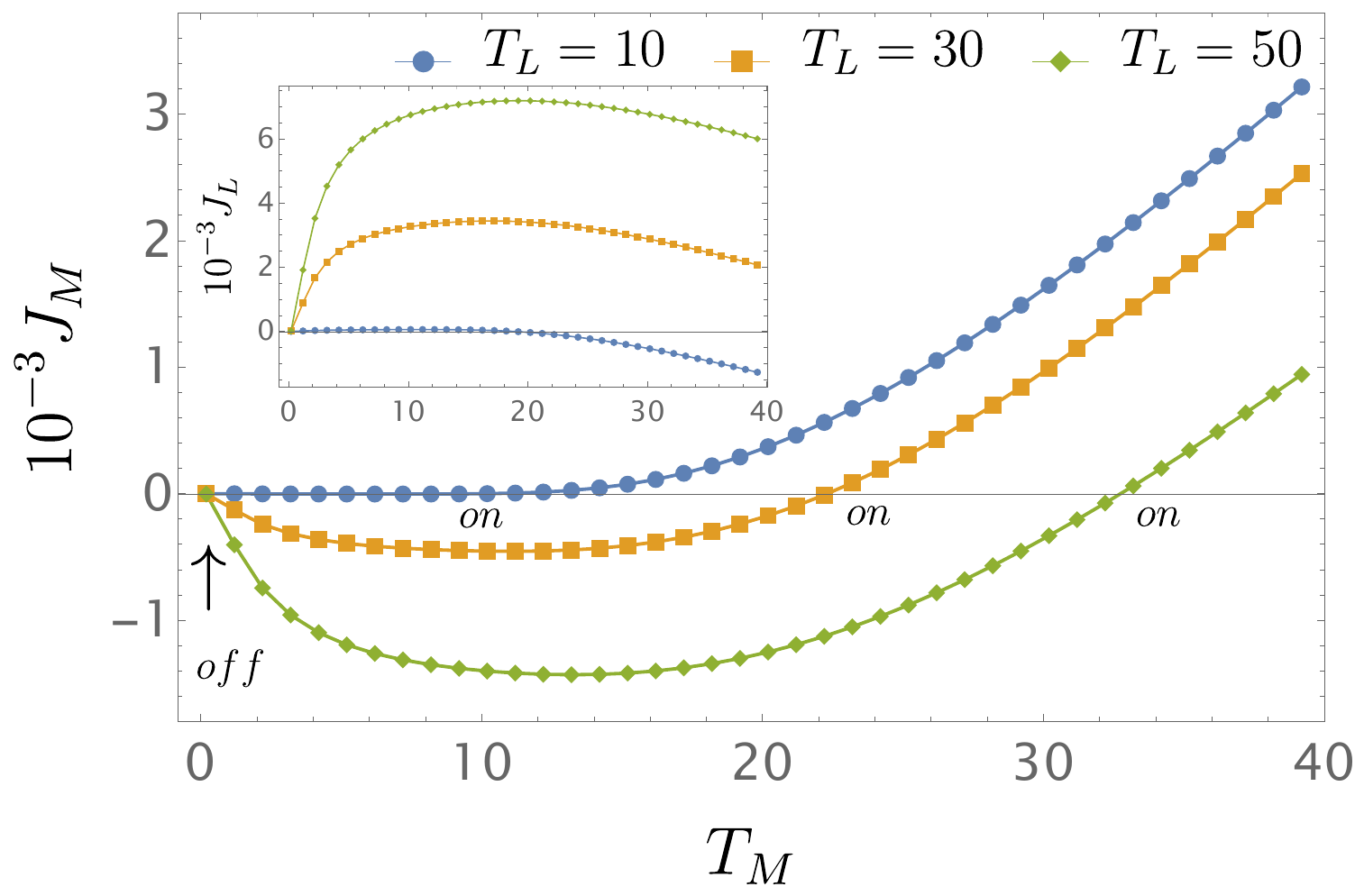}
\caption{Heat switch - Profile of $J_M$ and $J_L$ for different temperatures $T_{L}$. All remaining parameters are identical to those in Fig. \ref{Fig1}, implying  $||\hat{H}||=103.2\omega_{L}$. `On' and `off' marks indicate the position of two operational temperatures of the heat switch for given $T_{L}$.}
\label{FigSwitch}
\end{figure}

Given the characteristics of heat current profiles of the thermal transistor (as shown in the inset $(a)$ of Fig. \ref{Fig1}), the main functioning region is defined by the temperature domain of the high amplification factors. However, one might also be interested in the \textit{on/off} aspect of the heat currents, i.e., notice that from temperatures $T_{M}$ around $T_{R}=0.2$ the left and right currents increase from negligible to significant values, while the middle current is close to zero (or exactly zero). In general, setting $T_{M}$ to values $T_{M}^{low}$ or $T_{M}^{high}>0$, such that $J_{M}(T_{M}^{low})=J_{M}(T_{M}^{high})\approx0$ with finite values of $J_{L, R}$, implements a heat switch. Fig. \ref{FigSwitch} illustrates this behavior for distinct $T_L$ and fixed $T_R$. It is possible to observe the values of heat switch currents can be significantly increased for temperatures of the order $T_{L}\approx ||\hat{H}||$. Thus, a higher temperature results in a stronger $L$ bath influence, making the heat switch more pronounced.
It is worth emphasizing that the observation of the transistor effect is not uniquely defined for a specific set of parameters of this architecture. In fact, it can be realized in different regions of the coupling-parameter space \cite{yang2020quantum} (see Appendix \ref{Coupling strengths} for more detail). Along these lines, once the local system's energy levels are fixed, one can tune the couplings according to specific needs and eventual experimental constraints. For instance, the working regime of the switch and amplifier can be straightforwardly modified by considering alternative values of $\chi_{ij}^{R, L}$, either stronger or weaker. However, it is important to highlight that, despite this freedom, the hypotheses for deriving Eq. \eqref{Master Equation} might break down for some particular parameter values, i.e., one should be careful and guarantee the validity of all working hypotheses, namely, non-degenerate energy gaps, Markov, and full secular approximations.

\begin{figure}[H]
\center
\includegraphics[width=8.5cm]{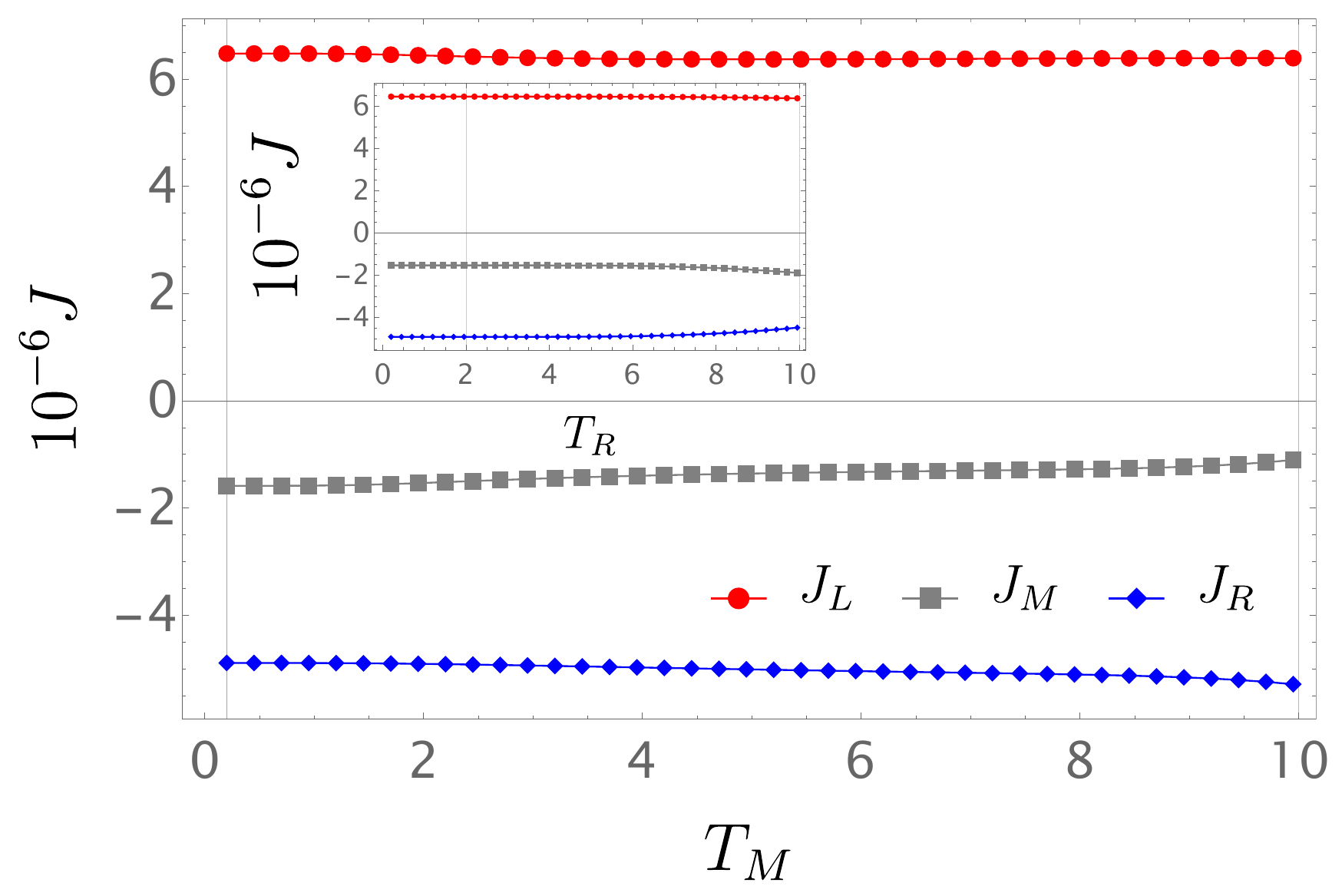}
\caption{Heat stabilizer regime. Heat currents for changing $T_{M}$ with $T_{R}=0.2$ and $T_{L}=10$ (main plot), and $T_{R}$ with $T_{M}=2$ and $T_{L}=10$. Vertical blue, red, and grey lines represent the fixed temperatures of the right, left, and middle reservoirs, respectively. Parameters include $\chi_{01}^{L}=20\omega_{L}$, $\chi_{02}^{L}=25\omega_{L}$, and $\chi_{02}^{R}=20\omega_{L}$. All other parameters are consistent with those in Fig. \ref{Fig1}.}
\label{Fig6_2}
\end{figure}

Finally, the heat current behavior depicted in Fig. \ref{Fig1} exhibits stability over variations in $T_{R}$ for any $T_{M}$ within the specified range, 
consistently with observations by \cite{guo2018quantum, PhysRevE.99.032112}. However, for a more detailed exploration of the internal coupling dependency, let us consider alternative strengths. By adjusting parameters such as $\chi_{ij}^{L, R}$ to different values, one can significantly alter the device's operation.
In this regard, Fig. \ref{Fig6_2} illustrates that, in this new regime, the previous heat amplification mechanism transforms into a heat stabilization effect for both $T_{M}$ and $T_{R}$. This behavior is characterized by low thermal sensitivity $\mathcal{S}_{L, M, R}$ in all terminals, resulting in high resilience to fluctuations in the reference temperature. Notably, within the considered domain, the terminal currents remain relatively stable, maintaining approximately constant values despite changes in the reference temperature $T_{M}$ (main figure) or $T_{R}$ (inset).
In the main plot of Fig. \ref{Fig6_2}, the average heats are approximately $\overline{J}_{L}\approx 6.40\times 10^{-6}$, $\overline{J}_{M}\approx -1.38\times 10^{-6}$, and $\overline{J}_{R}\approx -5.02\times 10^{-6}$, with respective standard deviations of $\sigma_{L} \approx 3.55 \times 10^{-8}$, $\sigma_{M} \approx 1.27\times 10^{-7} $, and $\sigma_{R} \approx 1.03 \times 10^{-7}$. In the inset, the values are $\overline{J}_{L}\approx 6.43\times 10^{-6}$, $\overline{J}_{M}\approx -1.60\times 10^{-6}$, and $\overline{J}_{R}\approx -4.83\times 10^{-6}$, with corresponding standard deviations of $\sigma_{L} \approx 2.17\times 10^{-8}$, $\sigma_{M} \approx 9.56 \times 10^{-8} $, and $\sigma_{R} \approx 1.17 \times 10^{-7}$.
\subsection{Dependency on the qutrit detuning}
\begin{figure}[H]
\centering
   \begin{subfigure}[b]{0.46\textwidth}
   \includegraphics[width=1\textwidth]{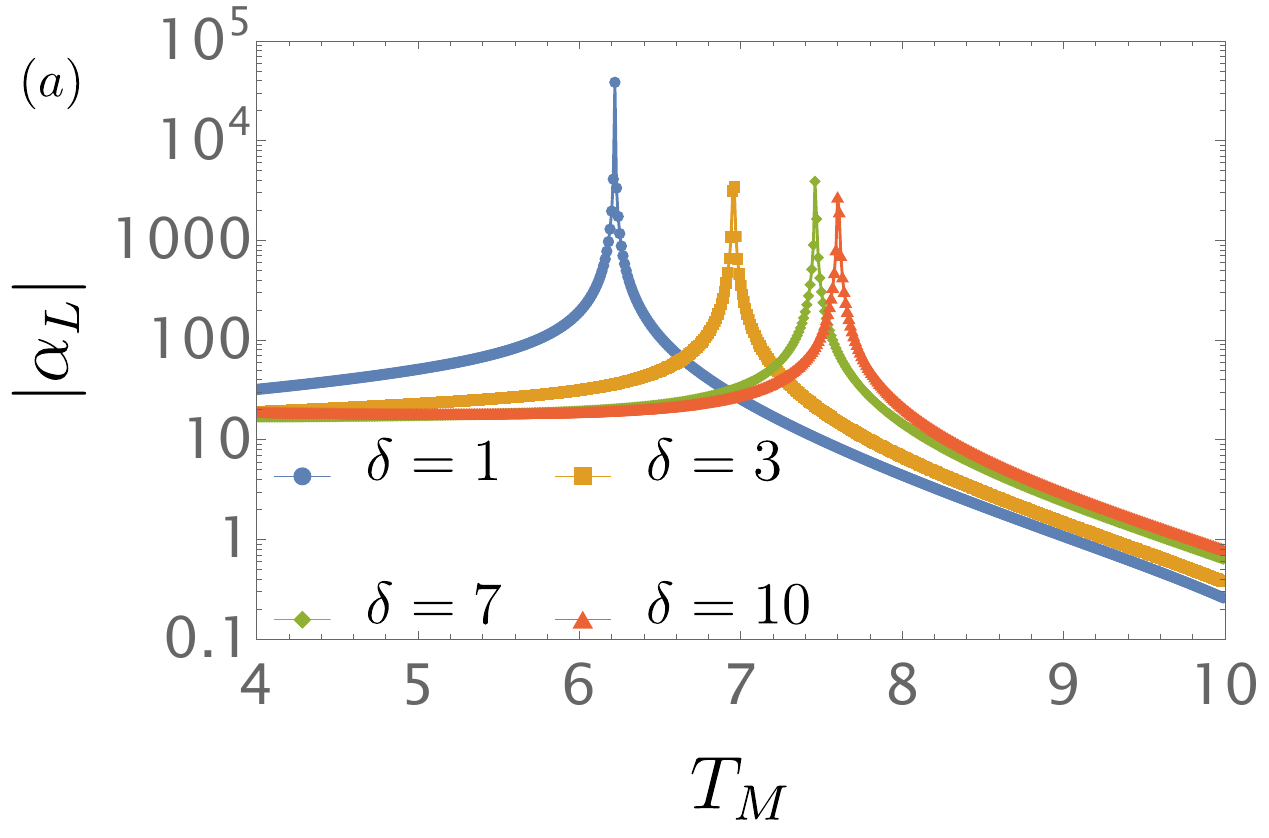}
   \label{Fig7} 
\end{subfigure}
\begin{subfigure}[b]{0.46\textwidth}
   \includegraphics[width=1\textwidth]{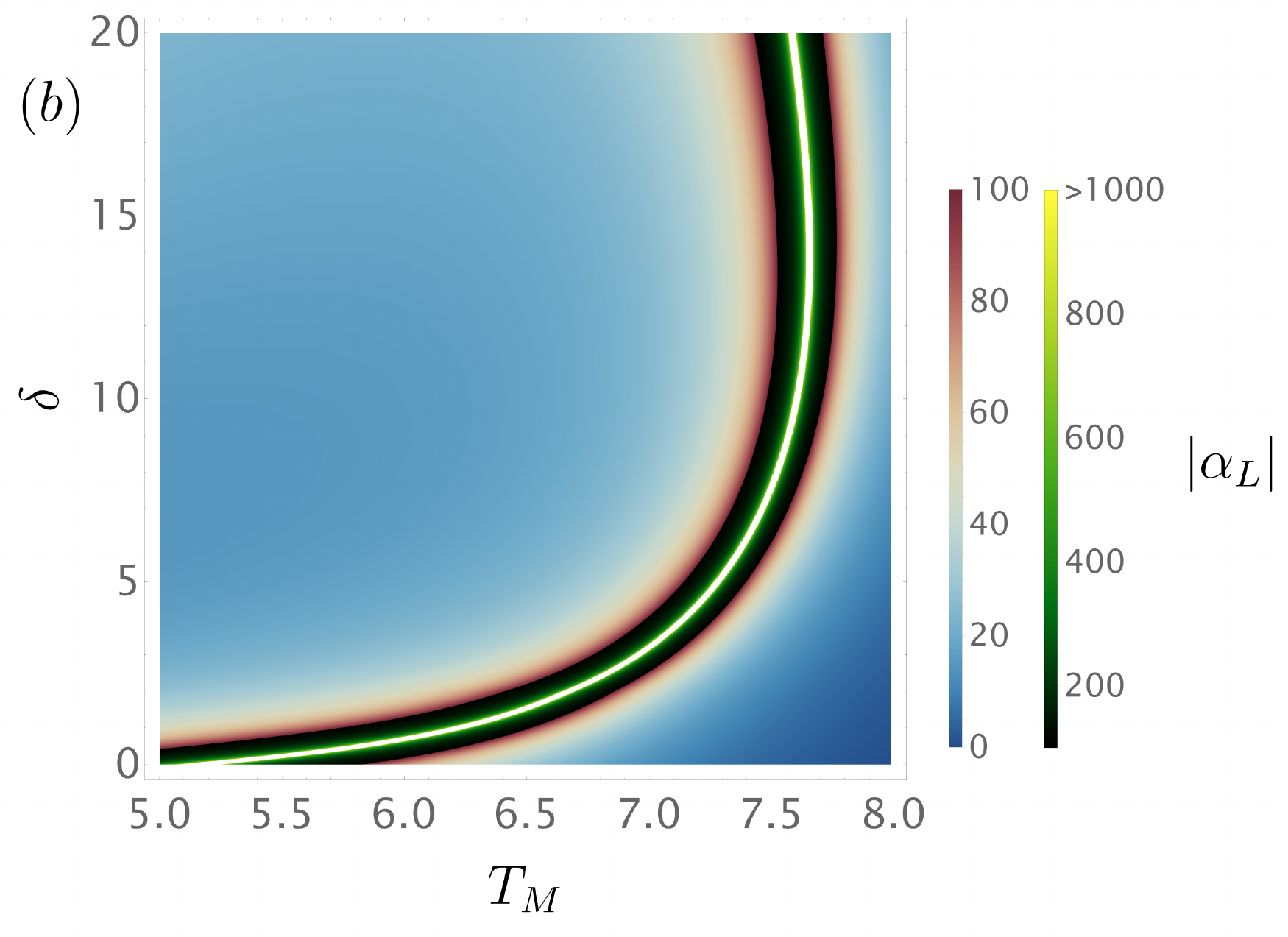}
   \label{Fig8}
\end{subfigure}
\caption{Modulation of the amplification window. $(a)$ Behavior of the amplification factor for terminal $L$ under various values of $\delta$; $(b)$ Density plot depicting $|\alpha_{L}|$ as a function of $\delta$ and $T_{M}$. The black-yellow region highlights the domain with high amplification, where $|\alpha_{L}|\geq100$. Parameters for the density plot: $\Delta T_{M}=10^{-2}$ and $\Delta \delta = 2.5 \times 10^{-2}$. All other parameters remain consistent with those in Fig. \ref{Fig1}.
}
\label{Fig7and8}
\end{figure}

\begin{figure}[H]
\center
\includegraphics[width=8.5cm]{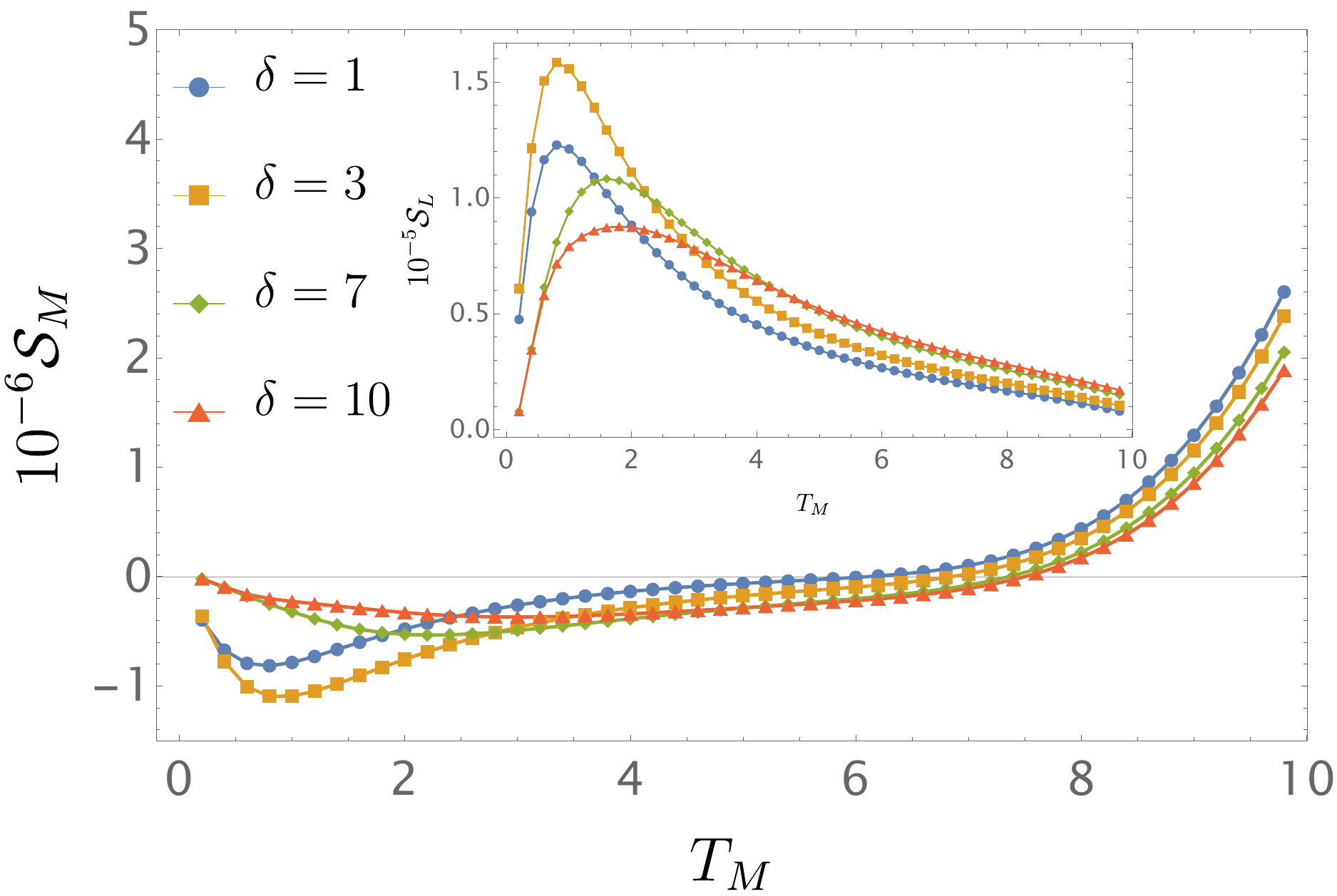}
\caption{Main plot: Differential thermal sensitivity $\mathcal{S}_{M}$ for terminal $M$. Inset: Differential thermal sensitivity $\mathcal{S}_{L}$ for terminal $L$, both against various values of $\delta$. All parameters are consistent with those in Fig. \ref{Fig7and8}.}
\label{Fig9}
\end{figure}

In this section, we will present the impact of the qutrit's third level on the system's functioning. This influence is characterized by the quantity $\delta$ measuring the gap between the two excited levels.
Fig. \ref{Fig7and8} presents the landscape of amplification factors for fixed internal couplings and frequencies, in the $T_{M}-\delta$ plane (b), and along selected snapshots of $\delta=1,3,7,10$ (a). According to the definition of the transistor effect, regions of high amplification factor are the ones for which the device may be used as a heat transistor, provided $J_{M}$ current is negligible. This condition is satisfied for the high amplification regions reported along the snapshots, and we conjecture that this can be extended to the whole plane. Similarly, we need to stress that while not all the reported points satisfy the necessary hypotheses underlying Eq. \eqref{Master Equation}, the set of points satisfying the hypothesis pertains to entire regions of parameters presented. We see that increasing qutrit's third-level gap between $\delta=1$ to $\delta=10$ results in a change of the high amplification regime in the approximate interval of $T_{M}\in[5.74, 7.71]$. This behavior is best assessed in the density plot Fig. \ref{Fig7and8}\textcolor{blue}{b} for $|\alpha_{L}|$. The black-yellow region highlights the domain with high amplification factors ($|\alpha_{L}|\geq100$).  Crucially, while one can also find such behavior for the qubit analog in response to changes of $\Omega$, the modulation window obtained for it is considerably narrower (see Appendix \ref{QUBIT} for more details).
Fig. \ref{Fig9} shows how the differential thermal sensitivities profiles of $L$ and $M$ depend on $\delta$. Shifting of high amplification regions with $\delta$ can be understood through the fact that modifying $\delta$ changes temperatures $T_{M}$ for which sensitivity $\mathcal{S}_{M}$ 
is null,
while accompanied by finite values of $\mathcal{S}_{L}$ and $\mathcal{S}_{R}$.
%
%
%
%

%
%
%
%
%
%
%
%
%
%
%
%
%
%
%
%
%
%
%
%
%
%
%
%

\begin{figure}[H]
\centering
   \begin{subfigure}[b]{0.4\textwidth}
   \includegraphics[width=1\textwidth]{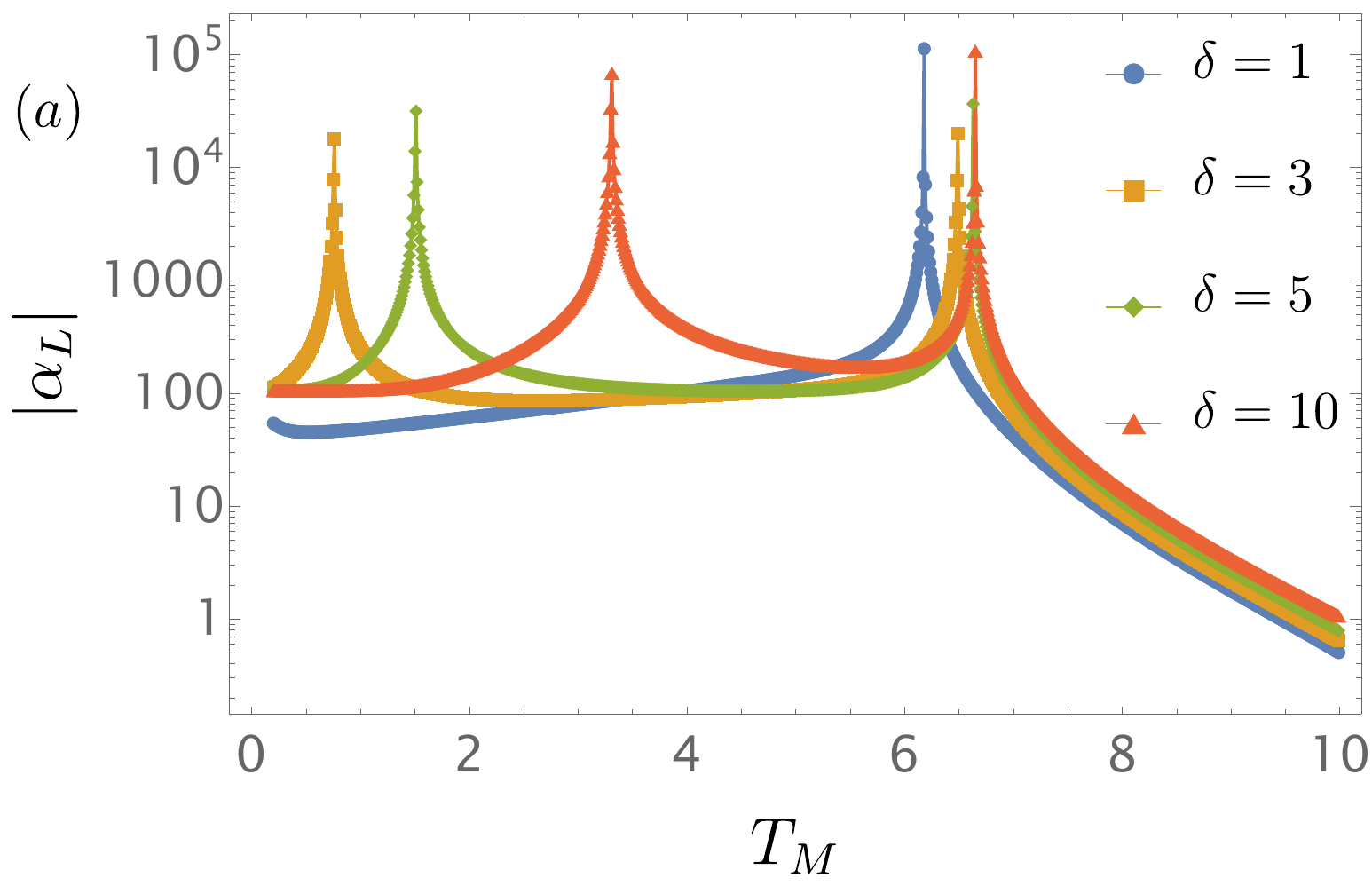}
   \label{Fig11_1} 
\end{subfigure}
\begin{subfigure}[b]{0.4\textwidth}
   \includegraphics[width=1\textwidth]{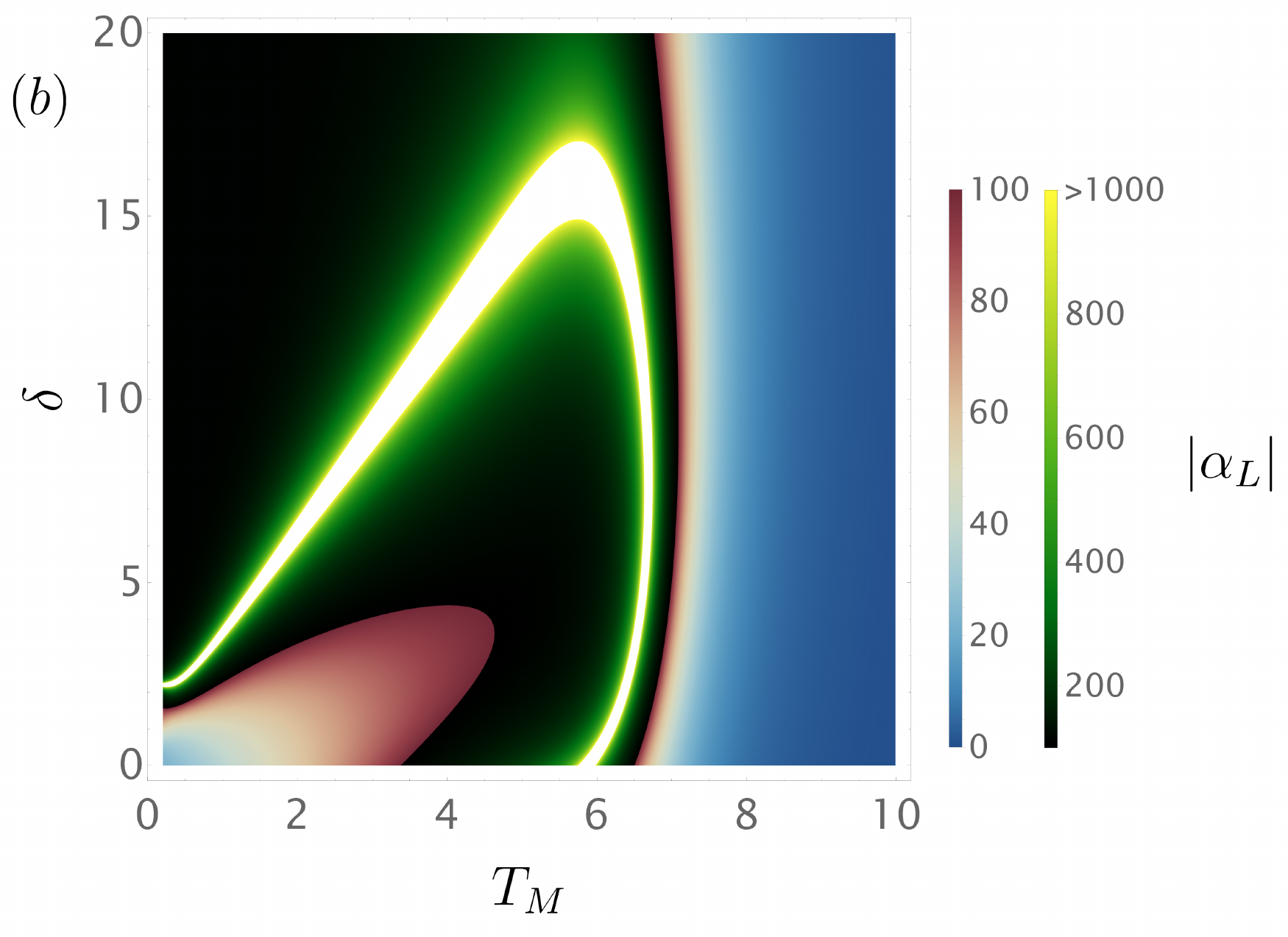}
   \label{Fig11_2}
\end{subfigure}
\caption{Amplification recurrence. $(a)$ Amplification factor for terminal $L$ under different detuning values, $\delta$; $(b)$ Density plot illustrating $|\alpha_{L}|$ in relation to $\delta$ and $T_{M}$. It were assumed $\chi_{02}^{L}=16.2\omega_{L}$, and $\Delta T_{M}=10^{-2}$, and $\Delta \delta = 2.5 \times 10^{-2}$ for the density plot. All other parameters are consistent with those in Fig. \ref{Fig1}.
}
\label{Fig11}
\end{figure}

%
%
%
As mentioned earlier, the transistor effect is characterized by the response of heat current to temperature variations, when the condition $\mathcal{S}_{M} \ll \mathcal{S}_{L, R}$, together with $J_{M}\approx 0$, is met. Notably, within certain regions of the parameter space, heat flux oscillations can occur, resulting in multiple high amplification zones and a recurrence of the transistor effect. Figure \ref{Fig11} depicts this behavior in more detail. The change of $|\alpha_{L}|$ in response to detuning modulation is demonstrated using the same parameter values as before, except for $\chi_{02}^{L}=16.2\omega_{L}$. Interestingly, such a small change from the previous value ($\chi_{02}^{L}=18\omega_{L}$) brings the system to a parameter space region where the high amplification sustains for a wide temperature interval (see Fig. \ref{FigCouplingstrengths}\textcolor{blue}{b} from Appendix \ref{Coupling strengths}). The corresponding density plot provides a more intricate view of this phenomenon. Generally, at low values of $\delta$, the characteristic transistor effect is evident, displaying a single peak around $T_{M}\approx 6$. As $\delta$ surpasses a critical value around $\delta\approx2\omega_{L}$, an additional peak emerges in a lower-temperature range. With further detuning, the high-temperature peak remains relatively stable, while the low-temperature peak gradually increases until both high-amplitude regions converge at approximately $T_{M}\approx 5$ and $\delta\approx17\omega_{L}$. 
Crucially, we observe that high detunings result in broad temperature windows for which amplification factors take large values ($\delta \gtrsim 5$). 

\begin{figure}[H]
\center
\includegraphics[width=8.5cm]{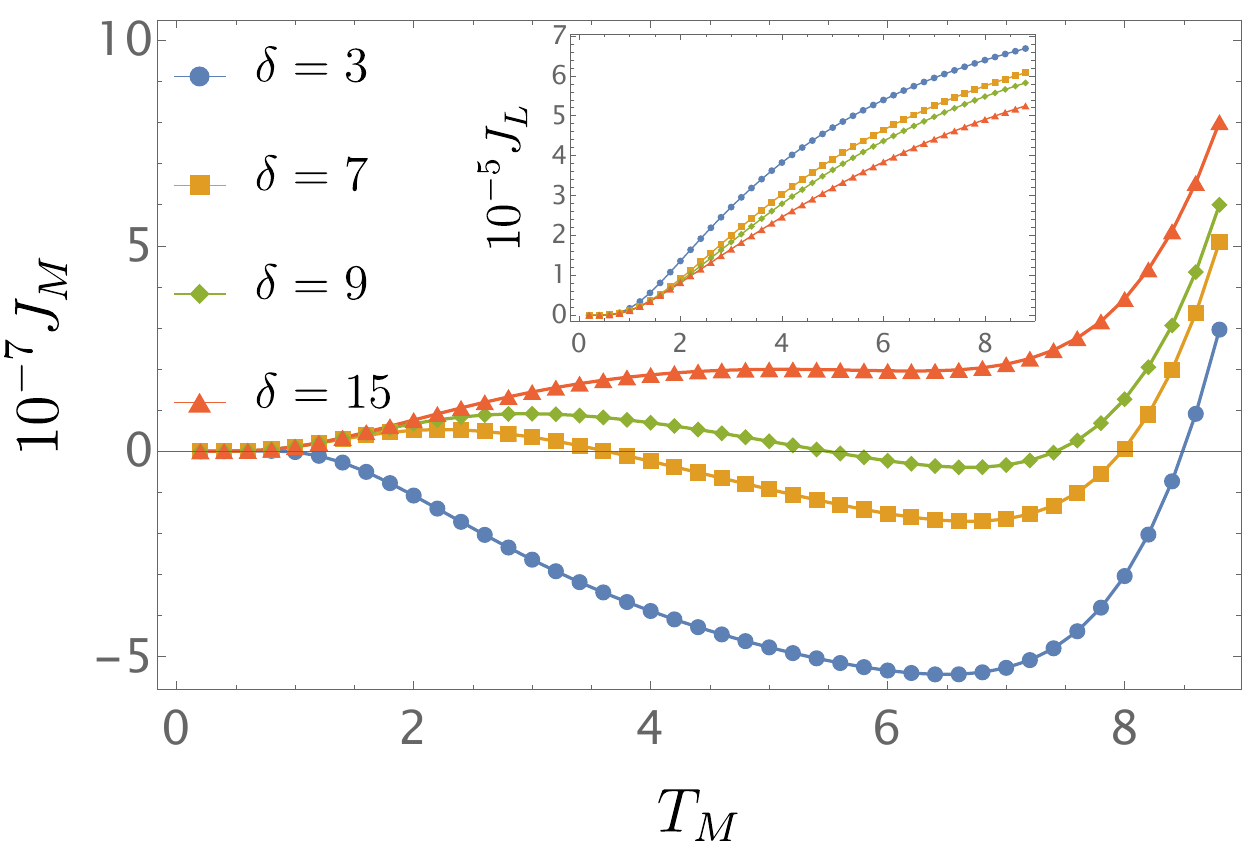}
\caption{Profile of $J_M$ (main) and $J_L$ (inset) for different detuning $\delta$. All remaining parameters are identical to those in Fig. \ref{Fig11}.}
\label{FigRecAndswi}
\end{figure}

Analyzing the snapshots for $\delta=3,7,9,15$ of the current-temperature dependence on Fig. \ref{FigRecAndswi}, we see that $J_{M}$ current can be minimized over a wide temperature window for proper adjustment of the energy gap (e.g. $\delta=9$), confined well in the regime of two orders of magnitudes smaller than the trans-system $J_{L, R}$ currents. On the technical note, we should add that 
by changing $\delta$ the local extremal points of $J_M$ with respect to $T_{M}$ converge into a saddle point, and 
for $\delta$ high enough, $J_{M}$ does not have any extremal points and high values of amplification factor correspond to quasi-linear strong amplification mode of the transistor. The extremal points are responsible for diverging amplification factors for $\delta \lesssim 18$.
Interestingly, the $\delta$ regime in which $J_{M}$ has multiple zero points allows for designing switch protocols working between different temperature values, i.e., for fixed setting and detuning one might attribute ``on'' states for any of the zero points displayed in Fig. \ref{FigRecAndswi}.

In short, both the high amplification region and the switch behavior exhibit dependence on the detuning. Specifically, non-trivial effects, such as multiple high amplification temperature intervals, can be observed within tuned parameter regions. Therefore, by dynamically controlling $\delta$, one can potentially leverage these characteristics to either modulate the device amplification window, and its on/off switch states or fine-adjust its operational regime based on specific requirements.
%
%
%
%
%
%
%
%
%
%
%
%
\subsection{Rectification}\label{Rectification}
\begin{figure}[H]
\centering
   \begin{subfigure}[b]{0.4\textwidth}
   \includegraphics[width=1\textwidth]{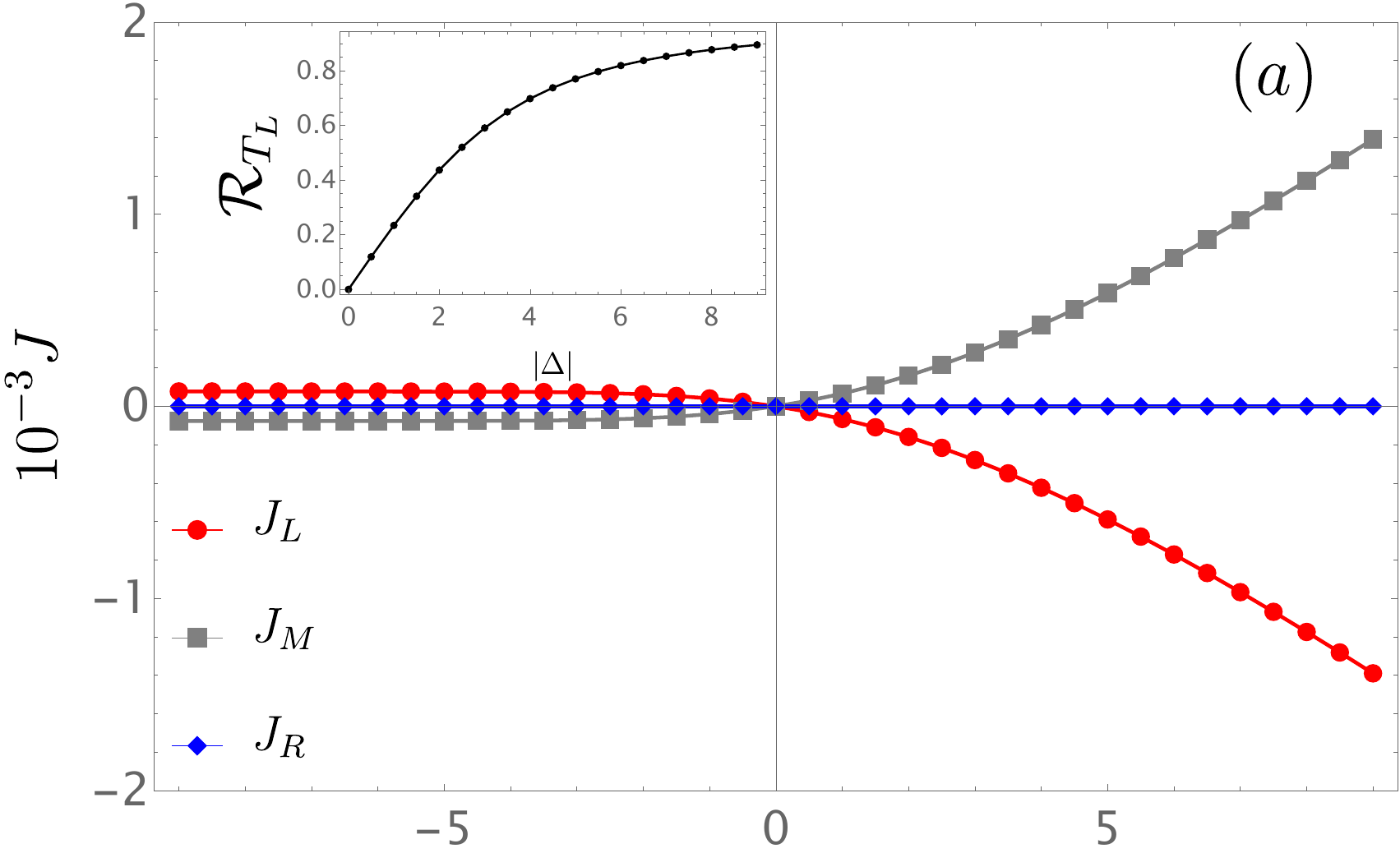}
   \label{Rect1} 
\end{subfigure}
\begin{subfigure}[b]{0.4\textwidth}
   \includegraphics[width=1\textwidth]{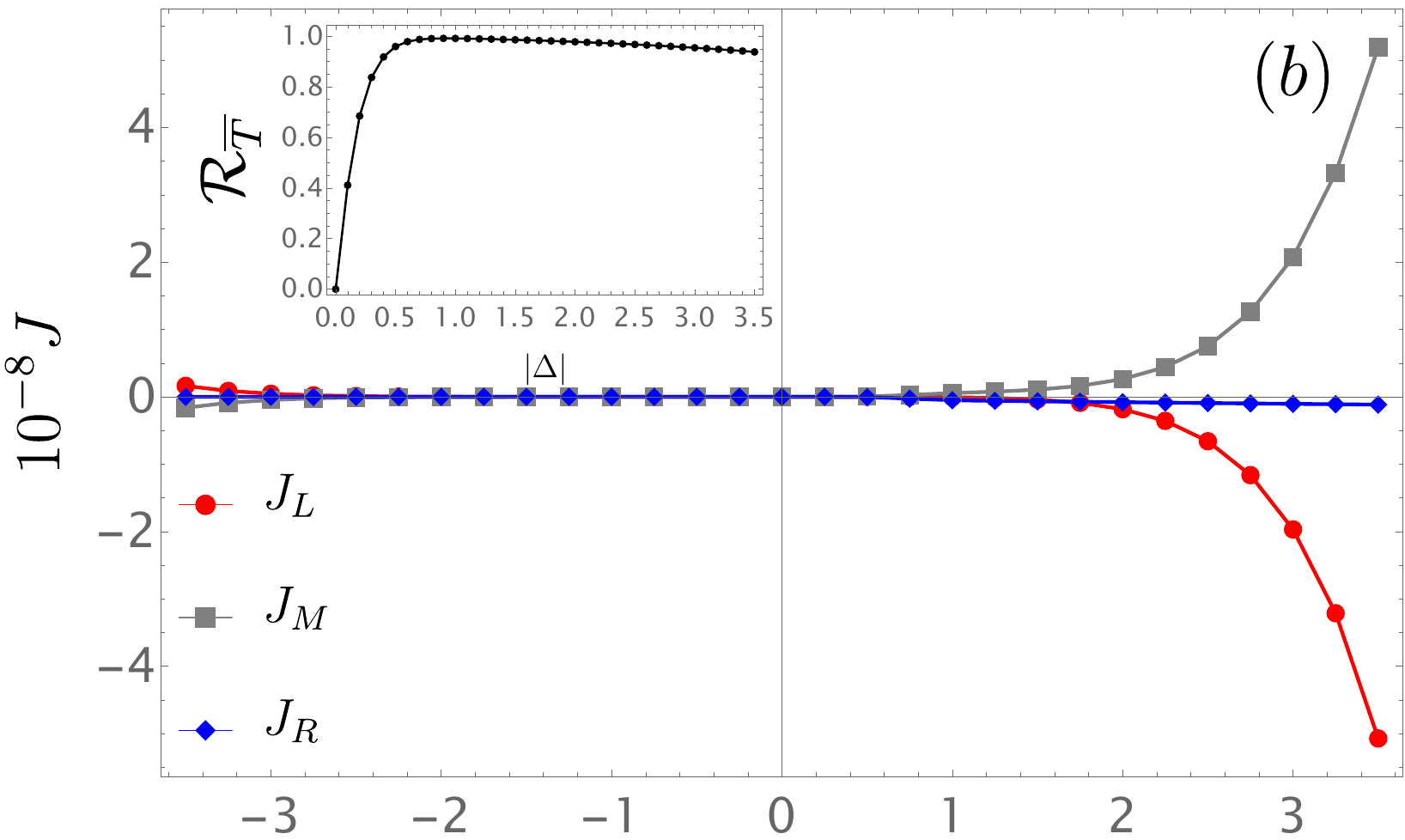}
   \label{Rect2}
\end{subfigure}
\begin{subfigure}[b]{0.4\textwidth}
   \includegraphics[width=1\textwidth]{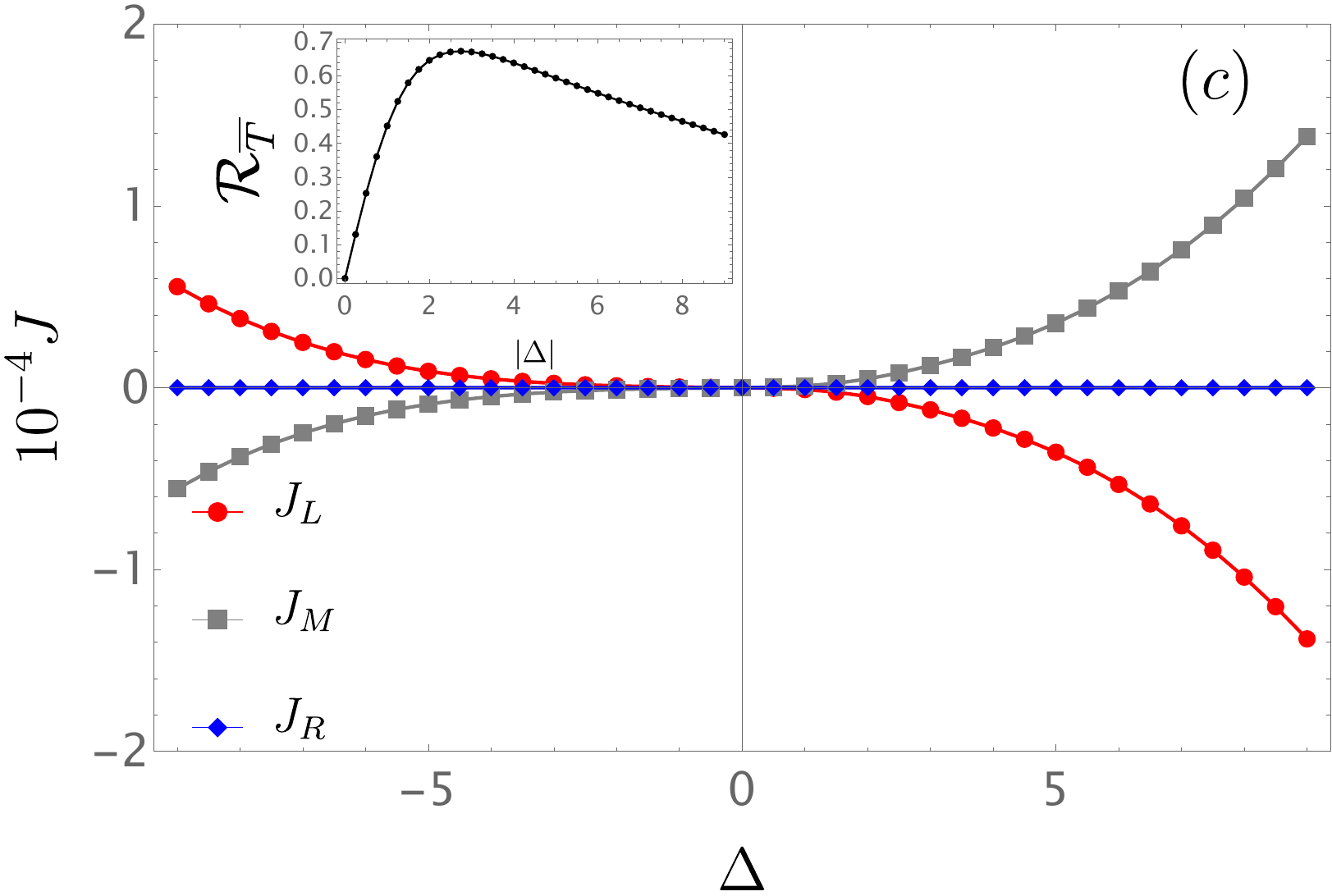}
   \label{Rect3}
\end{subfigure}

\caption{Main figures: Heat current directionality observed between the left and middle terminals of the heat-amplifier, with the right qubit deactivated. Insets: Thermal rectification factors. Parameters assumed are $\chi_{01}^{R}=0$ and $\chi_{02}^{R}=0.2\omega_{L}$, with all other parameters consistent with those in Fig. \ref{Fig1}. For $(a)$, the rectification factor $\mathcal{R}_{T_{L}}$ is depicted, with $T_{L}=10$, and $T_{M}=T_{L}+\Delta$. For $(b)$ and $(c)$, the rectification factor $\mathcal{R}_{\overline{T}}$ is used with $\overline{T}=2$ and $\overline{T}=5$, respectively.
}
\label{FigRect}
\end{figure}

A distinctive feature of a heat rectifier is the asymmetrical heat flow between two terminals under a reversed temperature bias. In an ideal scenario, all current is suppressed when the temperature gradient is reversed \cite{werlang2014optimal}, resulting in a maximum rectification factor, $\mathcal{R}=1$. The proposed architecture facilitates the effective decoupling of one of the TLSs by minimizing relevant internal couplings. Specifically, if $\hat{\chi}_{R}\approx 0$, the device operates in a two-terminal mode, functioning as a heat current rectifier. Main plots in Fig. \ref{FigRect} illustrate heat currents in the heat-amplifier once the decoupling is performed. The temperature difference, measured by $\Delta$ between the remaining reservoirs, reveals that, as anticipated, no current flows through terminal $R$ and no heat currents are observed when $\Delta=0$. Notably, a clear asymmetry between the $\Delta < 0$ and $\Delta > 0$ regimes is observed, highlighting characteristic rectification curves. With this effect vanishing for a setting with symmetric couplings, we attribute it to coupling asymmetry. Quantification of this behavior is detailed in the insets of Fig. \ref{FigRect}, depicting rectification factors for various $|\Delta|$. Deviating $T_{M}$ from a fixed value of $T_{L}$ reveals the unidirectional characteristic of the device working as a rectifier (Fig. \ref{FigRect}\textcolor{blue}{a}). Rectification characteristics improve in the regime of low temperatures, which we show by investigating heat currents for the average temperature of the reservoirs fixed (Fig. \ref{FigRect}\textcolor{blue}{b} and \textcolor{blue}{c}). At the price of decreasing heat currents, high rectification rates can be obtained for a wider range of temperatures, and for smaller temperature gradients. 

%
%
%
%
%
%
%
%
%
%
%
%
%
%
%
%
%
%
%
%
%
%
\section{Discussion and conclusion}\label{Discussion}
The imperfect realization of multi-level systems aiming to simulate two-level systems can lead to effective systems with richer energy configurations. In particular, taming the appearance of dark states has been a challenge in driven open quantum systems \cite{diehl2008quantum}. Nevertheless, their presence may also lead to applications in quantum thermal devices such as heat rectifiers \cite{PhysRevE.106.034116}. The simplest model where the interference plays a crucial role is observed in the so-called $V-$state, a three-level system in which the ground state interacts with two excited states nearly degenerate \cite{reynaud1988photon}. In the context of quantum thermal transistors, qutrits have been already considered in different settings \cite{su2018quantum,guo2018quantum,PhysRevE.99.042102,wang2020polaron,majland2020quantum,wang2022cycle} and, in particular, it was shown that the transistor effect for superconducting artificial atoms in a qubit-qutrit architecture strongly depends on its anharmonicity \cite{majland2020quantum}. In general, such inaccuracy in the physical description might either suppress the desired effect, leading to poor performance or hide some non-trivial features inherited by higher energetic complexity. This change can be even more dramatic if one is interested in strong internal coupling regimes.

Given the current high precision of controlling quantum systems, these questions hold practical relevance. Experiments probing quantum heat transport within condensed matter platforms, such as superconducting circuits \cite{gubaydullin2022photonic}, are established \cite{pekola2021colloquium}. These experiments allow the realization of quantum thermal devices, like valves \cite{ronzani2018tunable} and rectifiers \cite{scheibner2008quantum,senior2020heat}. Understanding the influence of additional energy levels is crucial for effectively controlling energetics for technological purposes. Therefore, having a comprehensive physical picture is essential for designing functional and efficient thermal devices applicable to realistic scenarios.

This paper examines how this can impact the performance of a multipurpose quantum thermal device. Our contribution builds upon earlier studies, expanding the scope by exploring a more intricate physical scenario depicted in Figure \ref{FigModel}. In this augmented configuration, the actively controlled subsystem incorporates a non-negligible third energy level. Significantly, we look into the nuanced impact of detuning between the two excited states on the behavior of this subsystem. This extension introduces an increased level of complexity to the system, enriching its dynamics and unveiling the intricate interplay between energy levels. The equation of motion of the system was obtained under the usual assumptions for deriving a global master equation, which provides the appropriate formalism for considering strong internal couplings \cite{cattaneo2019local, PhysRevE.99.032112, yang2020quantum}. Having in mind the assumed working hypotheses, it was possible to compute the heat currents flowing through the terminals and numerically simulate their behavior.

We demonstrate that, with appropriately tuned couplings, the strong and robust transistor effect can be achieved with the proposed setting over a wide temperature range $T_{M}\in[T_{R}, T_{L}]$, as shown in Fig. \ref{Fig1}. The thermal sensitivities of the terminals, and thus both amplification and switch behavior, depend significantly on the chosen detuning, parametrized by $\delta$. By changing its value, one can modify both the high heat current amplification region and the \textit{on/off} transition domain, as shown in Fig \ref{FigSwitch}. Also, we showed that the inclusion of the third level, not considered in previous works, leads to outperforming its qubit analog. This dependency can be exploited to engineer or modulate the system's working regime dynamically, adjusting the high amplification window. An amplification recurrence behavior is also observed, with two characteristic peaks of the transistor effect emerging, as shown in Fig. \ref{Fig11}. The positions of both peaks are highly dependent on the detuning value, converging for high enough $\delta$. To the best of our knowledge, this phenomenon has not been reported for qubits. It occurs due to intricate heat fluxes within the qutrit, leading to current oscillations with two local extremes. This is depicted in Fig. \ref{FigRecAndswi}. Such a characteristic can be utilized for providing different temperature working regimes for heat amplifiers, with no need for system parameters to be modified, and one can observe multiple $J_M$ zero points, providing distinct possibilities of on/off states of a switch. Finally, by effectively decoupling one of the TLS and turning our device into a two-terminal mode, we demonstrate in Fig. \ref{FigRect} that it can also function as a heat rectifier.

In summary, our work introduces a robust heat amplification system with a strongly coupled qubit-qutrit-qubit architecture, extending prior research in the field. Our characterization underscores the significant influence of the qutrit's detuning on the system's behavior, introducing a variety of effects. The ability to control this detuning offers the potential to dynamically modulate the device's working regime and heat flow characteristics, providing avenues for tailored optimization in quantum thermal applications. We find it essential for future research to address the following aspects: $(i)$ to comprehend the efficient engineering of general energetic structures and the spectral properties of reservoirs to execute specific tasks; $(ii)$ to investigate time-scales in which stationary states are achieved, and to explore the functioning of the device beyond stationary states.
%
%
%
%
%
\section{Acknowledgements}
AHAM acknowledges the support from the Polish National Science Centre grant
OPUS-21 (No: 2021/41/B/ST2/03207). BA acknowledges support from National Science Center, Poland within the QuantERA II Programme (No 2021/03/Y/ST2/00178, acronym ExTRaQT) that has received funding from the European Union’s Horizon 2020. 
%
%
%
%
%
%
\appendix
\section{Energy spectrum}\label{Energy spectrum}
The energy levels of the Hamiltonian shown in Eq. (\ref{Transistor Hamiltonian}) can be computed straightforwardly by $\hat{H}|l\rangle_{L}|j\rangle|r\rangle_{R}=E_{ljr}|l\rangle_{L}|j\rangle|r\rangle_{R}$. Below we present the spectrum for all values of $j$:
\begin{itemize}
    \item For $j=0$:
        \begin{equation}
            \begin{split}
            E_{g0g}&=\hbar\left(q_{0}^{L}+q_{0}^{R}\right)\\E_{g0e}&=\hbar\left(\omega_{R}+q_{0}^{L}-q_{0}^{R}\right)\\E_{e0g}&=\hbar\left(\omega_{L}-q_{0}^{L}+q_{0}^{R}\right)\\E_{e0e}&=\hbar\left(\omega_{L}+\omega_{R}-q_{0}^{L}-q_{0}^{R}\right).
         \end{split}
        \end{equation}
    \item For $j=1$:
    
        \begin{equation}
         \begin{split}
            E_{g1g}&=\hbar\left(\Omega+q_{1}^{L}+q_{1}^{R}\right)\\E_{g1e}&=\hbar\left(\Omega+\omega_{R}+q_{1}^{L}-q_{1}^{R}\right)\\E_{e1g}&=\hbar\left(\omega_{L}+\Omega-q_{1}^{L}+q_{1}^{R}\right)\\E_{e1e}&=\hbar\left(\omega_{L}+\Omega+\omega_{R}-q_{1}^{L}-q_{1}^{R}\right).
            \end{split}
        \end{equation}
    \item For $j=2$:  
        \begin{equation}
             \begin{split}
            E_{g2g}&=\hbar\left((\Omega+\delta)+q_{2}^{L}+q_{2}^{R}\right)\\E_{g2e}&=\hbar\left((\Omega+\delta)+\omega_{R}+q_{2}^{L}-q_{2}^{R}\right)\\E_{e2g}&=\hbar\left(\omega_{L}+(\Omega+\delta)-q_{2}^{L}+q_{2}^{R}\right)\\E_{e2e}&=\hbar\left(\omega_{L}+(\Omega+\delta)+\omega_{R}-q_{2}^{L}-q_{2}^{R}\right),
             \end{split}
        \end{equation}
\end{itemize}
where $q_{0}^{\alpha}=(\chi_{01}^{\alpha}+\chi_{02}^{\alpha})$, $q_{1}^{\alpha}=(\chi_{12}^{\alpha}-\chi_{01}^{\alpha})$ and $q_{2}^{\alpha}=(-\chi_{02}^{\alpha}-\chi_{12}^{\alpha})$.
\section{Bohr frequencies}\label{Bohr frequencies}
For the $\alpha-$th TLS, the relevant Bohr frequencies correspond to the transitions $|g\rangle_{\alpha}|j\rangle\leftrightarrow|e\rangle_{\alpha}|j\rangle$, given by
\begin{equation}\label{OmegaLR}
    \omega_{ge,j}^{\alpha}=\hbar\left(\omega_{\alpha}-2q_{j}^{\alpha}\right),\quad j=0,1,2\quad \&\quad \alpha=L,R
\end{equation}
where $q_{0,1,2}^{\alpha}$
encompasses the TLS-qutrit coupling strengths. Analogously, for $M$ we have the following transitions: 
\begin{enumerate}
    \item  For $|0\rangle\leftrightarrow|1\rangle$
\begin{equation}\label{OmegaM01}
    \begin{split}
        \omega_{01,ee}^{M}&=\hbar\left(\Omega+q_{0}^{L}+q_{0}^{R}-q_{1}^{L}-q_{1}^{R}\right)\\\omega_{01,ge}^{M}&=\hbar\left(\Omega-q_{0}^{L}+q_{0}^{R}+q_{1}^{L}-q_{1}^{R}\right)\\\omega_{01,eg}^{M}&=\hbar\left(\Omega+q_{0}^{L}-q_{0}^{R}-q_{1}^{L}+q_{1}^{R}\right)\\\omega_{01,gg}^{M}&=\hbar\left(\Omega-q_{0}^{L}-q_{0}^{R}+q_{1}^{L}+q_{1}^{R}\right)
    \end{split}
\end{equation}
\item For $|0\rangle\leftrightarrow|2\rangle$
\begin{equation}\label{OmegaM02}
    \begin{split}
        \omega_{02,ee}^{M}&=\hbar\left((\Omega+\delta)+q_{0}^{L}+q_{0}^{R}-q_{2}^{L}-q_{2}^{R}\right)\\\omega_{02,ge}^{M}&=\hbar\left((\Omega+\delta)-q_{0}^{L}+q_{0}^{R}+q_{2}^{L}-q_{2}^{R}\right)\\\omega_{02,eg}^{M}&=\hbar\left((\Omega+\delta)+q_{0}^{L}-q_{0}^{R}-q_{2}^{L}+q_{2}^{R}\right)\\\omega_{02,gg}^{M}&=\hbar\left((\Omega+\delta)-q_{0}^{L}-q_{0}^{R}+q_{2}^{L}+q_{2}^{R}\right)
    \end{split}
\end{equation}
\item For $|1\rangle\leftrightarrow|2\rangle$
\begin{equation}\label{OmegaM12}
    \begin{split}
        \omega_{12,ee}^{M}&=\hbar\left(\delta+q_{1}^{L}+q_{1}^{R}-q_{2}^{L}-q_{2}^{R}\right)\\\omega_{12,ge}^{M}&=\hbar\left(\delta-q_{1}^{L}+q_{1}^{R}+q_{2}^{L}-q_{2}^{R}\right)\\\omega_{12,eg}^{M}&=\hbar\left(\delta+q_{1}^{L}-q_{1}^{R}-q_{2}^{L}+q_{2}^{R}\right)\\\omega_{12,gg}^{M}&=\hbar\left(\delta-q_{1}^{L}-q_{1}^{R}+q_{2}^{L}+q_{2}^{R}\right)
    \end{split}
\end{equation}
\end{enumerate}
where $\omega_{ij,lr}^{M}$ are the energy gaps relative to the transitions $|l\rangle_{L}|i\rangle|r\rangle_{R}\leftrightarrow|l\rangle_{L}|j\rangle|r\rangle_{R}$.%
\section{Jump operators}\label{Jump operators}
The general structure of system-reservoir interactions reads $\hat{h}_{\alpha}=\sum_{l}\hat{S}_{l}^{\alpha}\otimes\hat{B}_{l}^{\alpha}$. Thus, given Eq. (\ref{BathInt}), one can identify a single $\hat{S}$ operator for each subsystem, s.t.,
\begin{equation}
    \begin{split}
        \hat{S}^{L,R}&=|e\rangle_{L,R}\langle g|_{L,R}+|g\rangle_{L,R}\langle e|_{L,R}\\\hat{S}^{M}&=\left(|1\rangle\langle0|+|0\rangle\langle1|\right)+\left(|2\rangle\langle0|+|0\rangle\langle2|\right)\\&+\left(|2\rangle\langle1|+|1\rangle\langle2|\right).
    \end{split}
\end{equation}
The jump operators are defined as
\begin{equation}
    \hat{S}_{\omega}^{\alpha}=\sum_{\epsilon^{\prime}-\epsilon=\omega}|\epsilon\rangle\langle\epsilon|\hat{S}^{\alpha}|\epsilon^{\prime}\rangle\langle\epsilon^{\prime}|
\end{equation}
for every fixed energy gap $\omega$, where $\{|\epsilon\rangle\}$ is the Hamiltonian basis, and $\hat{S}^{\alpha\dagger}_{\omega}=\hat{S}_{-\omega}^{\alpha}$. Since $\{|\epsilon\rangle=|E_{l,j,r}\rangle\}$, one has the following set of non-null coefficients
\begin{equation}
    \begin{split}
        \langle E_{l,j,r}|\hat{S}^{L}|E_{l^{\prime},j,r}\rangle&=\left(\delta_{l,e}\delta_{l^{\prime},g}+\delta_{l,g}\delta_{l^{\prime},e}\right)\\\langle E_{l,j,r}|\hat{S}^{M}|E_{l,j^{\prime},r}\rangle&=\left(\delta_{j,1}\delta_{j^{\prime},0}+\delta_{j,0}\delta_{j^{\prime},1}\right)\\&+\left(\delta_{j,2}\delta_{j^{\prime},0}+\delta_{j,0}\delta_{j^{\prime},2}\right)\\&+\left(\delta_{j,2}\delta_{j^{\prime},1}+\delta_{j,1}\delta_{j^{\prime},2}\right)\\\langle E_{l,j,r}|\hat{S}^{R}|E_{l,j,r^{\prime}}\rangle&=\left(\delta_{r,e}\delta_{r^{\prime},g}+\delta_{r,g}\delta_{r^{\prime},e}\right).
    \end{split}
\end{equation}
Thus, it is straightforward to compute the following set of jump operators:
\begin{equation}
    \begin{split}
        \hat{S}_{\omega_{ge,j}^{L}}^{L}&=|g\rangle_{L}\langle e|_{L}\otimes|j\rangle\langle j|\otimes\hat{1}_{R}\\\hat{S}_{\omega_{ge,j}^{R}}^{R}&=\hat{1}_{L}\otimes|j\rangle\langle j|\otimes|g\rangle_{R}\langle e|_{R},\qquad j=0,1,2
    \end{split}
\end{equation}
and
\begin{equation}
    \begin{split}
        \hat{S}_{\omega_{ij,ee}^{M}}^{M}&=|e\rangle_{L}\langle e|_{L}\otimes|i\rangle\langle j|\otimes|e\rangle_{R}\langle e|_{R}\\\hat{S}_{\omega_{ij,ge}^{M}}^{M}&=|g\rangle_{L}\langle g|_{L}\otimes|i\rangle\langle j|\otimes|e\rangle_{R}\langle e|_{R}\\\hat{S}_{\omega_{ij,eg}^{M}}^{M}&=|e\rangle_{L}\langle e|_{L}\otimes|i\rangle\langle j|\otimes|g\rangle_{R}\langle g|_{R}\\\hat{S}_{\omega_{ij,gg}^{M}}^{M}&=|g\rangle_{L}\langle g|_{L}\otimes|i\rangle\langle j|\otimes|g\rangle_{R}\langle g|_{R},
    \end{split}
\end{equation}
with $j>i$. The relevant energy gaps can be found in Appendix \ref{Bohr frequencies}, Eqs. \eqref{OmegaLR} - \eqref{OmegaM12}.
%
%
%
%
%
%
%

%
%
%
%
\section{High-temperature regime}\label{Asymtotics}
\begin{figure}[H]
\includegraphics[width=8.5cm]{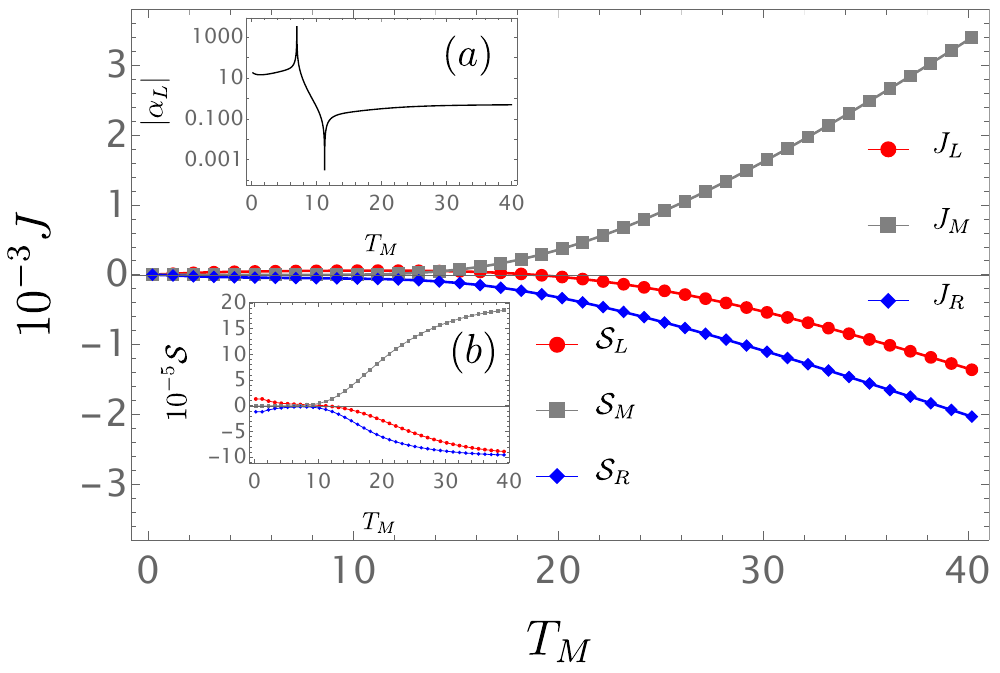}
\caption{High-temperature behavior. The main plot depicts the stationary heat currents $J_{\alpha}$ in relation to the temperature $T_{M}$. Inset: $(a)$ illustrates the amplification factor $|\alpha_{L}|$, showcasing the characteristic transistor effect; $(b)$ presents the differential thermal sensitivity for all terminals. All parameters are consistent with those in Fig. \ref{Fig1}.}
\label{FigNEW}
\end{figure}

In Section \ref{Thermal transistor behavior}, we demonstrated the functionality of the proposed architecture as a quantum thermal transistor within the temperature range $T_{M}\in[T_{R}, T_{L}]$. To provide a comprehensive understanding, we now explore its behavior for temperatures $T_{M}$ at a comparable order of magnitude as $T_{L}$. Figure \ref{FigNEW} illustrates the profiles of heat currents, amplification factor, and differential thermal sensitivities based on the parameters depicted in Figure \ref{Fig1}. Observing the results, it is evident that the currents exhibit a substantial increase with higher temperature gradients. Furthermore, when $T_{M}$ reaches a sufficiently high level, the heat fluxes manifest as $J_{M}>0$ and $J_{L, R}<0$.  This happens due to the boost in the $M$ terminal decay rates, Eq. \eqref{decayR}, resulting in the increase of $J_{M}$ and simultaneous decreasing of $J_{L, R}$, Eq. \eqref{tth}. It is worth mentioning that for fixed spectrum, temperature values are upper bound by the validity of the assumptions utilized for obtaining the master equation. In Figure \ref{FigNEW}\textcolor{blue}{a}, a significant drop in the amplification factor is observed immediately after the characteristic transistor effect peak. This decline is attributed to the region where $\mathcal{S}_{L}$ is approximately null, while $\mathcal{S}_{M}>\mathcal{S}_{L}$. A more detailed analysis of this behavior is presented in Figure \ref{FigNEW}\textcolor{blue}{b}.
\section{Comparison with the analogous qubit transistor}\label{QUBIT}
The three-qubit version of the quantum thermal transistor can be considered by replacing the middle qutrit subsystem with a two-level system. This can be achieved by effectively decoupling the third level from both the TLSs and the $M$ reservoir. The former is guaranteed once we consider $\chi^{L, R}_{02}=\chi^{L, R}_{12}=0$. To perform the latter we generalize the $M$ reservoir interaction and introduce the scalars $\nu_{ij}$ parametrizing the interaction strengths, such that $\hat{\sigma}_{M}^{x}=\sum_{j>i}\nu_{ij}\left(|i\rangle\langle j|+|j\rangle\langle i|\right)$. Hence, we can consider $\nu_{02}=\nu_{12}=0$ to disconnect the third energy level from the reservoir (Eq. \eqref{BathInt}).

\begin{figure}[H]
\center
\includegraphics[width=8.5cm]{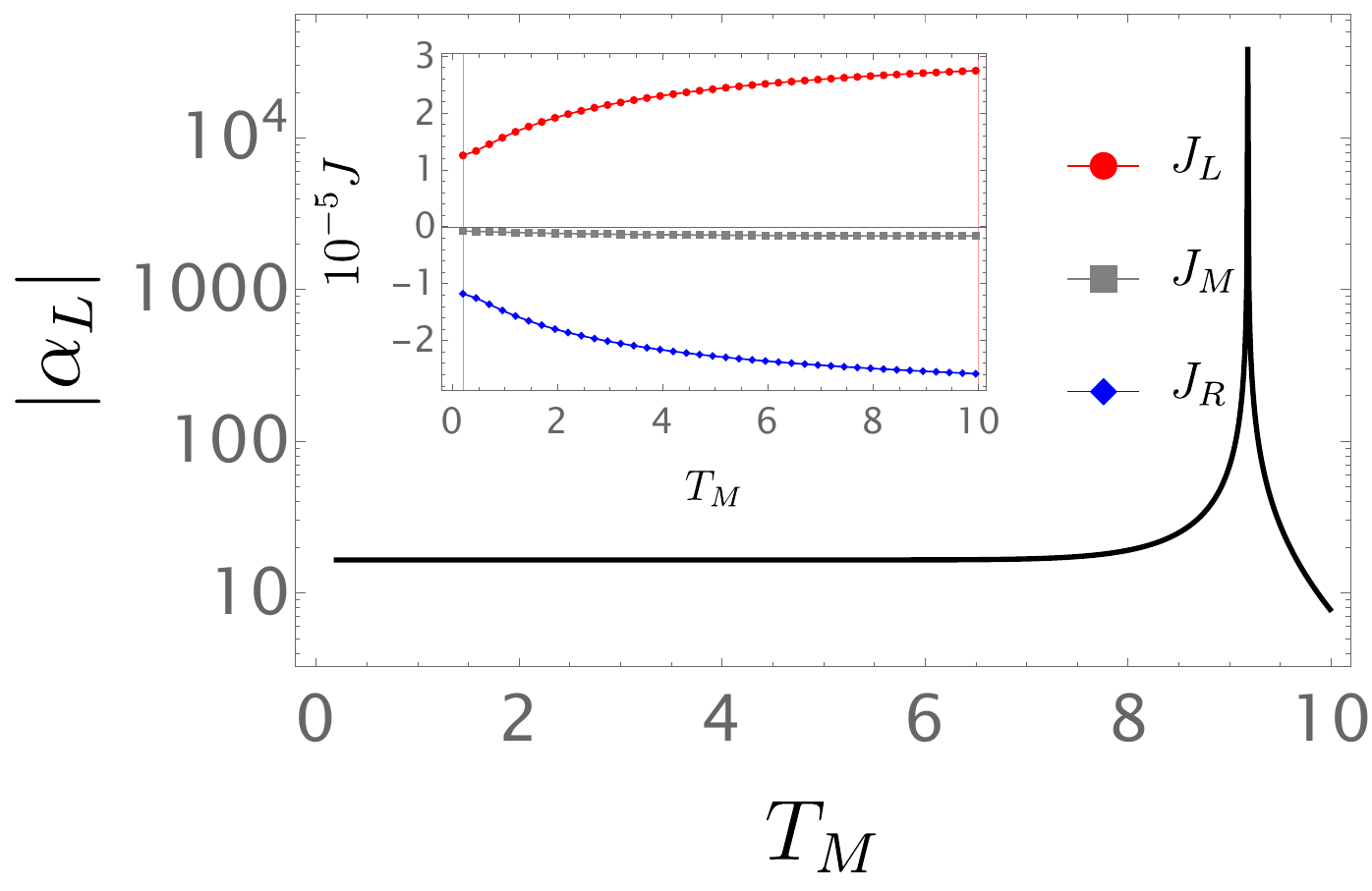}
\caption{$|\alpha_{L}|$ (main figure) and heat currents (inset) behavior for the qubit analog of the quantum thermal transistor. It was assumed 
$\nu_{01}=1$, $\chi^{L}_{01}=33.1\omega_{L}$, and $\chi^{R}_{01}=31.1\omega_{L}$. All remaining parameters as Fig. \ref{Fig1}.
}
\label{Qubit}
\end{figure}

\begin{figure}[H]
\centering
   \begin{subfigure}[b]{0.4\textwidth}
   \includegraphics[width=1\textwidth]{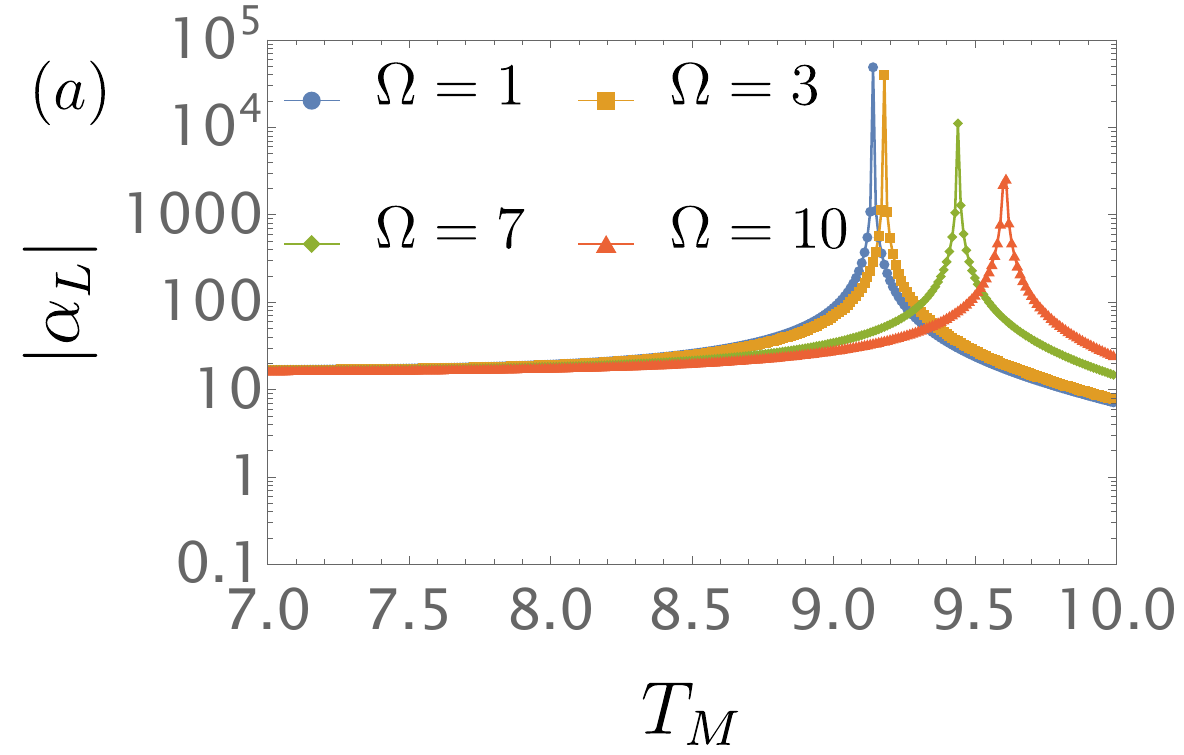}
\end{subfigure}
\begin{subfigure}[b]{0.4\textwidth}
   \includegraphics[width=1\textwidth]{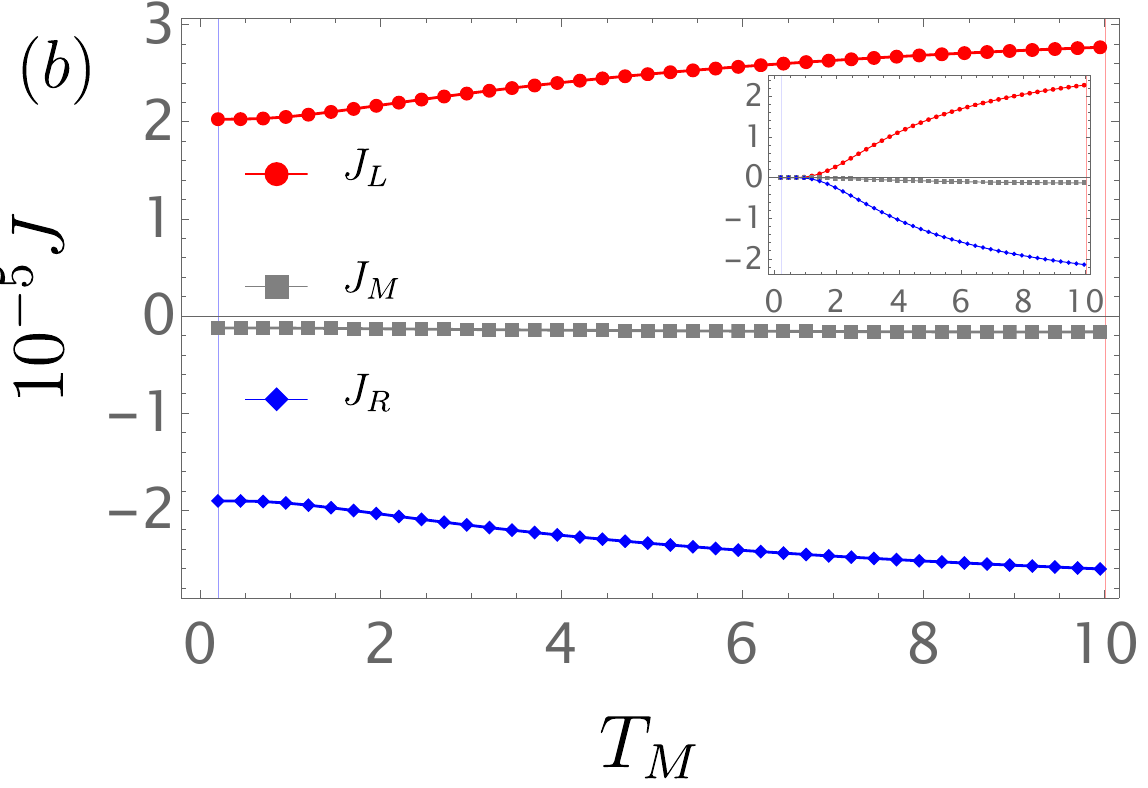}
\end{subfigure}
\begin{subfigure}[b]{0.4\textwidth}
   \includegraphics[width=1\textwidth]{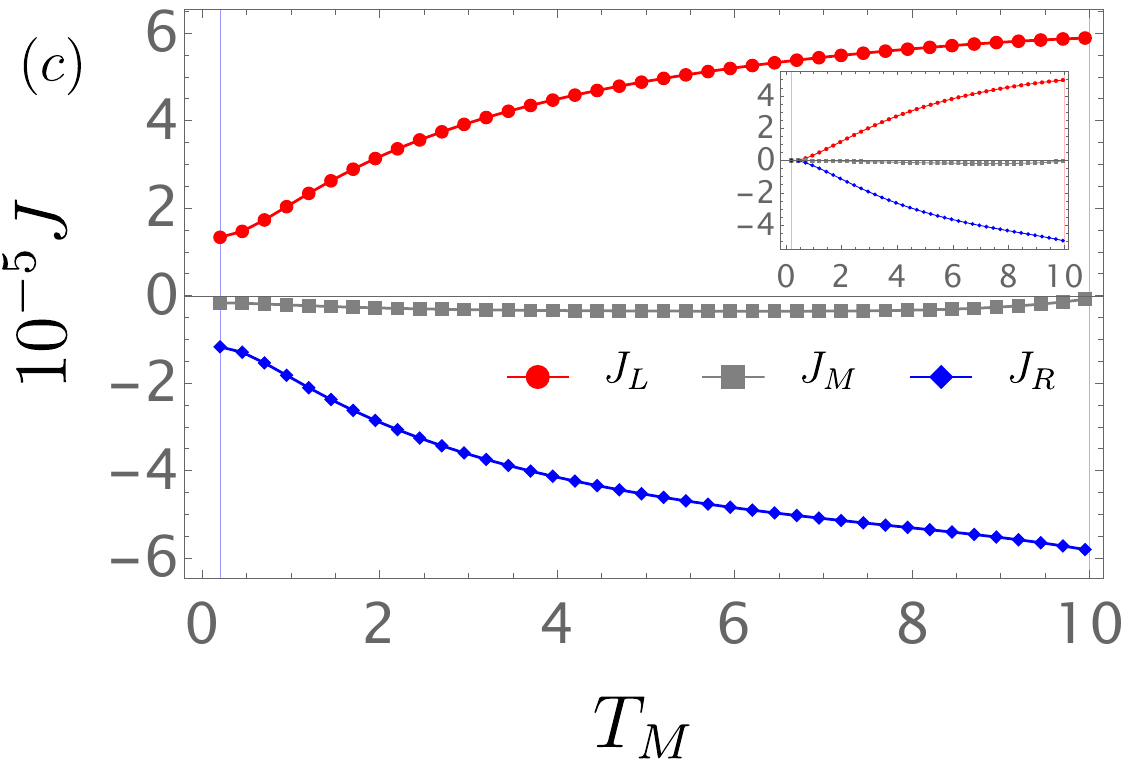}
\end{subfigure}
\caption{Behavior of $(a)$ $|\alpha_{L}|$ for different values of $\Omega$, and $(b)$ heat currents for $\Omega=\omega_{L}$ (main plot) and $\Omega=10\omega_{L}$ (inset) for the qubit analogous; $(c)$ Heat currents for the qutrit assuming $\delta=\omega_{L}$ (main plot) and $\delta=10\omega_{L}$ (inset). The assumed parameters for the qubit (qutrit) are consistent with those presented in Fig. \ref{Qubit} (Fig. \ref{Fig1}).
}
\label{QubitALPHA}
\end{figure}

Figure \ref{Qubit} depicts $|\alpha_L|$ in the main plot and presents the heat currents in the inset for the analogous qubit system, utilizing the same parameters as in Figure \ref{Fig1}. To maintain the system's energy scale and facilitate a fair comparison, the previous internal coupling values were redistributed among the remaining couplings, ensuring $\chi^{L}_{01}=33.1\omega_{L}$ and $\chi^{R}_{01}=31.1\omega_{L}$. Consequently, the transistor effect remains achievable under these conditions. However, it is noteworthy that the attained stationary heat currents are noticeably lower, approximately half, compared to those observed in the qutrit architecture, as shown in Figure \ref{Fig1}.
Fig. \ref{QubitALPHA}\textcolor{blue}{a} and \ref{QubitALPHA}\textcolor{blue}{b} show how the transistor effect and heat currents respond to different middle qubit gaps $\Omega$, respectively. Notably, adjusting this parameter allows for the modulation of the high amplification region, characterized by $|\alpha_L|\geq100$, within the approximate temperature range $T_{M}\in[9.02, 9.73]$. This interval contrasts sharply with the one observed for the qutrit, as shown in Figure \ref{Fig7and8}, which is given by $[5.74, 7.71]$. Consequently, considering that $\Delta T_{M}^{qutrit}\approx 1.97 > \Delta T_{M}^{qubit}\approx 0.71$, the qutrit architecture covers a larger working window, presenting a distinct advantage over its two-level counterpart.
Additionally, Figure \ref{QubitALPHA}\textcolor{blue}{c} displays the heat current profiles for the qutrit considered in Figure \ref{Fig1} for detuning values of $\delta=\omega_{L}$ in the main plot and $\delta=10\omega_{L}$ in the inset. 
%
%
\begin{figure*}
\centering
\begin{minipage}[b]{.32\textwidth}
\includegraphics[width=\linewidth]{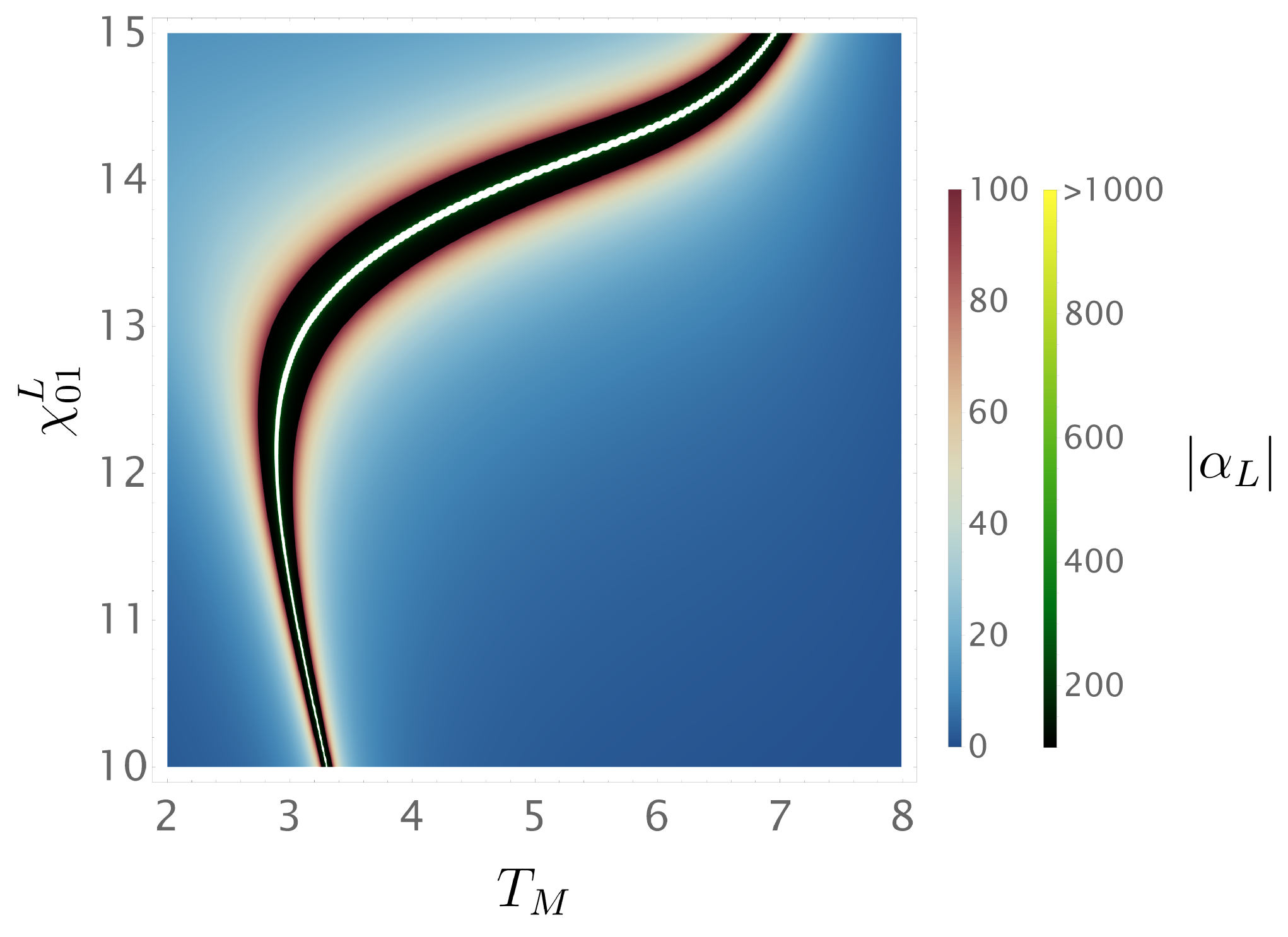}
\includegraphics[width=\linewidth]{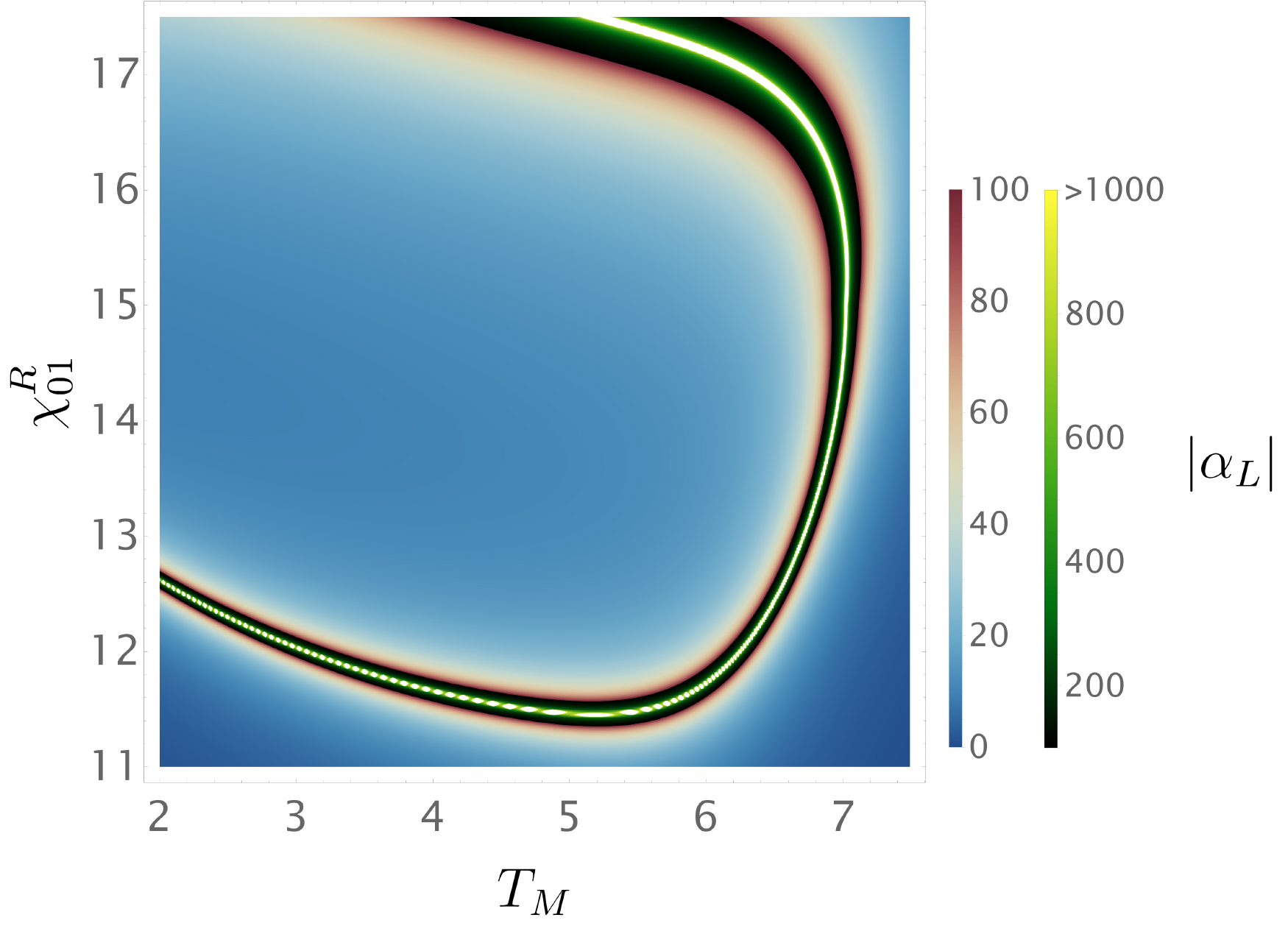}
   \subcaption{}
   \label{X01Density} 
\end{minipage}\quad
\begin{minipage}[b]{.32\textwidth}
\includegraphics[width=\linewidth]{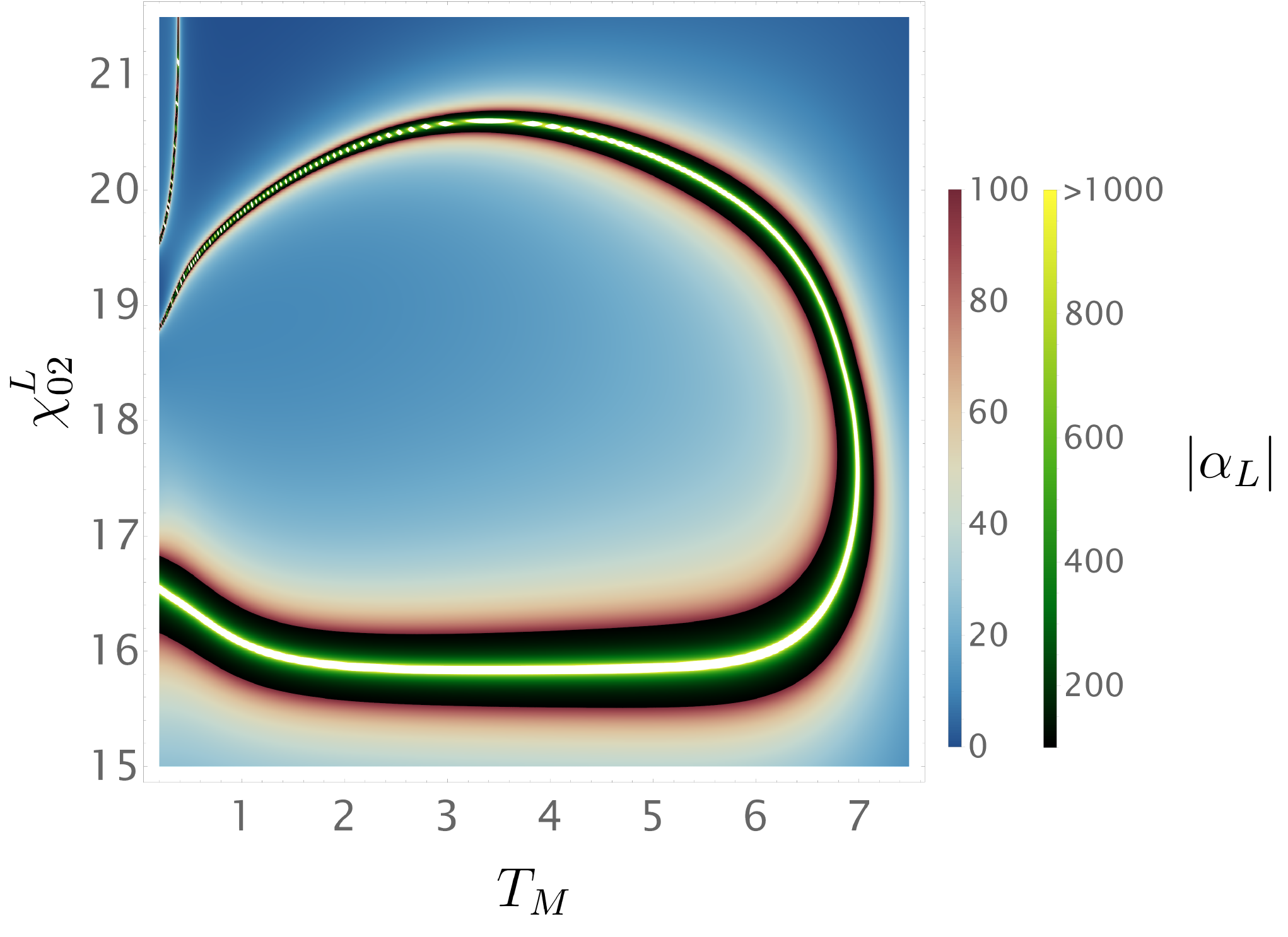}
\includegraphics[width=\linewidth]{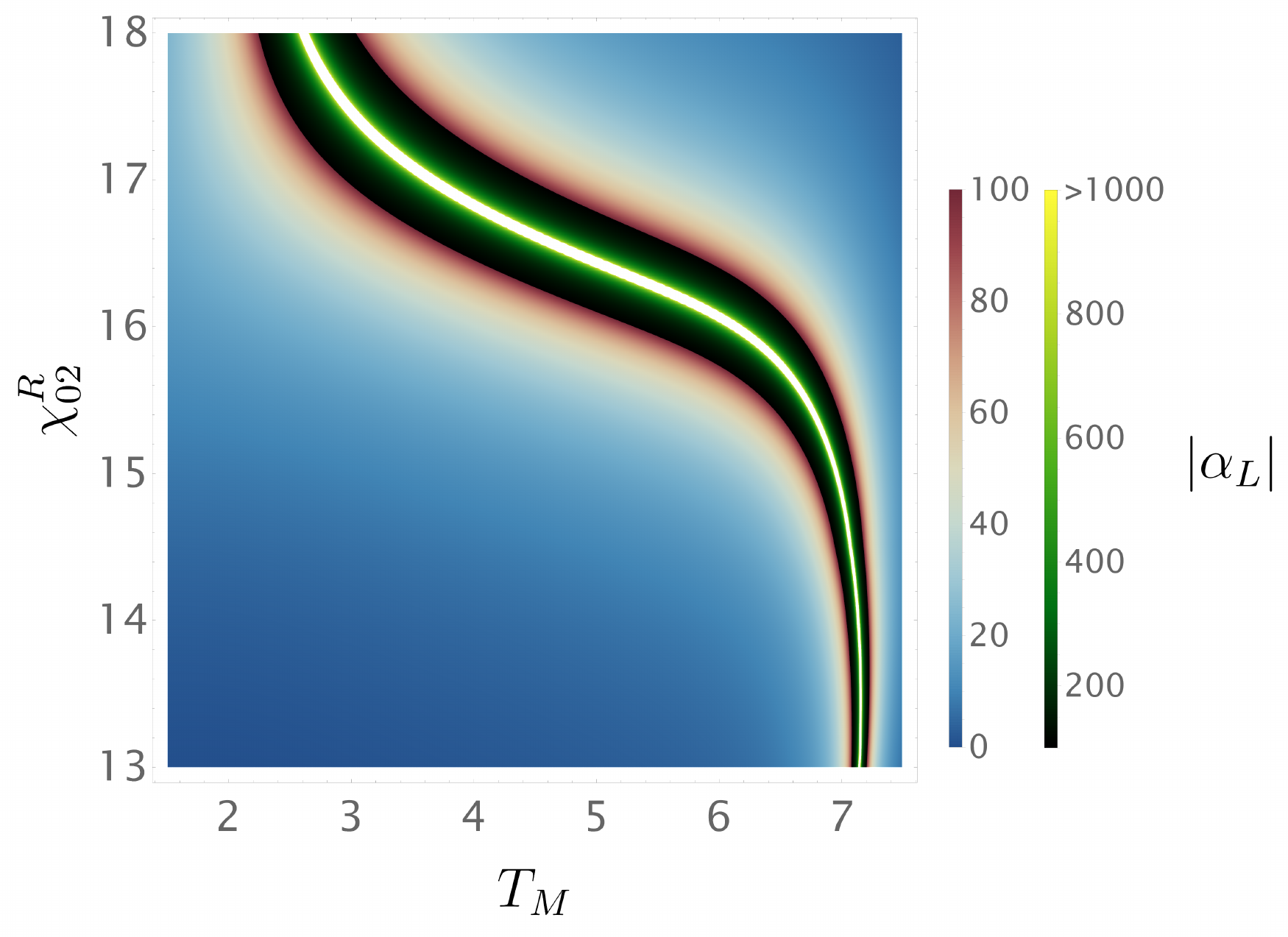}
   \subcaption{}
   \label{X02Density} 
\end{minipage}\quad
\begin{minipage}[b]{.32\textwidth}
\includegraphics[width=\linewidth]{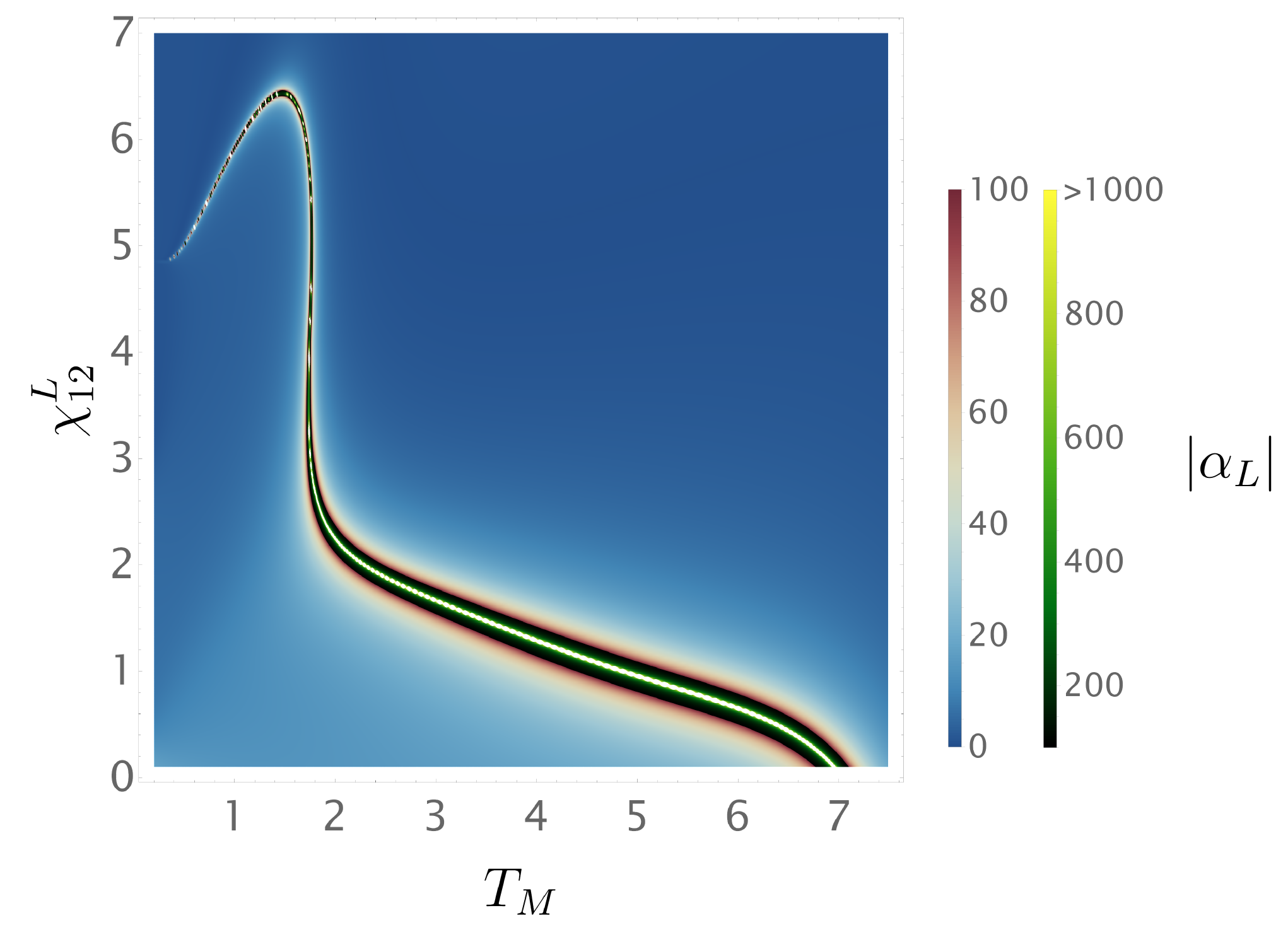}
\includegraphics[width=\linewidth]{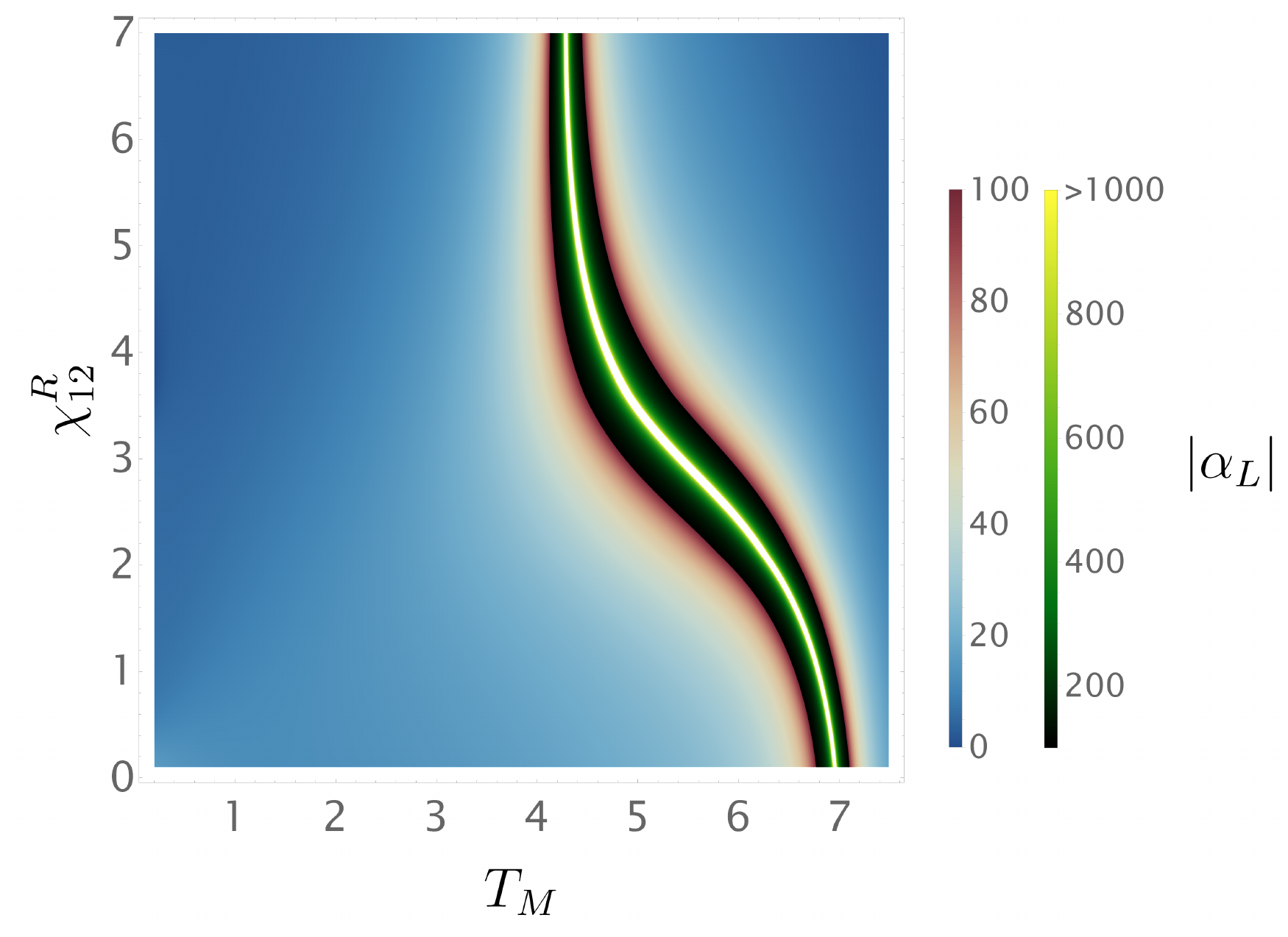}
   \subcaption{}
   \label{X12Density} 
\end{minipage}
\captionsetup{justification=justified}
\caption{Modulus of the amplification factor, $|\alpha_{L}|$, for different temperatures $T_{M}$. For each plot, one coupling constant was modified with all other parameters maintained constant. $(a)$ $\chi_{01}^{L, R}$; $(b)$ $\chi_{02}^{L, R}$; $(c)$ $\chi_{12}^{L, R}$. The same parameters assumed as in Fig \ref{Fig1}, and $\Delta T_{M}=10^{-2}$ and $\Delta \chi_{ij}^{L, R}=2.5 \times 10^{-2}$. \justifying}
\label{FigCouplingstrengths}
\end{figure*}
In the case of the qubit, elevating the qubit energy to a sufficiently high level results in the suppression of currents within a narrow temperature range. This suppression enhances the sensitivity of the currents, effectively transforming the setup into a switch-knob capable of toggling between selected low and high temperatures, where $J_{M}\approx 0$. Similarly, the same effect can be achieved by manipulating the highest energy level of the qutrit. However, the qutrit configuration provides an added advantage by yielding higher currents for a given high value of $T_{M}$: $|J_{L, R}^{qubit}|<|J_{L, R}^{qutrit}|$. This highlights the potential for leveraging additional energy levels in the transistor setting to significantly enhance its performance.
\section{Internal coupling strengths}\label{Coupling strengths}
The TLS-qutrit coupling is expressed as $\hat{V}=\hbar\hat{\sigma}_{L}^{z}\otimes\hat{\chi}_{L}+\hbar\hat{\chi}_{R}\otimes\hat{\sigma}_{R}^{z}$. As expected, the dynamics of the system are entirely influenced by the internal coupling strengths $\chi_{ij}^{L, R}$. These couplings modify the system's energy spectrum and, consequently, the relevant energy gaps (refer to Appendix \ref{Bohr frequencies}). This study focuses primarily on strong internal coupling regimes. Given the existence of six coupling parameters, the system's complexity significantly increases, making its energetic analysis more intricate.
It is crucial to note that the observation of the transistor effect is not uniquely determined for a specific set of parameters. Instead, various constrained regions of the parameter space ensure amplification. To provide further insight, Figure \ref{FigCouplingstrengths} illustrates how the modulus of the amplification factor $|\alpha_{L}|$ behaves when each $\chi_{ij}^{L, R}$ is individually modified at different temperatures $T_{M}$. Thus, in principle, if one can tune the coupling constants, Fig. \ref{FigCouplingstrengths} elucidate the optimal working region, i.e., one can identify the black/yellow areas as the ideal working regimes for a functioning quantum thermal transistor, in contrast, the blue/red areas illustrate the points to be avoided, where $|\alpha_{L}|<100$. Notably, for some particular coupling values one may achieve high amplification that sustains for a wide temperature interval, e.g., for $\chi_{02}^L\approx 16$ we observe $|\alpha_{L}|\geq100$ from $T_{M}\approx 0$ to $T_{M}\approx 7$ (Fig. \ref{FigCouplingstrengths}\textcolor{blue}{b}).

It is crucial to emphasize the following points: $(i)$ The density plots in Figure \ref{FigCouplingstrengths} were generated assuming $\Delta T_{M}=10^{-2}$ and $\Delta \chi_{ij}^{L, R}=2.5 \times 10^{-2}$. Consequently, for more accurate figures, finer resolutions would be necessary; $(ii)$ Not all points in Figure \ref{FigCouplingstrengths} correspond to valid parameter choices. It is imperative to ensure the validity of the assumptions necessary for deriving Eq. \eqref{Master Equation} at every point in the parameter space. These assumptions include the absence of gap degeneracy, adherence to the Markov approximation, and fulfillment of the full-secular approximation.
\bibliography{References}

\begin{thebibliography}{86}%
\makeatletter
\providecommand \@ifxundefined [1]{%
 \@ifx{#1\undefined}
}%
\providecommand \@ifnum [1]{%
 \ifnum #1\expandafter \@firstoftwo
 \else \expandafter \@secondoftwo
 \fi
}%
\providecommand \@ifx [1]{%
 \ifx #1\expandafter \@firstoftwo
 \else \expandafter \@secondoftwo
 \fi
}%
\providecommand \natexlab [1]{#1}%
\providecommand \enquote  [1]{``#1''}%
\providecommand \bibnamefont  [1]{#1}%
\providecommand \bibfnamefont [1]{#1}%
\providecommand \citenamefont [1]{#1}%
\providecommand \href@noop [0]{\@secondoftwo}%
\providecommand \href [0]{\begingroup \@sanitize@url \@href}%
\providecommand \@href[1]{\@@startlink{#1}\@@href}%
\providecommand \@@href[1]{\endgroup#1\@@endlink}%
\providecommand \@sanitize@url [0]{\catcode `\\12\catcode `\$12\catcode `\&12\catcode `\#12\catcode `\^12\catcode `\_12\catcode `\%12\relax}%
\providecommand \@@startlink[1]{}%
\providecommand \@@endlink[0]{}%
\providecommand \url  [0]{\begingroup\@sanitize@url \@url }%
\providecommand \@url [1]{\endgroup\@href {#1}{\urlprefix }}%
\providecommand \urlprefix  [0]{URL }%
\providecommand \Eprint [0]{\href }%
\providecommand \doibase [0]{https://doi.org/}%
\providecommand \selectlanguage [0]{\@gobble}%
\providecommand \bibinfo  [0]{\@secondoftwo}%
\providecommand \bibfield  [0]{\@secondoftwo}%
\providecommand \translation [1]{[#1]}%
\providecommand \BibitemOpen [0]{}%
\providecommand \bibitemStop [0]{}%
\providecommand \bibitemNoStop [0]{.\EOS\space}%
\providecommand \EOS [0]{\spacefactor3000\relax}%
\providecommand \BibitemShut  [1]{\csname bibitem#1\endcsname}%
\let\auto@bib@innerbib\@empty
\bibitem [{\citenamefont {Auff\`eves}(2022)}]{auffeves_qenergy}%
  \BibitemOpen
  \bibfield  {author} {\bibinfo {author} {\bibfnamefont {A.}~\bibnamefont {Auff\`eves}},\ }\bibfield  {title} {\bibinfo {title} {Quantum technologies need a quantum energy initiative},\ }\href {https://doi.org/10.1103/PRXQuantum.3.020101} {\bibfield  {journal} {\bibinfo  {journal} {PRX Quantum}\ }\textbf {\bibinfo {volume} {3}},\ \bibinfo {pages} {020101} (\bibinfo {year} {2022})}\BibitemShut {NoStop}%
\bibitem [{\citenamefont {Deutsch}(2020)}]{PRXQuantum.1.020101}%
  \BibitemOpen
  \bibfield  {author} {\bibinfo {author} {\bibfnamefont {I.~H.}\ \bibnamefont {Deutsch}},\ }\bibfield  {title} {\bibinfo {title} {Harnessing the power of the second quantum revolution},\ }\href {https://doi.org/10.1103/PRXQuantum.1.020101} {\bibfield  {journal} {\bibinfo  {journal} {PRX Quantum}\ }\textbf {\bibinfo {volume} {1}},\ \bibinfo {pages} {020101} (\bibinfo {year} {2020})}\BibitemShut {NoStop}%
\bibitem [{\citenamefont {Berger}\ \emph {et~al.}(2021)\citenamefont {Berger}, \citenamefont {Di~Paolo}, \citenamefont {Forrest}, \citenamefont {Hadfield}, \citenamefont {Sawaya}, \citenamefont {St{\k{e}}ch{\l}y},\ and\ \citenamefont {Thibault}}]{berger2021quantum}%
  \BibitemOpen
  \bibfield  {author} {\bibinfo {author} {\bibfnamefont {C.}~\bibnamefont {Berger}}, \bibinfo {author} {\bibfnamefont {A.}~\bibnamefont {Di~Paolo}}, \bibinfo {author} {\bibfnamefont {T.}~\bibnamefont {Forrest}}, \bibinfo {author} {\bibfnamefont {S.}~\bibnamefont {Hadfield}}, \bibinfo {author} {\bibfnamefont {N.}~\bibnamefont {Sawaya}}, \bibinfo {author} {\bibfnamefont {M.}~\bibnamefont {St{\k{e}}ch{\l}y}},\ and\ \bibinfo {author} {\bibfnamefont {K.}~\bibnamefont {Thibault}},\ }\bibfield  {title} {\bibinfo {title} {Quantum technologies for climate change: Preliminary assessment},\ }\href {https://doi.org/10.48550/arXiv.2107.05362} {\bibfield  {journal} {\bibinfo  {journal} {arXiv:2107.05362}\ } (\bibinfo {year} {2021})}\BibitemShut {NoStop}%
\bibitem [{\citenamefont {Zhu}\ \emph {et~al.}(2022)\citenamefont {Zhu}, \citenamefont {Xie}, \citenamefont {Jing}, \citenamefont {Yu}, \citenamefont {Yu}, \citenamefont {Zhang}, \citenamefont {Qin}, \citenamefont {Duan}, \citenamefont {Rong},\ and\ \citenamefont {Du}}]{PRXEnergy.1.033002}%
  \BibitemOpen
  \bibfield  {author} {\bibinfo {author} {\bibfnamefont {Y.}~\bibnamefont {Zhu}}, \bibinfo {author} {\bibfnamefont {Y.}~\bibnamefont {Xie}}, \bibinfo {author} {\bibfnamefont {K.}~\bibnamefont {Jing}}, \bibinfo {author} {\bibfnamefont {Z.}~\bibnamefont {Yu}}, \bibinfo {author} {\bibfnamefont {H.}~\bibnamefont {Yu}}, \bibinfo {author} {\bibfnamefont {W.}~\bibnamefont {Zhang}}, \bibinfo {author} {\bibfnamefont {X.}~\bibnamefont {Qin}}, \bibinfo {author} {\bibfnamefont {C.-K.}\ \bibnamefont {Duan}}, \bibinfo {author} {\bibfnamefont {X.}~\bibnamefont {Rong}},\ and\ \bibinfo {author} {\bibfnamefont {J.}~\bibnamefont {Du}},\ }\bibfield  {title} {\bibinfo {title} {Sunlight-driven quantum magnetometry},\ }\href {https://doi.org/10.1103/PRXEnergy.1.033002} {\bibfield  {journal} {\bibinfo  {journal} {PRX Energy}\ }\textbf {\bibinfo {volume} {1}},\ \bibinfo {pages} {033002} (\bibinfo {year} {2022})}\BibitemShut {NoStop}%
\bibitem [{\citenamefont {Werren}\ \emph {et~al.}(2023)\citenamefont {Werren}, \citenamefont {Brown},\ and\ \citenamefont {Gauger}}]{PRXEnergy.2.013002}%
  \BibitemOpen
  \bibfield  {author} {\bibinfo {author} {\bibfnamefont {N.}~\bibnamefont {Werren}}, \bibinfo {author} {\bibfnamefont {W.}~\bibnamefont {Brown}},\ and\ \bibinfo {author} {\bibfnamefont {E.~M.}\ \bibnamefont {Gauger}},\ }\bibfield  {title} {\bibinfo {title} {Light harvesting enhanced by quantum ratchet states},\ }\href {https://doi.org/10.1103/PRXEnergy.2.013002} {\bibfield  {journal} {\bibinfo  {journal} {PRX Energy}\ }\textbf {\bibinfo {volume} {2}},\ \bibinfo {pages} {013002} (\bibinfo {year} {2023})}\BibitemShut {NoStop}%
\bibitem [{\citenamefont {Moutinho}\ \emph {et~al.}(2023)\citenamefont {Moutinho}, \citenamefont {Pezzutto}, \citenamefont {Pratapsi}, \citenamefont {da~Silva}, \citenamefont {De~Franceschi}, \citenamefont {Bose}, \citenamefont {Costa},\ and\ \citenamefont {Omar}}]{PRXEnergy.2.033002}%
  \BibitemOpen
  \bibfield  {author} {\bibinfo {author} {\bibfnamefont {J.~a.~P.}\ \bibnamefont {Moutinho}}, \bibinfo {author} {\bibfnamefont {M.}~\bibnamefont {Pezzutto}}, \bibinfo {author} {\bibfnamefont {S.~S.}\ \bibnamefont {Pratapsi}}, \bibinfo {author} {\bibfnamefont {F.~F.}\ \bibnamefont {da~Silva}}, \bibinfo {author} {\bibfnamefont {S.}~\bibnamefont {De~Franceschi}}, \bibinfo {author} {\bibfnamefont {S.}~\bibnamefont {Bose}}, \bibinfo {author} {\bibfnamefont {A.~T.}\ \bibnamefont {Costa}},\ and\ \bibinfo {author} {\bibfnamefont {Y.}~\bibnamefont {Omar}},\ }\bibfield  {title} {\bibinfo {title} {Quantum dynamics for energetic advantage in a charge-based classical full adder},\ }\href {https://doi.org/10.1103/PRXEnergy.2.033002} {\bibfield  {journal} {\bibinfo  {journal} {PRX Energy}\ }\textbf {\bibinfo {volume} {2}},\ \bibinfo {pages} {033002} (\bibinfo {year} {2023})}\BibitemShut {NoStop}%
\bibitem [{\citenamefont {Senior}\ \emph {et~al.}(2020)\citenamefont {Senior}, \citenamefont {Gubaydullin}, \citenamefont {Karimi}, \citenamefont {Peltonen}, \citenamefont {Ankerhold},\ and\ \citenamefont {Pekola}}]{senior2020heat}%
  \BibitemOpen
  \bibfield  {author} {\bibinfo {author} {\bibfnamefont {J.}~\bibnamefont {Senior}}, \bibinfo {author} {\bibfnamefont {A.}~\bibnamefont {Gubaydullin}}, \bibinfo {author} {\bibfnamefont {B.}~\bibnamefont {Karimi}}, \bibinfo {author} {\bibfnamefont {J.~T.}\ \bibnamefont {Peltonen}}, \bibinfo {author} {\bibfnamefont {J.}~\bibnamefont {Ankerhold}},\ and\ \bibinfo {author} {\bibfnamefont {J.~P.}\ \bibnamefont {Pekola}},\ }\bibfield  {title} {\bibinfo {title} {Heat rectification via a superconducting artificial atom},\ }\href {https://doi.org/10.1038/s42005-020-0307-5} {\bibfield  {journal} {\bibinfo  {journal} {Communications Physics}\ }\textbf {\bibinfo {volume} {3}},\ \bibinfo {pages} {40} (\bibinfo {year} {2020})}\BibitemShut {NoStop}%
\bibitem [{\citenamefont {Wang}\ and\ \citenamefont {Li}(2007)}]{wang2007thermal}%
  \BibitemOpen
  \bibfield  {author} {\bibinfo {author} {\bibfnamefont {L.}~\bibnamefont {Wang}}\ and\ \bibinfo {author} {\bibfnamefont {B.}~\bibnamefont {Li}},\ }\bibfield  {title} {\bibinfo {title} {Thermal logic gates: Computation with phonons},\ }\href {https://doi.org/10.1103/PhysRevLett.99.177208} {\bibfield  {journal} {\bibinfo  {journal} {Phys. Rev. Lett.}\ }\textbf {\bibinfo {volume} {99}},\ \bibinfo {pages} {177208} (\bibinfo {year} {2007})}\BibitemShut {NoStop}%
\bibitem [{\citenamefont {Kathmann}\ \emph {et~al.}(2020)\citenamefont {Kathmann}, \citenamefont {Reina}, \citenamefont {Messina}, \citenamefont {Ben-Abdallah},\ and\ \citenamefont {Biehs}}]{Kathmann_2020}%
  \BibitemOpen
  \bibfield  {author} {\bibinfo {author} {\bibfnamefont {C.}~\bibnamefont {Kathmann}}, \bibinfo {author} {\bibfnamefont {M.}~\bibnamefont {Reina}}, \bibinfo {author} {\bibfnamefont {R.}~\bibnamefont {Messina}}, \bibinfo {author} {\bibfnamefont {P.}~\bibnamefont {Ben-Abdallah}},\ and\ \bibinfo {author} {\bibfnamefont {S.-A.}\ \bibnamefont {Biehs}},\ }\bibfield  {title} {\bibinfo {title} {Scalable radiative thermal logic gates based on nanoparticle networks},\ }\href {http://dx.doi.org/10.1038/s41598-020-60603-4} {\bibfield  {journal} {\bibinfo  {journal} {Scientific Reports}\ }\textbf {\bibinfo {volume} {10}},\ \bibinfo {pages} {3596} (\bibinfo {year} {2020})}\BibitemShut {NoStop}%
\bibitem [{\citenamefont {Lipka-Bartosik}\ \emph {et~al.}(2023)\citenamefont {Lipka-Bartosik}, \citenamefont {Perarnau-Llobet},\ and\ \citenamefont {Brunner}}]{lipkabartosik2023thermodynamic}%
  \BibitemOpen
  \bibfield  {author} {\bibinfo {author} {\bibfnamefont {P.}~\bibnamefont {Lipka-Bartosik}}, \bibinfo {author} {\bibfnamefont {M.}~\bibnamefont {Perarnau-Llobet}},\ and\ \bibinfo {author} {\bibfnamefont {N.}~\bibnamefont {Brunner}},\ }\bibfield  {title} {\bibinfo {title} {Thermodynamic computing via autonomous quantum thermal machines},\ }\href {https://doi.org/10.48550/arXiv.2308.15905} {\bibfield  {journal} {\bibinfo  {journal} {arXiv:2308.15905}\ } (\bibinfo {year} {2023})}\BibitemShut {NoStop}%
\bibitem [{\citenamefont {Conte}\ \emph {et~al.}(2019)\citenamefont {Conte}, \citenamefont {DeBenedictis}, \citenamefont {Ganesh}, \citenamefont {Hylton}, \citenamefont {Strachan}, \citenamefont {Williams}, \citenamefont {Alemi}, \citenamefont {Altenberg}, \citenamefont {Crooks}, \citenamefont {Crutchfield} \emph {et~al.}}]{conte2019thermodynamic}%
  \BibitemOpen
  \bibfield  {author} {\bibinfo {author} {\bibfnamefont {T.}~\bibnamefont {Conte}}, \bibinfo {author} {\bibfnamefont {E.}~\bibnamefont {DeBenedictis}}, \bibinfo {author} {\bibfnamefont {N.}~\bibnamefont {Ganesh}}, \bibinfo {author} {\bibfnamefont {T.}~\bibnamefont {Hylton}}, \bibinfo {author} {\bibfnamefont {J.~P.}\ \bibnamefont {Strachan}}, \bibinfo {author} {\bibfnamefont {R.~S.}\ \bibnamefont {Williams}}, \bibinfo {author} {\bibfnamefont {A.}~\bibnamefont {Alemi}}, \bibinfo {author} {\bibfnamefont {L.}~\bibnamefont {Altenberg}}, \bibinfo {author} {\bibfnamefont {G.}~\bibnamefont {Crooks}}, \bibinfo {author} {\bibfnamefont {J.}~\bibnamefont {Crutchfield}}, \emph {et~al.},\ }\bibfield  {title} {\bibinfo {title} {Thermodynamic computing},\ }\href {https://arxiv.org/abs/1911.01968} {\bibfield  {journal} {\bibinfo  {journal} {arXiv:1911.01968}\ } (\bibinfo {year} {2019})}\BibitemShut {NoStop}%
\bibitem [{\citenamefont {Niu}\ \emph {et~al.}(2023)\citenamefont {Niu}, \citenamefont {Sopp}, \citenamefont {Lodi}, \citenamefont {Gee}, \citenamefont {Kong}, \citenamefont {Pei}, \citenamefont {Gehring}, \citenamefont {N{\"a}gele}, \citenamefont {Lau}, \citenamefont {Ma}, \citenamefont {Liu}, \citenamefont {Narita}, \citenamefont {Mol}, \citenamefont {Burghard}, \citenamefont {M{\"u}llen}, \citenamefont {Mai}, \citenamefont {Feng},\ and\ \citenamefont {Bogani}}]{niu2023exceptionally}%
  \BibitemOpen
  \bibfield  {author} {\bibinfo {author} {\bibfnamefont {W.}~\bibnamefont {Niu}}, \bibinfo {author} {\bibfnamefont {S.}~\bibnamefont {Sopp}}, \bibinfo {author} {\bibfnamefont {A.}~\bibnamefont {Lodi}}, \bibinfo {author} {\bibfnamefont {A.}~\bibnamefont {Gee}}, \bibinfo {author} {\bibfnamefont {F.}~\bibnamefont {Kong}}, \bibinfo {author} {\bibfnamefont {T.}~\bibnamefont {Pei}}, \bibinfo {author} {\bibfnamefont {P.}~\bibnamefont {Gehring}}, \bibinfo {author} {\bibfnamefont {J.}~\bibnamefont {N{\"a}gele}}, \bibinfo {author} {\bibfnamefont {C.~S.}\ \bibnamefont {Lau}}, \bibinfo {author} {\bibfnamefont {J.}~\bibnamefont {Ma}}, \bibinfo {author} {\bibfnamefont {J.}~\bibnamefont {Liu}}, \bibinfo {author} {\bibfnamefont {A.}~\bibnamefont {Narita}}, \bibinfo {author} {\bibfnamefont {J.}~\bibnamefont {Mol}}, \bibinfo {author} {\bibfnamefont {M.}~\bibnamefont {Burghard}}, \bibinfo {author} {\bibfnamefont {K.}~\bibnamefont {M{\"u}llen}}, \bibinfo {author} {\bibfnamefont {Y.}~\bibnamefont {Mai}}, \bibinfo {author}
  {\bibfnamefont {X.}~\bibnamefont {Feng}},\ and\ \bibinfo {author} {\bibfnamefont {L.}~\bibnamefont {Bogani}},\ }\bibfield  {title} {\bibinfo {title} {Exceptionally clean single-electron transistors from solutions of molecular graphene nanoribbons},\ }\href {https://doi.org/10.1038/s41563-022-01460-6} {\bibfield  {journal} {\bibinfo  {journal} {Nature Materials}\ }\textbf {\bibinfo {volume} {22}},\ \bibinfo {pages} {180} (\bibinfo {year} {2023})}\BibitemShut {NoStop}%
\bibitem [{\citenamefont {Binder}\ \emph {et~al.}(2018)\citenamefont {Binder}, \citenamefont {Correa}, \citenamefont {Gogolin}, \citenamefont {Anders},\ and\ \citenamefont {Adesso}}]{binder2018thermodynamics}%
  \BibitemOpen
  \bibfield  {author} {\bibinfo {author} {\bibfnamefont {F.}~\bibnamefont {Binder}}, \bibinfo {author} {\bibfnamefont {L.~A.}\ \bibnamefont {Correa}}, \bibinfo {author} {\bibfnamefont {C.}~\bibnamefont {Gogolin}}, \bibinfo {author} {\bibfnamefont {J.}~\bibnamefont {Anders}},\ and\ \bibinfo {author} {\bibfnamefont {G.}~\bibnamefont {Adesso}},\ }\bibfield  {title} {\bibinfo {title} {Thermodynamics in the quantum regime},\ }\href {https://doi.org/https://doi.org/10.1007/978-3-319-99046-0} {\bibfield  {journal} {\bibinfo  {journal} {Fundamental Theories of Physics}\ }\textbf {\bibinfo {volume} {195}},\ \bibinfo {pages} {1} (\bibinfo {year} {2018})}\BibitemShut {NoStop}%
\bibitem [{\citenamefont {Pekola}(2015)}]{pekola2015towards}%
  \BibitemOpen
  \bibfield  {author} {\bibinfo {author} {\bibfnamefont {J.~P.}\ \bibnamefont {Pekola}},\ }\bibfield  {title} {\bibinfo {title} {Towards quantum thermodynamics in electronic circuits},\ }\href {https://doi.org/10.1038/nphys3169} {\bibfield  {journal} {\bibinfo  {journal} {Nature Physics}\ }\textbf {\bibinfo {volume} {11}},\ \bibinfo {pages} {118} (\bibinfo {year} {2015})}\BibitemShut {NoStop}%
\bibitem [{\citenamefont {Pekola}\ and\ \citenamefont {Khaymovich}(2019)}]{pekola2019thermodynamics}%
  \BibitemOpen
  \bibfield  {author} {\bibinfo {author} {\bibfnamefont {J.}~\bibnamefont {Pekola}}\ and\ \bibinfo {author} {\bibfnamefont {I.}~\bibnamefont {Khaymovich}},\ }\bibfield  {title} {\bibinfo {title} {Thermodynamics in single-electron circuits and superconducting qubits},\ }\href {https://doi.org/10.1146/annurev-conmatphys-033117-054120} {\bibfield  {journal} {\bibinfo  {journal} {Annual Review of Condensed Matter Physics}\ }\textbf {\bibinfo {volume} {10}},\ \bibinfo {pages} {193} (\bibinfo {year} {2019})}\BibitemShut {NoStop}%
\bibitem [{\citenamefont {Pekola}\ and\ \citenamefont {Karimi}(2021)}]{pekola2021colloquium}%
  \BibitemOpen
  \bibfield  {author} {\bibinfo {author} {\bibfnamefont {J.~P.}\ \bibnamefont {Pekola}}\ and\ \bibinfo {author} {\bibfnamefont {B.}~\bibnamefont {Karimi}},\ }\bibfield  {title} {\bibinfo {title} {Colloquium: Quantum heat transport in condensed matter systems},\ }\href {https://doi.org/10.1103/RevModPhys.93.041001} {\bibfield  {journal} {\bibinfo  {journal} {Rev. Mod. Phys.}\ }\textbf {\bibinfo {volume} {93}},\ \bibinfo {pages} {041001} (\bibinfo {year} {2021})}\BibitemShut {NoStop}%
\bibitem [{\citenamefont {Garrido}\ \emph {et~al.}(2001)\citenamefont {Garrido}, \citenamefont {Hurtado},\ and\ \citenamefont {Nadrowski}}]{garrido2001simple}%
  \BibitemOpen
  \bibfield  {author} {\bibinfo {author} {\bibfnamefont {P.~L.}\ \bibnamefont {Garrido}}, \bibinfo {author} {\bibfnamefont {P.~I.}\ \bibnamefont {Hurtado}},\ and\ \bibinfo {author} {\bibfnamefont {B.}~\bibnamefont {Nadrowski}},\ }\bibfield  {title} {\bibinfo {title} {Simple one-dimensional model of heat conduction which obeys fourier's law},\ }\href {https://doi.org/10.1103/PhysRevLett.86.5486} {\bibfield  {journal} {\bibinfo  {journal} {Phys. Rev. Lett.}\ }\textbf {\bibinfo {volume} {86}},\ \bibinfo {pages} {5486} (\bibinfo {year} {2001})}\BibitemShut {NoStop}%
\bibitem [{\citenamefont {Dubi}\ and\ \citenamefont {Di~Ventra}(2011)}]{dubi2011colloquium}%
  \BibitemOpen
  \bibfield  {author} {\bibinfo {author} {\bibfnamefont {Y.}~\bibnamefont {Dubi}}\ and\ \bibinfo {author} {\bibfnamefont {M.}~\bibnamefont {Di~Ventra}},\ }\bibfield  {title} {\bibinfo {title} {Colloquium: Heat flow and thermoelectricity in atomic and molecular junctions},\ }\href {https://doi.org/10.1103/RevModPhys.83.131} {\bibfield  {journal} {\bibinfo  {journal} {Rev. Mod. Phys.}\ }\textbf {\bibinfo {volume} {83}},\ \bibinfo {pages} {131} (\bibinfo {year} {2011})}\BibitemShut {NoStop}%
\bibitem [{\citenamefont {Bhattacharjee}\ and\ \citenamefont {Dutta}(2021)}]{bhattacharjee2021quantum}%
  \BibitemOpen
  \bibfield  {author} {\bibinfo {author} {\bibfnamefont {S.}~\bibnamefont {Bhattacharjee}}\ and\ \bibinfo {author} {\bibfnamefont {A.}~\bibnamefont {Dutta}},\ }\bibfield  {title} {\bibinfo {title} {Quantum thermal machines and batteries},\ }\href {https://doi.org/10.1140/epjb/s10051-021-00235-3} {\bibfield  {journal} {\bibinfo  {journal} {The European Physical Journal B}\ }\textbf {\bibinfo {volume} {94}},\ \bibinfo {pages} {239} (\bibinfo {year} {2021})}\BibitemShut {NoStop}%
\bibitem [{\citenamefont {Myers}\ \emph {et~al.}(2022)\citenamefont {Myers}, \citenamefont {Abah},\ and\ \citenamefont {Deffner}}]{myers2022quantum}%
  \BibitemOpen
  \bibfield  {author} {\bibinfo {author} {\bibfnamefont {N.~M.}\ \bibnamefont {Myers}}, \bibinfo {author} {\bibfnamefont {O.}~\bibnamefont {Abah}},\ and\ \bibinfo {author} {\bibfnamefont {S.}~\bibnamefont {Deffner}},\ }\bibfield  {title} {\bibinfo {title} {{Quantum thermodynamic devices: From theoretical proposals to experimental reality}},\ }\href {https://doi.org/10.1116/5.0083192} {\bibfield  {journal} {\bibinfo  {journal} {AVS Quantum Science}\ }\textbf {\bibinfo {volume} {4}},\ \bibinfo {pages} {027101} (\bibinfo {year} {2022})}\BibitemShut {NoStop}%
\bibitem [{\citenamefont {Ronzani}\ \emph {et~al.}(2018)\citenamefont {Ronzani}, \citenamefont {Karimi}, \citenamefont {Senior}, \citenamefont {Chang}, \citenamefont {Peltonen}, \citenamefont {Chen},\ and\ \citenamefont {Pekola}}]{ronzani2018tunable}%
  \BibitemOpen
  \bibfield  {author} {\bibinfo {author} {\bibfnamefont {A.}~\bibnamefont {Ronzani}}, \bibinfo {author} {\bibfnamefont {B.}~\bibnamefont {Karimi}}, \bibinfo {author} {\bibfnamefont {J.}~\bibnamefont {Senior}}, \bibinfo {author} {\bibfnamefont {Y.-C.}\ \bibnamefont {Chang}}, \bibinfo {author} {\bibfnamefont {J.~T.}\ \bibnamefont {Peltonen}}, \bibinfo {author} {\bibfnamefont {C.}~\bibnamefont {Chen}},\ and\ \bibinfo {author} {\bibfnamefont {J.~P.}\ \bibnamefont {Pekola}},\ }\bibfield  {title} {\bibinfo {title} {Tunable photonic heat transport in a quantum heat valve},\ }\href {https://doi.org/10.1038/s41567-018-0199-4} {\bibfield  {journal} {\bibinfo  {journal} {Nature Physics}\ }\textbf {\bibinfo {volume} {14}},\ \bibinfo {pages} {991} (\bibinfo {year} {2018})}\BibitemShut {NoStop}%
\bibitem [{\citenamefont {Dutta}\ \emph {et~al.}(2020)\citenamefont {Dutta}, \citenamefont {Majidi}, \citenamefont {Talarico}, \citenamefont {Lo~Gullo}, \citenamefont {Courtois},\ and\ \citenamefont {Winkelmann}}]{PRL_heatValve}%
  \BibitemOpen
  \bibfield  {author} {\bibinfo {author} {\bibfnamefont {B.}~\bibnamefont {Dutta}}, \bibinfo {author} {\bibfnamefont {D.}~\bibnamefont {Majidi}}, \bibinfo {author} {\bibfnamefont {N.~W.}\ \bibnamefont {Talarico}}, \bibinfo {author} {\bibfnamefont {N.}~\bibnamefont {Lo~Gullo}}, \bibinfo {author} {\bibfnamefont {H.}~\bibnamefont {Courtois}},\ and\ \bibinfo {author} {\bibfnamefont {C.~B.}\ \bibnamefont {Winkelmann}},\ }\bibfield  {title} {\bibinfo {title} {Single-quantum-dot heat valve},\ }\href {https://doi.org/10.1103/PhysRevLett.125.237701} {\bibfield  {journal} {\bibinfo  {journal} {Phys. Rev. Lett.}\ }\textbf {\bibinfo {volume} {125}},\ \bibinfo {pages} {237701} (\bibinfo {year} {2020})}\BibitemShut {NoStop}%
\bibitem [{\citenamefont {Terraneo}\ \emph {et~al.}(2002)\citenamefont {Terraneo}, \citenamefont {Peyrard},\ and\ \citenamefont {Casati}}]{terraneo2002controlling}%
  \BibitemOpen
  \bibfield  {author} {\bibinfo {author} {\bibfnamefont {M.}~\bibnamefont {Terraneo}}, \bibinfo {author} {\bibfnamefont {M.}~\bibnamefont {Peyrard}},\ and\ \bibinfo {author} {\bibfnamefont {G.}~\bibnamefont {Casati}},\ }\bibfield  {title} {\bibinfo {title} {Controlling the energy flow in nonlinear lattices: A model for a thermal rectifier},\ }\href {https://doi.org/10.1103/PhysRevLett.88.094302} {\bibfield  {journal} {\bibinfo  {journal} {Phys. Rev. Lett.}\ }\textbf {\bibinfo {volume} {88}},\ \bibinfo {pages} {094302} (\bibinfo {year} {2002})}\BibitemShut {NoStop}%
\bibitem [{\citenamefont {Scheibner}\ \emph {et~al.}(2008)\citenamefont {Scheibner}, \citenamefont {König}, \citenamefont {Reuter}, \citenamefont {Wieck}, \citenamefont {Gould}, \citenamefont {Buhmann},\ and\ \citenamefont {Molenkamp}}]{scheibner2008quantum}%
  \BibitemOpen
  \bibfield  {author} {\bibinfo {author} {\bibfnamefont {R.}~\bibnamefont {Scheibner}}, \bibinfo {author} {\bibfnamefont {M.}~\bibnamefont {König}}, \bibinfo {author} {\bibfnamefont {D.}~\bibnamefont {Reuter}}, \bibinfo {author} {\bibfnamefont {A.~D.}\ \bibnamefont {Wieck}}, \bibinfo {author} {\bibfnamefont {C.}~\bibnamefont {Gould}}, \bibinfo {author} {\bibfnamefont {H.}~\bibnamefont {Buhmann}},\ and\ \bibinfo {author} {\bibfnamefont {L.~W.}\ \bibnamefont {Molenkamp}},\ }\bibfield  {title} {\bibinfo {title} {Quantum dot as thermal rectifier},\ }\href {https://doi.org/10.1088/1367-2630/10/8/083016} {\bibfield  {journal} {\bibinfo  {journal} {New Journal of Physics}\ }\textbf {\bibinfo {volume} {10}},\ \bibinfo {pages} {083016} (\bibinfo {year} {2008})}\BibitemShut {NoStop}%
\bibitem [{\citenamefont {Werlang}\ \emph {et~al.}(2014)\citenamefont {Werlang}, \citenamefont {Marchiori}, \citenamefont {Cornelio},\ and\ \citenamefont {Valente}}]{werlang2014optimal}%
  \BibitemOpen
  \bibfield  {author} {\bibinfo {author} {\bibfnamefont {T.}~\bibnamefont {Werlang}}, \bibinfo {author} {\bibfnamefont {M.~A.}\ \bibnamefont {Marchiori}}, \bibinfo {author} {\bibfnamefont {M.~F.}\ \bibnamefont {Cornelio}},\ and\ \bibinfo {author} {\bibfnamefont {D.}~\bibnamefont {Valente}},\ }\bibfield  {title} {\bibinfo {title} {Optimal rectification in the ultrastrong coupling regime},\ }\href {https://doi.org/10.1103/PhysRevE.89.062109} {\bibfield  {journal} {\bibinfo  {journal} {Phys. Rev. E}\ }\textbf {\bibinfo {volume} {89}},\ \bibinfo {pages} {062109} (\bibinfo {year} {2014})}\BibitemShut {NoStop}%
\bibitem [{\citenamefont {Marcos-Vicioso}\ \emph {et~al.}(2018)\citenamefont {Marcos-Vicioso}, \citenamefont {L\'opez-Jurado}, \citenamefont {Ruiz-Garcia},\ and\ \citenamefont {S\'anchez}}]{Markos_2018}%
  \BibitemOpen
  \bibfield  {author} {\bibinfo {author} {\bibfnamefont {A.}~\bibnamefont {Marcos-Vicioso}}, \bibinfo {author} {\bibfnamefont {C.}~\bibnamefont {L\'opez-Jurado}}, \bibinfo {author} {\bibfnamefont {M.}~\bibnamefont {Ruiz-Garcia}},\ and\ \bibinfo {author} {\bibfnamefont {R.}~\bibnamefont {S\'anchez}},\ }\bibfield  {title} {\bibinfo {title} {Thermal rectification with interacting electronic channels: Exploiting degeneracy, quantum superpositions, and interference},\ }\href {https://doi.org/10.1103/PhysRevB.98.035414} {\bibfield  {journal} {\bibinfo  {journal} {Phys. Rev. B}\ }\textbf {\bibinfo {volume} {98}},\ \bibinfo {pages} {035414} (\bibinfo {year} {2018})}\BibitemShut {NoStop}%
\bibitem [{\citenamefont {Karg\ifmmode \imath \else~\i \fi{}}\ \emph {et~al.}(2019)\citenamefont {Karg\ifmmode \imath \else~\i \fi{}}, \citenamefont {Naseem}, \citenamefont {Opatrn\'y}, \citenamefont {M\"ustecapl\ifmmode \imath \else \i \fi{}o\ifmmode~\breve{g}\else \u{g}\fi{}lu},\ and\ \citenamefont {Kurizki}}]{PhysRevE.99.042121}%
  \BibitemOpen
  \bibfield  {author} {\bibinfo {author} {\bibfnamefont {C.}~\bibnamefont {Karg\ifmmode \imath \else~\i \fi{}}}, \bibinfo {author} {\bibfnamefont {M.~T.}\ \bibnamefont {Naseem}}, \bibinfo {author} {\bibfnamefont {T.~c.~v.}\ \bibnamefont {Opatrn\'y}}, \bibinfo {author} {\bibfnamefont {O.~E.}\ \bibnamefont {M\"ustecapl\ifmmode \imath \else \i \fi{}o\ifmmode~\breve{g}\else \u{g}\fi{}lu}},\ and\ \bibinfo {author} {\bibfnamefont {G.}~\bibnamefont {Kurizki}},\ }\bibfield  {title} {\bibinfo {title} {Quantum optical two-atom thermal diode},\ }\href {https://doi.org/10.1103/PhysRevE.99.042121} {\bibfield  {journal} {\bibinfo  {journal} {Phys. Rev. E}\ }\textbf {\bibinfo {volume} {99}},\ \bibinfo {pages} {042121} (\bibinfo {year} {2019})}\BibitemShut {NoStop}%
\bibitem [{\citenamefont {Poulsen}\ and\ \citenamefont {Zinner}(2022)}]{PhysRevE.106.034116}%
  \BibitemOpen
  \bibfield  {author} {\bibinfo {author} {\bibfnamefont {K.}~\bibnamefont {Poulsen}}\ and\ \bibinfo {author} {\bibfnamefont {N.~T.}\ \bibnamefont {Zinner}},\ }\bibfield  {title} {\bibinfo {title} {Dark-state-induced heat rectification},\ }\href {https://doi.org/10.1103/PhysRevE.106.034116} {\bibfield  {journal} {\bibinfo  {journal} {Phys. Rev. E}\ }\textbf {\bibinfo {volume} {106}},\ \bibinfo {pages} {034116} (\bibinfo {year} {2022})}\BibitemShut {NoStop}%
\bibitem [{\citenamefont {Upadhyay}\ \emph {et~al.}(2021)\citenamefont {Upadhyay}, \citenamefont {Naseem}, \citenamefont {Marathe},\ and\ \citenamefont {M\"ustecapl\ifmmode \imath \else \i \fi{}o\ifmmode~\breve{g}\else \u{g}\fi{}lu}}]{PhysRevE.104.054137}%
  \BibitemOpen
  \bibfield  {author} {\bibinfo {author} {\bibfnamefont {V.}~\bibnamefont {Upadhyay}}, \bibinfo {author} {\bibfnamefont {M.~T.}\ \bibnamefont {Naseem}}, \bibinfo {author} {\bibfnamefont {R.}~\bibnamefont {Marathe}},\ and\ \bibinfo {author} {\bibfnamefont {O.~E.}\ \bibnamefont {M\"ustecapl\ifmmode \imath \else \i \fi{}o\ifmmode~\breve{g}\else \u{g}\fi{}lu}},\ }\bibfield  {title} {\bibinfo {title} {Heat rectification by two qubits coupled with dzyaloshinskii-moriya interaction},\ }\href {https://doi.org/10.1103/PhysRevE.104.054137} {\bibfield  {journal} {\bibinfo  {journal} {Phys. Rev. E}\ }\textbf {\bibinfo {volume} {104}},\ \bibinfo {pages} {054137} (\bibinfo {year} {2021})}\BibitemShut {NoStop}%
\bibitem [{\citenamefont {Poulsen}\ \emph {et~al.}(2022)\citenamefont {Poulsen}, \citenamefont {Santos}, \citenamefont {Kristensen},\ and\ \citenamefont {Zinner}}]{poulsen2022entanglement}%
  \BibitemOpen
  \bibfield  {author} {\bibinfo {author} {\bibfnamefont {K.}~\bibnamefont {Poulsen}}, \bibinfo {author} {\bibfnamefont {A.~C.}\ \bibnamefont {Santos}}, \bibinfo {author} {\bibfnamefont {L.~B.}\ \bibnamefont {Kristensen}},\ and\ \bibinfo {author} {\bibfnamefont {N.~T.}\ \bibnamefont {Zinner}},\ }\bibfield  {title} {\bibinfo {title} {Entanglement-enhanced quantum rectification},\ }\href {https://doi.org/10.1103/PhysRevA.105.052605} {\bibfield  {journal} {\bibinfo  {journal} {Phys. Rev. A}\ }\textbf {\bibinfo {volume} {105}},\ \bibinfo {pages} {052605} (\bibinfo {year} {2022})}\BibitemShut {NoStop}%
\bibitem [{\citenamefont {Salimi}\ and\ \citenamefont {Khorashad}(2021)}]{ahmadi2021irreversible}%
  \BibitemOpen
  \bibfield  {author} {\bibinfo {author} {\bibfnamefont {S.}~\bibnamefont {Salimi}}\ and\ \bibinfo {author} {\bibfnamefont {A.}~\bibnamefont {Khorashad}},\ }\bibfield  {title} {\bibinfo {title} {Irreversible work and maxwell demon in terms of quantum thermodynamic force},\ }\href {https://doi.org/10.1038/s41598-021-81737-z} {\bibfield  {journal} {\bibinfo  {journal} {Scientific reports}\ }\textbf {\bibinfo {volume} {11}},\ \bibinfo {pages} {2301} (\bibinfo {year} {2021})}\BibitemShut {NoStop}%
\bibitem [{\citenamefont {Scovil}\ and\ \citenamefont {Schulz-DuBois}(1959)}]{PhysRevLett.2.262}%
  \BibitemOpen
  \bibfield  {author} {\bibinfo {author} {\bibfnamefont {H.~E.~D.}\ \bibnamefont {Scovil}}\ and\ \bibinfo {author} {\bibfnamefont {E.~O.}\ \bibnamefont {Schulz-DuBois}},\ }\bibfield  {title} {\bibinfo {title} {Three-level masers as heat engines},\ }\href {https://doi.org/10.1103/PhysRevLett.2.262} {\bibfield  {journal} {\bibinfo  {journal} {Phys. Rev. Lett.}\ }\textbf {\bibinfo {volume} {2}},\ \bibinfo {pages} {262} (\bibinfo {year} {1959})}\BibitemShut {NoStop}%
\bibitem [{\citenamefont {Quan}\ \emph {et~al.}(2007)\citenamefont {Quan}, \citenamefont {Liu}, \citenamefont {Sun},\ and\ \citenamefont {Nori}}]{quan2007quantum}%
  \BibitemOpen
  \bibfield  {author} {\bibinfo {author} {\bibfnamefont {H.~T.}\ \bibnamefont {Quan}}, \bibinfo {author} {\bibfnamefont {Y.-x.}\ \bibnamefont {Liu}}, \bibinfo {author} {\bibfnamefont {C.~P.}\ \bibnamefont {Sun}},\ and\ \bibinfo {author} {\bibfnamefont {F.}~\bibnamefont {Nori}},\ }\bibfield  {title} {\bibinfo {title} {Quantum thermodynamic cycles and quantum heat engines},\ }\href {https://doi.org/10.1103/PhysRevE.76.031105} {\bibfield  {journal} {\bibinfo  {journal} {Phys. Rev. E}\ }\textbf {\bibinfo {volume} {76}},\ \bibinfo {pages} {031105} (\bibinfo {year} {2007})}\BibitemShut {NoStop}%
\bibitem [{\citenamefont {Quan}(2009)}]{quan2009quantum}%
  \BibitemOpen
  \bibfield  {author} {\bibinfo {author} {\bibfnamefont {H.~T.}\ \bibnamefont {Quan}},\ }\bibfield  {title} {\bibinfo {title} {Quantum thermodynamic cycles and quantum heat engines. ii.},\ }\href {https://doi.org/10.1103/PhysRevE.79.041129} {\bibfield  {journal} {\bibinfo  {journal} {Phys. Rev. E}\ }\textbf {\bibinfo {volume} {79}},\ \bibinfo {pages} {041129} (\bibinfo {year} {2009})}\BibitemShut {NoStop}%
\bibitem [{\citenamefont {Roßnagel}\ \emph {et~al.}(2016)\citenamefont {Roßnagel}, \citenamefont {Dawkins}, \citenamefont {Tolazzi}, \citenamefont {Abah}, \citenamefont {Lutz}, \citenamefont {Schmidt-Kaler},\ and\ \citenamefont {Singer}}]{rossnagel2016single}%
  \BibitemOpen
  \bibfield  {author} {\bibinfo {author} {\bibfnamefont {J.}~\bibnamefont {Roßnagel}}, \bibinfo {author} {\bibfnamefont {S.~T.}\ \bibnamefont {Dawkins}}, \bibinfo {author} {\bibfnamefont {K.~N.}\ \bibnamefont {Tolazzi}}, \bibinfo {author} {\bibfnamefont {O.}~\bibnamefont {Abah}}, \bibinfo {author} {\bibfnamefont {E.}~\bibnamefont {Lutz}}, \bibinfo {author} {\bibfnamefont {F.}~\bibnamefont {Schmidt-Kaler}},\ and\ \bibinfo {author} {\bibfnamefont {K.}~\bibnamefont {Singer}},\ }\bibfield  {title} {\bibinfo {title} {A single-atom heat engine},\ }\href {https://doi.org/10.1126/science.aad6320} {\bibfield  {journal} {\bibinfo  {journal} {Science}\ }\textbf {\bibinfo {volume} {352}},\ \bibinfo {pages} {325} (\bibinfo {year} {2016})}\BibitemShut {NoStop}%
\bibitem [{\citenamefont {Peterson}\ \emph {et~al.}(2019)\citenamefont {Peterson}, \citenamefont {Batalh\~ao}, \citenamefont {Herrera}, \citenamefont {Souza}, \citenamefont {Sarthour}, \citenamefont {Oliveira},\ and\ \citenamefont {Serra}}]{peterson2019experimental}%
  \BibitemOpen
  \bibfield  {author} {\bibinfo {author} {\bibfnamefont {J.~P.~S.}\ \bibnamefont {Peterson}}, \bibinfo {author} {\bibfnamefont {T.~B.}\ \bibnamefont {Batalh\~ao}}, \bibinfo {author} {\bibfnamefont {M.}~\bibnamefont {Herrera}}, \bibinfo {author} {\bibfnamefont {A.~M.}\ \bibnamefont {Souza}}, \bibinfo {author} {\bibfnamefont {R.~S.}\ \bibnamefont {Sarthour}}, \bibinfo {author} {\bibfnamefont {I.~S.}\ \bibnamefont {Oliveira}},\ and\ \bibinfo {author} {\bibfnamefont {R.~M.}\ \bibnamefont {Serra}},\ }\bibfield  {title} {\bibinfo {title} {Experimental characterization of a spin quantum heat engine},\ }\href {https://doi.org/10.1103/PhysRevLett.123.240601} {\bibfield  {journal} {\bibinfo  {journal} {Phys. Rev. Lett.}\ }\textbf {\bibinfo {volume} {123}},\ \bibinfo {pages} {240601} (\bibinfo {year} {2019})}\BibitemShut {NoStop}%
\bibitem [{\citenamefont {{\L{}}obejko}\ \emph {et~al.}(2020)\citenamefont {{\L{}}obejko}, \citenamefont {Mazurek},\ and\ \citenamefont {Horodecki}}]{lobejko2020thermodynamics}%
  \BibitemOpen
  \bibfield  {author} {\bibinfo {author} {\bibfnamefont {M.}~\bibnamefont {{\L{}}obejko}}, \bibinfo {author} {\bibfnamefont {P.}~\bibnamefont {Mazurek}},\ and\ \bibinfo {author} {\bibfnamefont {M.}~\bibnamefont {Horodecki}},\ }\bibfield  {title} {\bibinfo {title} {Thermodynamics of {M}inimal {C}oupling {Q}uantum {H}eat {E}ngines},\ }\href {https://doi.org/10.22331/q-2020-12-23-375} {\bibfield  {journal} {\bibinfo  {journal} {{Quantum}}\ }\textbf {\bibinfo {volume} {4}},\ \bibinfo {pages} {375} (\bibinfo {year} {2020})}\BibitemShut {NoStop}%
\bibitem [{\citenamefont {Alicki}\ and\ \citenamefont {Fannes}(2013)}]{alicki2013entanglement}%
  \BibitemOpen
  \bibfield  {author} {\bibinfo {author} {\bibfnamefont {R.}~\bibnamefont {Alicki}}\ and\ \bibinfo {author} {\bibfnamefont {M.}~\bibnamefont {Fannes}},\ }\bibfield  {title} {\bibinfo {title} {Entanglement boost for extractable work from ensembles of quantum batteries},\ }\href {https://doi.org/10.1103/PhysRevE.87.042123} {\bibfield  {journal} {\bibinfo  {journal} {Phys. Rev. E}\ }\textbf {\bibinfo {volume} {87}},\ \bibinfo {pages} {042123} (\bibinfo {year} {2013})}\BibitemShut {NoStop}%
\bibitem [{\citenamefont {Binder}\ \emph {et~al.}(2015)\citenamefont {Binder}, \citenamefont {Vinjanampathy}, \citenamefont {Modi},\ and\ \citenamefont {Goold}}]{binder2015quantacell}%
  \BibitemOpen
  \bibfield  {author} {\bibinfo {author} {\bibfnamefont {F.~C.}\ \bibnamefont {Binder}}, \bibinfo {author} {\bibfnamefont {S.}~\bibnamefont {Vinjanampathy}}, \bibinfo {author} {\bibfnamefont {K.}~\bibnamefont {Modi}},\ and\ \bibinfo {author} {\bibfnamefont {J.}~\bibnamefont {Goold}},\ }\bibfield  {title} {\bibinfo {title} {Quantacell: powerful charging of quantum batteries},\ }\href {https://doi.org/10.1088/1367-2630/17/7/075015} {\bibfield  {journal} {\bibinfo  {journal} {New Journal of Physics}\ }\textbf {\bibinfo {volume} {17}},\ \bibinfo {pages} {075015} (\bibinfo {year} {2015})}\BibitemShut {NoStop}%
\bibitem [{\citenamefont {Campaioli}\ \emph {et~al.}(2018)\citenamefont {Campaioli}, \citenamefont {Pollock},\ and\ \citenamefont {Vinjanampathy}}]{campaioli2018quantum}%
  \BibitemOpen
  \bibfield  {author} {\bibinfo {author} {\bibfnamefont {F.}~\bibnamefont {Campaioli}}, \bibinfo {author} {\bibfnamefont {F.~A.}\ \bibnamefont {Pollock}},\ and\ \bibinfo {author} {\bibfnamefont {S.}~\bibnamefont {Vinjanampathy}},\ }\bibinfo {title} {Quantum batteries},\ in\ \href {https://doi.org/10.1007/978-3-319-99046-0_8} {\emph {\bibinfo {booktitle} {Thermodynamics in the Quantum Regime: Fundamental Aspects and New Directions}}},\ \bibinfo {editor} {edited by\ \bibinfo {editor} {\bibfnamefont {F.}~\bibnamefont {Binder}}, \bibinfo {editor} {\bibfnamefont {L.~A.}\ \bibnamefont {Correa}}, \bibinfo {editor} {\bibfnamefont {C.}~\bibnamefont {Gogolin}}, \bibinfo {editor} {\bibfnamefont {J.}~\bibnamefont {Anders}},\ and\ \bibinfo {editor} {\bibfnamefont {G.}~\bibnamefont {Adesso}}}\ (\bibinfo  {publisher} {Springer International Publishing},\ \bibinfo {address} {Cham},\ \bibinfo {year} {2018})\ pp.\ \bibinfo {pages} {207--225}\BibitemShut {NoStop}%
\bibitem [{\citenamefont {Lipka-Bartosik}\ \emph {et~al.}(2021)\citenamefont {Lipka-Bartosik}, \citenamefont {Mazurek},\ and\ \citenamefont {Horodecki}}]{lipka2021second}%
  \BibitemOpen
  \bibfield  {author} {\bibinfo {author} {\bibfnamefont {P.}~\bibnamefont {Lipka-Bartosik}}, \bibinfo {author} {\bibfnamefont {P.}~\bibnamefont {Mazurek}},\ and\ \bibinfo {author} {\bibfnamefont {M.}~\bibnamefont {Horodecki}},\ }\bibfield  {title} {\bibinfo {title} {Second law of thermodynamics for batteries with vacuum state},\ }\href {https://doi.org/10.22331/q-2021-03-10-408} {\bibfield  {journal} {\bibinfo  {journal} {{Quantum}}\ }\textbf {\bibinfo {volume} {5}},\ \bibinfo {pages} {408} (\bibinfo {year} {2021})}\BibitemShut {NoStop}%
\bibitem [{\citenamefont {Cruz}\ \emph {et~al.}(2022)\citenamefont {Cruz}, \citenamefont {Anka}, \citenamefont {Reis}, \citenamefont {Bachelard},\ and\ \citenamefont {Santos}}]{cruz2022quantum}%
  \BibitemOpen
  \bibfield  {author} {\bibinfo {author} {\bibfnamefont {C.}~\bibnamefont {Cruz}}, \bibinfo {author} {\bibfnamefont {M.~F.}\ \bibnamefont {Anka}}, \bibinfo {author} {\bibfnamefont {M.~S.}\ \bibnamefont {Reis}}, \bibinfo {author} {\bibfnamefont {R.}~\bibnamefont {Bachelard}},\ and\ \bibinfo {author} {\bibfnamefont {A.~C.}\ \bibnamefont {Santos}},\ }\bibfield  {title} {\bibinfo {title} {Quantum battery based on quantum discord at room temperature},\ }\href {https://doi.org/10.1088/2058-9565/ac57f3} {\bibfield  {journal} {\bibinfo  {journal} {Quantum Science and Technology}\ }\textbf {\bibinfo {volume} {7}},\ \bibinfo {pages} {025020} (\bibinfo {year} {2022})}\BibitemShut {NoStop}%
\bibitem [{\citenamefont {Rodriguez}\ \emph {et~al.}(2022)\citenamefont {Rodriguez}, \citenamefont {Ahmadi}, \citenamefont {Suarez}, \citenamefont {Mazurek}, \citenamefont {Barzanjeh},\ and\ \citenamefont {Horodecki}}]{rodriguez2023optimal}%
  \BibitemOpen
  \bibfield  {author} {\bibinfo {author} {\bibfnamefont {R.}~\bibnamefont {Rodriguez}}, \bibinfo {author} {\bibfnamefont {B.}~\bibnamefont {Ahmadi}}, \bibinfo {author} {\bibfnamefont {G.}~\bibnamefont {Suarez}}, \bibinfo {author} {\bibfnamefont {P.}~\bibnamefont {Mazurek}}, \bibinfo {author} {\bibfnamefont {S.}~\bibnamefont {Barzanjeh}},\ and\ \bibinfo {author} {\bibfnamefont {P.}~\bibnamefont {Horodecki}},\ }\bibfield  {title} {\bibinfo {title} {Optimal quantum control of charging quantum batteries},\ }\href {https://arxiv.org/abs/2207.00094} {\bibfield  {journal} {\bibinfo  {journal} {arXiv:2207.00094}\ } (\bibinfo {year} {2022})}\BibitemShut {NoStop}%
\bibitem [{\citenamefont {Rodriguez}\ \emph {et~al.}(2023)\citenamefont {Rodriguez}, \citenamefont {Ahmadi}, \citenamefont {Mazurek}, \citenamefont {Barzanjeh}, \citenamefont {Alicki},\ and\ \citenamefont {Horodecki}}]{rodriguez2023catalysis}%
  \BibitemOpen
  \bibfield  {author} {\bibinfo {author} {\bibfnamefont {R.}~\bibnamefont {Rodriguez}}, \bibinfo {author} {\bibfnamefont {B.}~\bibnamefont {Ahmadi}}, \bibinfo {author} {\bibfnamefont {P.}~\bibnamefont {Mazurek}}, \bibinfo {author} {\bibfnamefont {S.}~\bibnamefont {Barzanjeh}}, \bibinfo {author} {\bibfnamefont {R.}~\bibnamefont {Alicki}},\ and\ \bibinfo {author} {\bibfnamefont {P.}~\bibnamefont {Horodecki}},\ }\bibfield  {title} {\bibinfo {title} {Catalysis in charging quantum batteries},\ }\href {https://doi.org/10.1103/PhysRevA.107.042419} {\bibfield  {journal} {\bibinfo  {journal} {Physical Review A}\ }\textbf {\bibinfo {volume} {107}},\ \bibinfo {pages} {042419} (\bibinfo {year} {2023})}\BibitemShut {NoStop}%
\bibitem [{\citenamefont {Ahmadi}\ \emph {et~al.}(2024)\citenamefont {Ahmadi}, \citenamefont {Mazurek}, \citenamefont {Horodecki},\ and\ \citenamefont {Barzanjeh}}]{ahmadi2024nonreciprocal}%
  \BibitemOpen
  \bibfield  {author} {\bibinfo {author} {\bibfnamefont {B.}~\bibnamefont {Ahmadi}}, \bibinfo {author} {\bibfnamefont {P.}~\bibnamefont {Mazurek}}, \bibinfo {author} {\bibfnamefont {P.}~\bibnamefont {Horodecki}},\ and\ \bibinfo {author} {\bibfnamefont {S.}~\bibnamefont {Barzanjeh}},\ }\bibfield  {title} {\bibinfo {title} {Nonreciprocal quantum batteries},\ }\href {https://arxiv.org/abs/2401.05090} {\bibfield  {journal} {\bibinfo  {journal} {arXiv:2401.05090}\ } (\bibinfo {year} {2024})}\BibitemShut {NoStop}%
\bibitem [{\citenamefont {Quach}\ \emph {et~al.}(2022)\citenamefont {Quach}, \citenamefont {McGhee}, \citenamefont {Ganzer}, \citenamefont {Rouse}, \citenamefont {Lovett}, \citenamefont {Gauger}, \citenamefont {Keeling}, \citenamefont {Cerullo}, \citenamefont {Lidzey},\ and\ \citenamefont {Virgili}}]{quach2022superabsorption}%
  \BibitemOpen
  \bibfield  {author} {\bibinfo {author} {\bibfnamefont {J.~Q.}\ \bibnamefont {Quach}}, \bibinfo {author} {\bibfnamefont {K.~E.}\ \bibnamefont {McGhee}}, \bibinfo {author} {\bibfnamefont {L.}~\bibnamefont {Ganzer}}, \bibinfo {author} {\bibfnamefont {D.~M.}\ \bibnamefont {Rouse}}, \bibinfo {author} {\bibfnamefont {B.~W.}\ \bibnamefont {Lovett}}, \bibinfo {author} {\bibfnamefont {E.~M.}\ \bibnamefont {Gauger}}, \bibinfo {author} {\bibfnamefont {J.}~\bibnamefont {Keeling}}, \bibinfo {author} {\bibfnamefont {G.}~\bibnamefont {Cerullo}}, \bibinfo {author} {\bibfnamefont {D.~G.}\ \bibnamefont {Lidzey}},\ and\ \bibinfo {author} {\bibfnamefont {T.}~\bibnamefont {Virgili}},\ }\bibfield  {title} {\bibinfo {title} {Superabsorption in an organic microcavity: Toward a quantum battery},\ }\href {https://doi.org/10.1126/sciadv.abk3160} {\bibfield  {journal} {\bibinfo  {journal} {Science advances}\ }\textbf {\bibinfo {volume} {8}},\ \bibinfo {pages} {eabk3160} (\bibinfo {year} {2022})}\BibitemShut {NoStop}%
\bibitem [{\citenamefont {Joshi}\ and\ \citenamefont {Mahesh}(2022)}]{PhysRevA.106.042601}%
  \BibitemOpen
  \bibfield  {author} {\bibinfo {author} {\bibfnamefont {J.}~\bibnamefont {Joshi}}\ and\ \bibinfo {author} {\bibfnamefont {T.~S.}\ \bibnamefont {Mahesh}},\ }\bibfield  {title} {\bibinfo {title} {Experimental investigation of a quantum battery using star-topology nmr spin systems},\ }\href {https://doi.org/10.1103/PhysRevA.106.042601} {\bibfield  {journal} {\bibinfo  {journal} {Phys. Rev. A}\ }\textbf {\bibinfo {volume} {106}},\ \bibinfo {pages} {042601} (\bibinfo {year} {2022})}\BibitemShut {NoStop}%
\bibitem [{\citenamefont {Kamin}\ \emph {et~al.}(2020)\citenamefont {Kamin}, \citenamefont {Tabesh}, \citenamefont {Salimi}, \citenamefont {Kheirandish},\ and\ \citenamefont {Santos}}]{kamin2020non}%
  \BibitemOpen
  \bibfield  {author} {\bibinfo {author} {\bibfnamefont {F.}~\bibnamefont {Kamin}}, \bibinfo {author} {\bibfnamefont {F.}~\bibnamefont {Tabesh}}, \bibinfo {author} {\bibfnamefont {S.}~\bibnamefont {Salimi}}, \bibinfo {author} {\bibfnamefont {F.}~\bibnamefont {Kheirandish}},\ and\ \bibinfo {author} {\bibfnamefont {A.~C.}\ \bibnamefont {Santos}},\ }\bibfield  {title} {\bibinfo {title} {Non-markovian effects on charging and self-discharging process of quantum batteries},\ }\href {https://doi.org/10.1088/1367-2630/ab9ee2} {\bibfield  {journal} {\bibinfo  {journal} {New Journal of Physics}\ }\textbf {\bibinfo {volume} {22}},\ \bibinfo {pages} {083007} (\bibinfo {year} {2020})}\BibitemShut {NoStop}%
\bibitem [{\citenamefont {Kamin}\ \emph {et~al.}(2024)\citenamefont {Kamin}, \citenamefont {Salimi},\ and\ \citenamefont {Arjmandi}}]{kamin2023steady}%
  \BibitemOpen
  \bibfield  {author} {\bibinfo {author} {\bibfnamefont {F.~H.}\ \bibnamefont {Kamin}}, \bibinfo {author} {\bibfnamefont {S.}~\bibnamefont {Salimi}},\ and\ \bibinfo {author} {\bibfnamefont {M.~B.}\ \bibnamefont {Arjmandi}},\ }\bibfield  {title} {\bibinfo {title} {Steady-state charging of quantum batteries via dissipative ancillas},\ }\href {https://doi.org/10.1103/PhysRevA.109.022226} {\bibfield  {journal} {\bibinfo  {journal} {Phys. Rev. A}\ }\textbf {\bibinfo {volume} {109}},\ \bibinfo {pages} {022226} (\bibinfo {year} {2024})}\BibitemShut {NoStop}%
\bibitem [{\citenamefont {Li}\ \emph {et~al.}(2006)\citenamefont {Li}, \citenamefont {Wang},\ and\ \citenamefont {Casati}}]{10.1063/1.2191730}%
  \BibitemOpen
  \bibfield  {author} {\bibinfo {author} {\bibfnamefont {B.}~\bibnamefont {Li}}, \bibinfo {author} {\bibfnamefont {L.}~\bibnamefont {Wang}},\ and\ \bibinfo {author} {\bibfnamefont {G.}~\bibnamefont {Casati}},\ }\bibfield  {title} {\bibinfo {title} {{Negative differential thermal resistance and thermal transistor}},\ }\href {https://doi.org/10.1063/1.2191730} {\bibfield  {journal} {\bibinfo  {journal} {Applied Physics Letters}\ }\textbf {\bibinfo {volume} {88}},\ \bibinfo {pages} {143501} (\bibinfo {year} {2006})}\BibitemShut {NoStop}%
\bibitem [{\citenamefont {Joulain}\ \emph {et~al.}(2016)\citenamefont {Joulain}, \citenamefont {Drevillon}, \citenamefont {Ezzahri},\ and\ \citenamefont {Ordonez-Miranda}}]{joulain2016quantum}%
  \BibitemOpen
  \bibfield  {author} {\bibinfo {author} {\bibfnamefont {K.}~\bibnamefont {Joulain}}, \bibinfo {author} {\bibfnamefont {J.}~\bibnamefont {Drevillon}}, \bibinfo {author} {\bibfnamefont {Y.}~\bibnamefont {Ezzahri}},\ and\ \bibinfo {author} {\bibfnamefont {J.}~\bibnamefont {Ordonez-Miranda}},\ }\bibfield  {title} {\bibinfo {title} {Quantum thermal transistor},\ }\href {https://doi.org/10.1103/PhysRevLett.116.200601} {\bibfield  {journal} {\bibinfo  {journal} {Phys. Rev. Lett.}\ }\textbf {\bibinfo {volume} {116}},\ \bibinfo {pages} {200601} (\bibinfo {year} {2016})}\BibitemShut {NoStop}%
\bibitem [{\citenamefont {Guo}\ \emph {et~al.}(2019)\citenamefont {Guo}, \citenamefont {Liu},\ and\ \citenamefont {Yu}}]{PhysRevE.99.032112}%
  \BibitemOpen
  \bibfield  {author} {\bibinfo {author} {\bibfnamefont {B.-q.}\ \bibnamefont {Guo}}, \bibinfo {author} {\bibfnamefont {T.}~\bibnamefont {Liu}},\ and\ \bibinfo {author} {\bibfnamefont {C.-s.}\ \bibnamefont {Yu}},\ }\bibfield  {title} {\bibinfo {title} {Multifunctional quantum thermal device utilizing three qubits},\ }\href {https://doi.org/10.1103/PhysRevE.99.032112} {\bibfield  {journal} {\bibinfo  {journal} {Phys. Rev. E}\ }\textbf {\bibinfo {volume} {99}},\ \bibinfo {pages} {032112} (\bibinfo {year} {2019})}\BibitemShut {NoStop}%
\bibitem [{\citenamefont {Wang}\ \emph {et~al.}(2018)\citenamefont {Wang}, \citenamefont {Chen}, \citenamefont {Sun},\ and\ \citenamefont {Ren}}]{PhysRevA.97.052112}%
  \BibitemOpen
  \bibfield  {author} {\bibinfo {author} {\bibfnamefont {C.}~\bibnamefont {Wang}}, \bibinfo {author} {\bibfnamefont {X.-M.}\ \bibnamefont {Chen}}, \bibinfo {author} {\bibfnamefont {K.-W.}\ \bibnamefont {Sun}},\ and\ \bibinfo {author} {\bibfnamefont {J.}~\bibnamefont {Ren}},\ }\bibfield  {title} {\bibinfo {title} {Heat amplification and negative differential thermal conductance in a strongly coupled nonequilibrium spin-boson system},\ }\href {https://doi.org/10.1103/PhysRevA.97.052112} {\bibfield  {journal} {\bibinfo  {journal} {Phys. Rev. A}\ }\textbf {\bibinfo {volume} {97}},\ \bibinfo {pages} {052112} (\bibinfo {year} {2018})}\BibitemShut {NoStop}%
\bibitem [{\citenamefont {Yang}\ and\ \citenamefont {Tan}(2020)}]{yang2020quantum}%
  \BibitemOpen
  \bibfield  {author} {\bibinfo {author} {\bibfnamefont {H.-F.}\ \bibnamefont {Yang}}\ and\ \bibinfo {author} {\bibfnamefont {Y.-G.}\ \bibnamefont {Tan}},\ }\bibfield  {title} {\bibinfo {title} {Quantum thermal transistor: a unified method from weak to strong internal coupling},\ }\href {https://doi.org/10.1088/1361-6455/abade1} {\bibfield  {journal} {\bibinfo  {journal} {Journal of Physics B: Atomic, Molecular and Optical Physics}\ }\textbf {\bibinfo {volume} {53}},\ \bibinfo {pages} {205504} (\bibinfo {year} {2020})}\BibitemShut {NoStop}%
\bibitem [{\citenamefont {Wijesekara}\ \emph {et~al.}(2020)\citenamefont {Wijesekara}, \citenamefont {Gunapala}, \citenamefont {Stockman},\ and\ \citenamefont {Premaratne}}]{PhysRevB.101.245402}%
  \BibitemOpen
  \bibfield  {author} {\bibinfo {author} {\bibfnamefont {R.~T.}\ \bibnamefont {Wijesekara}}, \bibinfo {author} {\bibfnamefont {S.~D.}\ \bibnamefont {Gunapala}}, \bibinfo {author} {\bibfnamefont {M.~I.}\ \bibnamefont {Stockman}},\ and\ \bibinfo {author} {\bibfnamefont {M.}~\bibnamefont {Premaratne}},\ }\bibfield  {title} {\bibinfo {title} {Optically controlled quantum thermal gate},\ }\href {https://doi.org/10.1103/PhysRevB.101.245402} {\bibfield  {journal} {\bibinfo  {journal} {Phys. Rev. B}\ }\textbf {\bibinfo {volume} {101}},\ \bibinfo {pages} {245402} (\bibinfo {year} {2020})}\BibitemShut {NoStop}%
\bibitem [{\citenamefont {Wijesekara}\ \emph {et~al.}(2021)\citenamefont {Wijesekara}, \citenamefont {Gunapala},\ and\ \citenamefont {Premaratne}}]{PhysRevB.104.045405}%
  \BibitemOpen
  \bibfield  {author} {\bibinfo {author} {\bibfnamefont {R.~T.}\ \bibnamefont {Wijesekara}}, \bibinfo {author} {\bibfnamefont {S.~D.}\ \bibnamefont {Gunapala}},\ and\ \bibinfo {author} {\bibfnamefont {M.}~\bibnamefont {Premaratne}},\ }\bibfield  {title} {\bibinfo {title} {Darlington pair of quantum thermal transistors},\ }\href {https://doi.org/10.1103/PhysRevB.104.045405} {\bibfield  {journal} {\bibinfo  {journal} {Phys. Rev. B}\ }\textbf {\bibinfo {volume} {104}},\ \bibinfo {pages} {045405} (\bibinfo {year} {2021})}\BibitemShut {NoStop}%
\bibitem [{\citenamefont {Liu}\ \emph {et~al.}(2022)\citenamefont {Liu}, \citenamefont {Yu},\ and\ \citenamefont {Yu}}]{liu2021common}%
  \BibitemOpen
  \bibfield  {author} {\bibinfo {author} {\bibfnamefont {Y.-Q.}\ \bibnamefont {Liu}}, \bibinfo {author} {\bibfnamefont {D.-H.}\ \bibnamefont {Yu}},\ and\ \bibinfo {author} {\bibfnamefont {C.-S.}\ \bibnamefont {Yu}},\ }\bibfield  {title} {\bibinfo {title} {Common environmental effects on quantum thermal transistor},\ }\href {https://www.mdpi.com/1099-4300/24/1/32} {\bibfield  {journal} {\bibinfo  {journal} {Entropy}\ }\textbf {\bibinfo {volume} {24}} (\bibinfo {year} {2022})}\BibitemShut {NoStop}%
\bibitem [{\citenamefont {Ghosh}\ \emph {et~al.}(2021)\citenamefont {Ghosh}, \citenamefont {Ghoshal},\ and\ \citenamefont {Sen}}]{ghosh2021quantum}%
  \BibitemOpen
  \bibfield  {author} {\bibinfo {author} {\bibfnamefont {R.}~\bibnamefont {Ghosh}}, \bibinfo {author} {\bibfnamefont {A.}~\bibnamefont {Ghoshal}},\ and\ \bibinfo {author} {\bibfnamefont {U.}~\bibnamefont {Sen}},\ }\bibfield  {title} {\bibinfo {title} {Quantum thermal transistors: Operation characteristics in steady state versus transient regimes},\ }\href {https://doi.org/10.1103/PhysRevA.103.052613} {\bibfield  {journal} {\bibinfo  {journal} {Phys. Rev. A}\ }\textbf {\bibinfo {volume} {103}},\ \bibinfo {pages} {052613} (\bibinfo {year} {2021})}\BibitemShut {NoStop}%
\bibitem [{\citenamefont {Mandarino}\ \emph {et~al.}(2021)\citenamefont {Mandarino}, \citenamefont {Joulain}, \citenamefont {G\'omez},\ and\ \citenamefont {Bellomo}}]{mandarino2021thermal}%
  \BibitemOpen
  \bibfield  {author} {\bibinfo {author} {\bibfnamefont {A.}~\bibnamefont {Mandarino}}, \bibinfo {author} {\bibfnamefont {K.}~\bibnamefont {Joulain}}, \bibinfo {author} {\bibfnamefont {M.~D.}\ \bibnamefont {G\'omez}},\ and\ \bibinfo {author} {\bibfnamefont {B.}~\bibnamefont {Bellomo}},\ }\bibfield  {title} {\bibinfo {title} {Thermal transistor effect in quantum systems},\ }\href {https://doi.org/10.1103/PhysRevApplied.16.034026} {\bibfield  {journal} {\bibinfo  {journal} {Phys. Rev. Appl.}\ }\textbf {\bibinfo {volume} {16}},\ \bibinfo {pages} {034026} (\bibinfo {year} {2021})}\BibitemShut {NoStop}%
\bibitem [{\citenamefont {Du}\ \emph {et~al.}(2019)\citenamefont {Du}, \citenamefont {Shen}, \citenamefont {Su},\ and\ \citenamefont {Chen}}]{du2019quantum}%
  \BibitemOpen
  \bibfield  {author} {\bibinfo {author} {\bibfnamefont {J.}~\bibnamefont {Du}}, \bibinfo {author} {\bibfnamefont {W.}~\bibnamefont {Shen}}, \bibinfo {author} {\bibfnamefont {S.}~\bibnamefont {Su}},\ and\ \bibinfo {author} {\bibfnamefont {J.}~\bibnamefont {Chen}},\ }\bibfield  {title} {\bibinfo {title} {Quantum thermal management devices based on strong coupling qubits},\ }\href {https://doi.org/10.1103/PhysRevE.99.062123} {\bibfield  {journal} {\bibinfo  {journal} {Phys. Rev. E}\ }\textbf {\bibinfo {volume} {99}},\ \bibinfo {pages} {062123} (\bibinfo {year} {2019})}\BibitemShut {NoStop}%
\bibitem [{\citenamefont {Guo}\ \emph {et~al.}(2018)\citenamefont {Guo}, \citenamefont {Liu},\ and\ \citenamefont {Yu}}]{guo2018quantum}%
  \BibitemOpen
  \bibfield  {author} {\bibinfo {author} {\bibfnamefont {B.-q.}\ \bibnamefont {Guo}}, \bibinfo {author} {\bibfnamefont {T.}~\bibnamefont {Liu}},\ and\ \bibinfo {author} {\bibfnamefont {C.-s.}\ \bibnamefont {Yu}},\ }\bibfield  {title} {\bibinfo {title} {Quantum thermal transistor based on qubit-qutrit coupling},\ }\href {https://doi.org/10.1103/PhysRevE.98.022118} {\bibfield  {journal} {\bibinfo  {journal} {Phys. Rev. E}\ }\textbf {\bibinfo {volume} {98}},\ \bibinfo {pages} {022118} (\bibinfo {year} {2018})}\BibitemShut {NoStop}%
\bibitem [{\citenamefont {Su}\ \emph {et~al.}(2018)\citenamefont {Su}, \citenamefont {Zhang}, \citenamefont {Andresen},\ and\ \citenamefont {Chen}}]{su2018quantum}%
  \BibitemOpen
  \bibfield  {author} {\bibinfo {author} {\bibfnamefont {S.}~\bibnamefont {Su}}, \bibinfo {author} {\bibfnamefont {Y.}~\bibnamefont {Zhang}}, \bibinfo {author} {\bibfnamefont {B.}~\bibnamefont {Andresen}},\ and\ \bibinfo {author} {\bibfnamefont {J.}~\bibnamefont {Chen}},\ }\bibfield  {title} {\bibinfo {title} {Quantum coherence thermal transistors},\ }\href {https://doi.org/10.48550/arXiv.1811.02400} {\bibfield  {journal} {\bibinfo  {journal} {arXiv:1811.02400}\ } (\bibinfo {year} {2018})}\BibitemShut {NoStop}%
\bibitem [{\citenamefont {Wang}\ \emph {et~al.}(2019)\citenamefont {Wang}, \citenamefont {Xu}, \citenamefont {Liu},\ and\ \citenamefont {Gao}}]{PhysRevE.99.042102}%
  \BibitemOpen
  \bibfield  {author} {\bibinfo {author} {\bibfnamefont {C.}~\bibnamefont {Wang}}, \bibinfo {author} {\bibfnamefont {D.}~\bibnamefont {Xu}}, \bibinfo {author} {\bibfnamefont {H.}~\bibnamefont {Liu}},\ and\ \bibinfo {author} {\bibfnamefont {X.}~\bibnamefont {Gao}},\ }\bibfield  {title} {\bibinfo {title} {Thermal rectification and heat amplification in a nonequilibrium v-type three-level system},\ }\href {https://doi.org/10.1103/PhysRevE.99.042102} {\bibfield  {journal} {\bibinfo  {journal} {Phys. Rev. E}\ }\textbf {\bibinfo {volume} {99}},\ \bibinfo {pages} {042102} (\bibinfo {year} {2019})}\BibitemShut {NoStop}%
\bibitem [{\citenamefont {Wang}\ and\ \citenamefont {Xu}(2020)}]{wang2020polaron}%
  \BibitemOpen
  \bibfield  {author} {\bibinfo {author} {\bibfnamefont {C.}~\bibnamefont {Wang}}\ and\ \bibinfo {author} {\bibfnamefont {D.-Z.}\ \bibnamefont {Xu}},\ }\bibfield  {title} {\bibinfo {title} {A polaron theory of quantum thermal transistor in nonequilibrium three-level systems},\ }\href {https://doi.org/10.1088/1674-1056/ab973b} {\bibfield  {journal} {\bibinfo  {journal} {Chinese Physics B}\ }\textbf {\bibinfo {volume} {29}},\ \bibinfo {pages} {080504} (\bibinfo {year} {2020})}\BibitemShut {NoStop}%
\bibitem [{\citenamefont {Majland}\ \emph {et~al.}(2020)\citenamefont {Majland}, \citenamefont {Christensen},\ and\ \citenamefont {Zinner}}]{majland2020quantum}%
  \BibitemOpen
  \bibfield  {author} {\bibinfo {author} {\bibfnamefont {M.}~\bibnamefont {Majland}}, \bibinfo {author} {\bibfnamefont {K.~S.}\ \bibnamefont {Christensen}},\ and\ \bibinfo {author} {\bibfnamefont {N.~T.}\ \bibnamefont {Zinner}},\ }\bibfield  {title} {\bibinfo {title} {Quantum thermal transistor in superconducting circuits},\ }\href {https://doi.org/10.1103/PhysRevB.101.184510} {\bibfield  {journal} {\bibinfo  {journal} {Phys. Rev. B}\ }\textbf {\bibinfo {volume} {101}},\ \bibinfo {pages} {184510} (\bibinfo {year} {2020})}\BibitemShut {NoStop}%
\bibitem [{\citenamefont {Wang}\ \emph {et~al.}(2022)\citenamefont {Wang}, \citenamefont {Wang}, \citenamefont {Wang},\ and\ \citenamefont {Ren}}]{wang2022cycle}%
  \BibitemOpen
  \bibfield  {author} {\bibinfo {author} {\bibfnamefont {L.}~\bibnamefont {Wang}}, \bibinfo {author} {\bibfnamefont {Z.}~\bibnamefont {Wang}}, \bibinfo {author} {\bibfnamefont {C.}~\bibnamefont {Wang}},\ and\ \bibinfo {author} {\bibfnamefont {J.}~\bibnamefont {Ren}},\ }\bibfield  {title} {\bibinfo {title} {Cycle flux ranking of network analysis in quantum thermal devices},\ }\href {https://doi.org/10.1103/PhysRevLett.128.067701} {\bibfield  {journal} {\bibinfo  {journal} {Phys. Rev. Lett.}\ }\textbf {\bibinfo {volume} {128}},\ \bibinfo {pages} {067701} (\bibinfo {year} {2022})}\BibitemShut {NoStop}%
\bibitem [{\citenamefont {Boukobza}\ and\ \citenamefont {Tannor}(2007)}]{PhysRevLett.98.240601}%
  \BibitemOpen
  \bibfield  {author} {\bibinfo {author} {\bibfnamefont {E.}~\bibnamefont {Boukobza}}\ and\ \bibinfo {author} {\bibfnamefont {D.~J.}\ \bibnamefont {Tannor}},\ }\bibfield  {title} {\bibinfo {title} {Three-level systems as amplifiers and attenuators: A thermodynamic analysis},\ }\href {https://doi.org/10.1103/PhysRevLett.98.240601} {\bibfield  {journal} {\bibinfo  {journal} {Phys. Rev. Lett.}\ }\textbf {\bibinfo {volume} {98}},\ \bibinfo {pages} {240601} (\bibinfo {year} {2007})}\BibitemShut {NoStop}%
\bibitem [{\citenamefont {Li}\ \emph {et~al.}(2017)\citenamefont {Li}, \citenamefont {Kim}, \citenamefont {Agarwal},\ and\ \citenamefont {Scully}}]{PhysRevA.96.063806}%
  \BibitemOpen
  \bibfield  {author} {\bibinfo {author} {\bibfnamefont {S.-W.}\ \bibnamefont {Li}}, \bibinfo {author} {\bibfnamefont {M.~B.}\ \bibnamefont {Kim}}, \bibinfo {author} {\bibfnamefont {G.~S.}\ \bibnamefont {Agarwal}},\ and\ \bibinfo {author} {\bibfnamefont {M.~O.}\ \bibnamefont {Scully}},\ }\bibfield  {title} {\bibinfo {title} {Quantum statistics of a single-atom {S}covil--{S}chulz-{D}ubois heat engine},\ }\href {https://doi.org/10.1103/PhysRevA.96.063806} {\bibfield  {journal} {\bibinfo  {journal} {Phys. Rev. A}\ }\textbf {\bibinfo {volume} {96}},\ \bibinfo {pages} {063806} (\bibinfo {year} {2017})}\BibitemShut {NoStop}%
\bibitem [{\citenamefont {Díaz}\ and\ \citenamefont {Sánchez}(2021)}]{diaz2021qutrit}%
  \BibitemOpen
  \bibfield  {author} {\bibinfo {author} {\bibfnamefont {I.}~\bibnamefont {Díaz}}\ and\ \bibinfo {author} {\bibfnamefont {R.}~\bibnamefont {Sánchez}},\ }\bibfield  {title} {\bibinfo {title} {The qutrit as a heat diode and circulator},\ }\href {https://doi.org/10.1088/1367-2630/ac4211} {\bibfield  {journal} {\bibinfo  {journal} {New Journal of Physics}\ }\textbf {\bibinfo {volume} {23}},\ \bibinfo {pages} {125006} (\bibinfo {year} {2021})}\BibitemShut {NoStop}%
\bibitem [{\citenamefont {Galve}\ \emph {et~al.}(2017)\citenamefont {Galve}, \citenamefont {Mandarino}, \citenamefont {Paris}, \citenamefont {Benedetti},\ and\ \citenamefont {Zambrini}}]{galve2017microscopic}%
  \BibitemOpen
  \bibfield  {author} {\bibinfo {author} {\bibfnamefont {F.}~\bibnamefont {Galve}}, \bibinfo {author} {\bibfnamefont {A.}~\bibnamefont {Mandarino}}, \bibinfo {author} {\bibfnamefont {M.~G.~A.}\ \bibnamefont {Paris}}, \bibinfo {author} {\bibfnamefont {C.}~\bibnamefont {Benedetti}},\ and\ \bibinfo {author} {\bibfnamefont {R.}~\bibnamefont {Zambrini}},\ }\bibfield  {title} {\bibinfo {title} {Microscopic description for the emergence of collective dissipation in extended quantum systems},\ }\href {https://doi.org/10.1038/srep42050} {\bibfield  {journal} {\bibinfo  {journal} {Scientific Reports}\ }\textbf {\bibinfo {volume} {7}},\ \bibinfo {pages} {42050} (\bibinfo {year} {2017})}\BibitemShut {NoStop}%
\bibitem [{\citenamefont {Mandarino}(2022)}]{mandarino2022quantum}%
  \BibitemOpen
  \bibfield  {author} {\bibinfo {author} {\bibfnamefont {A.}~\bibnamefont {Mandarino}},\ }\bibfield  {title} {\bibinfo {title} {Quantum thermal amplifiers with engineered dissipation},\ }\href {https://www.mdpi.com/1099-4300/24/8/1031} {\bibfield  {journal} {\bibinfo  {journal} {Entropy}\ }\textbf {\bibinfo {volume} {24}},\ \bibinfo {pages} {1031} (\bibinfo {year} {2022})}\BibitemShut {NoStop}%
\bibitem [{\citenamefont {Gorini}\ \emph {et~al.}(1976)\citenamefont {Gorini}, \citenamefont {Kossakowski},\ and\ \citenamefont {Sudarshan}}]{gorini1976completely}%
  \BibitemOpen
  \bibfield  {author} {\bibinfo {author} {\bibfnamefont {V.}~\bibnamefont {Gorini}}, \bibinfo {author} {\bibfnamefont {A.}~\bibnamefont {Kossakowski}},\ and\ \bibinfo {author} {\bibfnamefont {E.~C.~G.}\ \bibnamefont {Sudarshan}},\ }\bibfield  {title} {\bibinfo {title} {{Completely positive dynamical semigroups of N‐level systems}},\ }\href {https://doi.org/10.1063/1.522979} {\bibfield  {journal} {\bibinfo  {journal} {Journal of Mathematical Physics}\ }\textbf {\bibinfo {volume} {17}},\ \bibinfo {pages} {821} (\bibinfo {year} {1976})}\BibitemShut {NoStop}%
\bibitem [{\citenamefont {Breuer}\ and\ \citenamefont {Petruccione}(2007)}]{Breuer}%
  \BibitemOpen
  \bibfield  {author} {\bibinfo {author} {\bibfnamefont {H.-P.}\ \bibnamefont {Breuer}}\ and\ \bibinfo {author} {\bibfnamefont {F.}~\bibnamefont {Petruccione}},\ }\href {https://doi.org/10.1093/acprof:oso/9780199213900.001.0001} {\emph {\bibinfo {title} {{The Theory of Open Quantum Systems}}}}\ (\bibinfo  {publisher} {Oxford University Press},\ \bibinfo {year} {2007})\BibitemShut {NoStop}%
\bibitem [{\citenamefont {Cattaneo}\ \emph {et~al.}(2019)\citenamefont {Cattaneo}, \citenamefont {Giorgi}, \citenamefont {Maniscalco},\ and\ \citenamefont {Zambrini}}]{cattaneo2019local}%
  \BibitemOpen
  \bibfield  {author} {\bibinfo {author} {\bibfnamefont {M.}~\bibnamefont {Cattaneo}}, \bibinfo {author} {\bibfnamefont {G.~L.}\ \bibnamefont {Giorgi}}, \bibinfo {author} {\bibfnamefont {S.}~\bibnamefont {Maniscalco}},\ and\ \bibinfo {author} {\bibfnamefont {R.}~\bibnamefont {Zambrini}},\ }\bibfield  {title} {\bibinfo {title} {Local versus global master equation with common and separate baths: superiority of the global approach in partial secular approximation},\ }\href {https://doi.org/10.1088/1367-2630/ab54ac} {\bibfield  {journal} {\bibinfo  {journal} {New Journal of Physics}\ }\textbf {\bibinfo {volume} {21}},\ \bibinfo {pages} {113045} (\bibinfo {year} {2019})}\BibitemShut {NoStop}%
\bibitem [{\citenamefont {Alicki}(1979)}]{Alicki}%
  \BibitemOpen
  \bibfield  {author} {\bibinfo {author} {\bibfnamefont {R.}~\bibnamefont {Alicki}},\ }\bibfield  {title} {\bibinfo {title} {The quantum open system as a model of the heat engine},\ }\href {https://doi.org/10.1088/0305-4470/12/5/007} {\bibfield  {journal} {\bibinfo  {journal} {Journal of Physics A: Mathematical and General}\ }\textbf {\bibinfo {volume} {12}},\ \bibinfo {pages} {L103} (\bibinfo {year} {1979})}\BibitemShut {NoStop}%
\bibitem [{\citenamefont {Gemmer}\ \emph {et~al.}(2009)\citenamefont {Gemmer}, \citenamefont {Michel},\ and\ \citenamefont {Mahler}}]{Gemmer}%
  \BibitemOpen
  \bibfield  {author} {\bibinfo {author} {\bibfnamefont {J.}~\bibnamefont {Gemmer}}, \bibinfo {author} {\bibfnamefont {M.}~\bibnamefont {Michel}},\ and\ \bibinfo {author} {\bibfnamefont {G.}~\bibnamefont {Mahler}},\ }\href {https://doi.org/https://doi.org/10.1007/978-3-540-70510-9} {\emph {\bibinfo {title} {Quantum thermodynamics: Emergence of thermodynamic behavior within composite quantum systems}}},\ Vol.\ \bibinfo {volume} {784}\ (\bibinfo  {publisher} {Springer},\ \bibinfo {year} {2009})\BibitemShut {NoStop}%
\bibitem [{\citenamefont {Ahmadi}\ \emph {et~al.}(2023)\citenamefont {Ahmadi}, \citenamefont {Salimi},\ and\ \citenamefont {Khorashad}}]{ahmadi2023work}%
  \BibitemOpen
  \bibfield  {author} {\bibinfo {author} {\bibfnamefont {B.}~\bibnamefont {Ahmadi}}, \bibinfo {author} {\bibfnamefont {S.}~\bibnamefont {Salimi}},\ and\ \bibinfo {author} {\bibfnamefont {A.}~\bibnamefont {Khorashad}},\ }\bibfield  {title} {\bibinfo {title} {On the contribution of work or heat in exchanged energy via interaction in open bipartite quantum systems},\ }\href {https://doi.org/https://doi.org/10.1038/s41598-022-27156-0} {\bibfield  {journal} {\bibinfo  {journal} {Scientific Reports}\ }\textbf {\bibinfo {volume} {13}},\ \bibinfo {pages} {160} (\bibinfo {year} {2023})}\BibitemShut {NoStop}%
\bibitem [{\citenamefont {Yang}\ \emph {et~al.}(2019)\citenamefont {Yang}, \citenamefont {Elouard}, \citenamefont {Splettstoesser}, \citenamefont {Sothmann}, \citenamefont {S\'anchez},\ and\ \citenamefont {Jordan}}]{yang2019thermal}%
  \BibitemOpen
  \bibfield  {author} {\bibinfo {author} {\bibfnamefont {J.}~\bibnamefont {Yang}}, \bibinfo {author} {\bibfnamefont {C.}~\bibnamefont {Elouard}}, \bibinfo {author} {\bibfnamefont {J.}~\bibnamefont {Splettstoesser}}, \bibinfo {author} {\bibfnamefont {B.}~\bibnamefont {Sothmann}}, \bibinfo {author} {\bibfnamefont {R.}~\bibnamefont {S\'anchez}},\ and\ \bibinfo {author} {\bibfnamefont {A.~N.}\ \bibnamefont {Jordan}},\ }\bibfield  {title} {\bibinfo {title} {Thermal transistor and thermometer based on coulomb-coupled conductors},\ }\href {https://doi.org/10.1103/PhysRevB.100.045418} {\bibfield  {journal} {\bibinfo  {journal} {Phys. Rev. B}\ }\textbf {\bibinfo {volume} {100}},\ \bibinfo {pages} {045418} (\bibinfo {year} {2019})}\BibitemShut {NoStop}%
\bibitem [{\citenamefont {Ekanayake}\ \emph {et~al.}(2023)\citenamefont {Ekanayake}, \citenamefont {Gunapala},\ and\ \citenamefont {Premaratne}}]{PhysRevB.108.235421}%
  \BibitemOpen
  \bibfield  {author} {\bibinfo {author} {\bibfnamefont {U.~N.}\ \bibnamefont {Ekanayake}}, \bibinfo {author} {\bibfnamefont {S.~D.}\ \bibnamefont {Gunapala}},\ and\ \bibinfo {author} {\bibfnamefont {M.}~\bibnamefont {Premaratne}},\ }\bibfield  {title} {\bibinfo {title} {Stochastic model of noise for a quantum thermal transistor},\ }\href {https://doi.org/10.1103/PhysRevB.108.235421} {\bibfield  {journal} {\bibinfo  {journal} {Phys. Rev. B}\ }\textbf {\bibinfo {volume} {108}},\ \bibinfo {pages} {235421} (\bibinfo {year} {2023})}\BibitemShut {NoStop}%
\bibitem [{\citenamefont {Sánchez}\ \emph {et~al.}(2015)\citenamefont {Sánchez}, \citenamefont {Sothmann},\ and\ \citenamefont {Jordan}}]{Sánchez_2015}%
  \BibitemOpen
  \bibfield  {author} {\bibinfo {author} {\bibfnamefont {R.}~\bibnamefont {Sánchez}}, \bibinfo {author} {\bibfnamefont {B.}~\bibnamefont {Sothmann}},\ and\ \bibinfo {author} {\bibfnamefont {A.~N.}\ \bibnamefont {Jordan}},\ }\bibfield  {title} {\bibinfo {title} {Heat diode and engine based on quantum hall edge states},\ }\href {https://doi.org/10.1088/1367-2630/17/7/075006} {\bibfield  {journal} {\bibinfo  {journal} {New Journal of Physics}\ }\textbf {\bibinfo {volume} {17}},\ \bibinfo {pages} {075006} (\bibinfo {year} {2015})}\BibitemShut {NoStop}%
\bibitem [{\citenamefont {Tesser}\ \emph {et~al.}(2022)\citenamefont {Tesser}, \citenamefont {Bhandari}, \citenamefont {Erdman}, \citenamefont {Paladino}, \citenamefont {Fazio},\ and\ \citenamefont {Taddei}}]{Tesser_2022}%
  \BibitemOpen
  \bibfield  {author} {\bibinfo {author} {\bibfnamefont {L.}~\bibnamefont {Tesser}}, \bibinfo {author} {\bibfnamefont {B.}~\bibnamefont {Bhandari}}, \bibinfo {author} {\bibfnamefont {P.~A.}\ \bibnamefont {Erdman}}, \bibinfo {author} {\bibfnamefont {E.}~\bibnamefont {Paladino}}, \bibinfo {author} {\bibfnamefont {R.}~\bibnamefont {Fazio}},\ and\ \bibinfo {author} {\bibfnamefont {F.}~\bibnamefont {Taddei}},\ }\bibfield  {title} {\bibinfo {title} {Heat rectification through single and coupled quantum dots},\ }\href {https://doi.org/10.1088/1367-2630/ac53b8} {\bibfield  {journal} {\bibinfo  {journal} {New Journal of Physics}\ }\textbf {\bibinfo {volume} {24}},\ \bibinfo {pages} {035001} (\bibinfo {year} {2022})}\BibitemShut {NoStop}%
\bibitem [{\citenamefont {Palafox}\ \emph {et~al.}(2022)\citenamefont {Palafox}, \citenamefont {Rom{\'a}n-Ancheyta}, \citenamefont {{\c{C}}akmak},\ and\ \citenamefont {M{\"u}stecapl{\i}o{\u{g}}lu}}]{Palafox_2022}%
  \BibitemOpen
  \bibfield  {author} {\bibinfo {author} {\bibfnamefont {S.}~\bibnamefont {Palafox}}, \bibinfo {author} {\bibfnamefont {R.}~\bibnamefont {Rom{\'a}n-Ancheyta}}, \bibinfo {author} {\bibfnamefont {B.}~\bibnamefont {{\c{C}}akmak}},\ and\ \bibinfo {author} {\bibfnamefont {{\"O}.~E.}\ \bibnamefont {M{\"u}stecapl{\i}o{\u{g}}lu}},\ }\bibfield  {title} {\bibinfo {title} {Heat transport and rectification via quantum statistical and coherence asymmetries},\ }\href {https://doi.org/10.1103/PhysRevE.106.054114} {\bibfield  {journal} {\bibinfo  {journal} {Phys. Rev. E}\ }\textbf {\bibinfo {volume} {106}},\ \bibinfo {pages} {054114} (\bibinfo {year} {2022})}\BibitemShut {NoStop}%
\bibitem [{\citenamefont {Haikka}\ and\ \citenamefont {Maniscalco}(2010)}]{PhysRevA.81.052103}%
  \BibitemOpen
  \bibfield  {author} {\bibinfo {author} {\bibfnamefont {P.}~\bibnamefont {Haikka}}\ and\ \bibinfo {author} {\bibfnamefont {S.}~\bibnamefont {Maniscalco}},\ }\bibfield  {title} {\bibinfo {title} {Non-markovian dynamics of a damped driven two-state system},\ }\href {https://doi.org/10.1103/PhysRevA.81.052103} {\bibfield  {journal} {\bibinfo  {journal} {Phys. Rev. A}\ }\textbf {\bibinfo {volume} {81}},\ \bibinfo {pages} {052103} (\bibinfo {year} {2010})}\BibitemShut {NoStop}%
\bibitem [{\citenamefont {Diehl}\ \emph {et~al.}(2008)\citenamefont {Diehl}, \citenamefont {Micheli}, \citenamefont {Kantian}, \citenamefont {Kraus}, \citenamefont {B{\"u}chler},\ and\ \citenamefont {Zoller}}]{diehl2008quantum}%
  \BibitemOpen
  \bibfield  {author} {\bibinfo {author} {\bibfnamefont {S.}~\bibnamefont {Diehl}}, \bibinfo {author} {\bibfnamefont {A.}~\bibnamefont {Micheli}}, \bibinfo {author} {\bibfnamefont {A.}~\bibnamefont {Kantian}}, \bibinfo {author} {\bibfnamefont {B.}~\bibnamefont {Kraus}}, \bibinfo {author} {\bibfnamefont {H.~P.}\ \bibnamefont {B{\"u}chler}},\ and\ \bibinfo {author} {\bibfnamefont {P.}~\bibnamefont {Zoller}},\ }\bibfield  {title} {\bibinfo {title} {Quantum states and phases in driven open quantum systems with cold atoms},\ }\href {https://doi.org/10.1038/nphys1073} {\bibfield  {journal} {\bibinfo  {journal} {Nature Physics}\ }\textbf {\bibinfo {volume} {4}},\ \bibinfo {pages} {878} (\bibinfo {year} {2008})}\BibitemShut {NoStop}%
\bibitem [{\citenamefont {Reynaud}\ \emph {et~al.}(1988)\citenamefont {Reynaud}, \citenamefont {Dalibard},\ and\ \citenamefont {Cohen-Tannoudji}}]{reynaud1988photon}%
  \BibitemOpen
  \bibfield  {author} {\bibinfo {author} {\bibfnamefont {S.}~\bibnamefont {Reynaud}}, \bibinfo {author} {\bibfnamefont {J.}~\bibnamefont {Dalibard}},\ and\ \bibinfo {author} {\bibfnamefont {C.}~\bibnamefont {Cohen-Tannoudji}},\ }\bibfield  {title} {\bibinfo {title} {Photon statistics and quantum jumps: the picture of the dressed atom radiative cascade},\ }\href {https://doi.org/10.1109/3.979} {\bibfield  {journal} {\bibinfo  {journal} {IEEE Journal of Quantum Electronics}\ }\textbf {\bibinfo {volume} {24}},\ \bibinfo {pages} {1395} (\bibinfo {year} {1988})}\BibitemShut {NoStop}%
\bibitem [{\citenamefont {Gubaydullin}\ \emph {et~al.}(2022)\citenamefont {Gubaydullin}, \citenamefont {Thomas}, \citenamefont {Golubev}, \citenamefont {Lvov}, \citenamefont {Peltonen},\ and\ \citenamefont {Pekola}}]{gubaydullin2022photonic}%
  \BibitemOpen
  \bibfield  {author} {\bibinfo {author} {\bibfnamefont {A.}~\bibnamefont {Gubaydullin}}, \bibinfo {author} {\bibfnamefont {G.}~\bibnamefont {Thomas}}, \bibinfo {author} {\bibfnamefont {D.~S.}\ \bibnamefont {Golubev}}, \bibinfo {author} {\bibfnamefont {D.}~\bibnamefont {Lvov}}, \bibinfo {author} {\bibfnamefont {J.~T.}\ \bibnamefont {Peltonen}},\ and\ \bibinfo {author} {\bibfnamefont {J.~P.}\ \bibnamefont {Pekola}},\ }\bibfield  {title} {\bibinfo {title} {Photonic heat transport in three terminal superconducting circuit},\ }\href {https://doi.org/10.1038/s41467-022-29078-x} {\bibfield  {journal} {\bibinfo  {journal} {Nature Communications}\ }\textbf {\bibinfo {volume} {13}},\ \bibinfo {pages} {1552} (\bibinfo {year} {2022})}\BibitemShut {NoStop}%
\end{thebibliography}%

\end{document}